\documentclass[preprint, 12pt, numafflabel]{elsarticle}

\usepackage{array}
\usepackage[title, titletoc]{appendix}
\setcitestyle{square}
\usepackage{adjustbox}
\usepackage{tabularx}
\usepackage{afterpage}

\makeatletter
\g@addto@macro\@floatboxreset\centering

\makeatother
\usepackage{fancyhdr}
\usepackage[utf8]{inputenc}
\usepackage{graphicx}
\usepackage{amsmath}
\usepackage{subcaption}
\usepackage{upgreek}
\usepackage{booktabs}
\usepackage{multirow}

\usepackage{pdflscape}
\usepackage{rotating}

\usepackage{hyperref}
\hypersetup{colorlinks,allcolors=black}
\usepackage[top=2cm, bottom=2cm, left=2cm, right=2cm]{geometry}

\newcommand{\exvivo}{\textit{ex vivo}}
\newcommand{\Exvivo}{\textit{Ex vivo}}
\newcommand{\invivo}{\textit{in vivo}}

\newcommand{\includepageasasubfigure}[1]{\begin{subfigure}{\linewidth}
    \includegraphics[width=\textwidth,page=#1]{images/materials_and_methods/ALL_SAMPLES.pdf}                
\end{subfigure}}

\usepackage{amssymb}

\usepackage{lineno}
\begin{document}
    \journal{Journal of Bone and Mineral Research}
    \begin{frontmatter}

        \title{Ultrasound imaging of cortical bone: cortex geometry and measurement of porosity based on wave speed for bone remodeling estimation}

 \affiliation[lib]{organization={Sorbonne Universite, INSERM, CNRS, Laboratoire d'Imagerie Biomedicale, LIB, F-75006}, city={Paris}, country={France}}

        \affiliation[delft]{organization={Department of Imaging Physics, Delft University of Technology}, country={The Netherlands}}
        \author[lib]{Amadou S. DIA}

        \author[lib,delft]{Guillaume RENAUD}
        \author[lib]{Christine CHAPPARD}
\author[lib]{Quentin GRIMAL\texorpdfstring{\footnote{Corrresponding author: quentin.grimal@bsorbonne-universite.fr}{}}}

    \end{frontmatter}

It has been suggested that US imaging can be utilized to assess cortical bone health, which is of particular interest due to its major role in bone mechanical stability.
Intracortical US imaging extends B-mode imaging into bone using a dedicated image reconstruction algorithm that corrects for refraction at the bone-soft tissue interfaces. It has shown promising results in a few healthy, predominantly young adults, providing anatomical images of the cortex (periosteal and endosteal surfaces) along with estimations of US wave speed.
However, its reliability in older or osteoporotic bones remains uncertain. In this study, we critically assessed the performance of intracortical US imaging ex vivo in bones with various microstructural patterns, including bones exhibiting signs of unbalanced intracortical remodeling.
We analyzed factors influencing US image quality, particularly endosteal surface reconstruction, as well as the accuracy of wave speed estimation and its relationship with porosity.
We imaged 20 regions of interest from the femoral diaphysis of five elderly donors using a 2.5 MHz US transducer. The reconstructed US images were compared to site-matched high-resolution micro-CT (HR-$\upmu$CT) images. In samples with moderate porosity, the endosteal surface was accurately identified, and thickness estimates from US and HR-$\upmu$CT differed by less than 10\%.
In highly remodeled bones with increased porosity, the reconstructed endosteal surface appeared less bright and was located above the cortex region containing resorption cavities.
We observed a decrease in US wave speed with increasing cortical porosity, aligning well with literature data, suggesting that the method could discriminate between bones with low porosity (less than 5\%) and those with moderate to high porosity (greater than ~10\%).
This study paves the way for the application of US imaging in diagnosing cortical bone health, particularly for detecting increased cortical porosity and reduced cortical thickness.

\section*{Key words}
Osteoporosis, Cortical bone, Microstructure, Cortical Porosity, Cortical Thickness, Ultrasound Imaging, Endosteal surface
\section*{Lay summary}
While conventional ultrasound (US) imaging cannot visualize the inside of bones, Intracortical US imaging uses a specialized reconstruction algorithm to examine the cortex of long bones. Previously validated in young adults, this technique provides anatomical images of the cortex along with estimates of ultrasound wave speed. In this study, we demonstrate that Intracortical US imaging can assess bone geometry and cortical thickness in elderly donors with signs of osteoporosis. Additionally, it can differentiate between low-porosity and high-porosity bones, offering valuable insights into bone health. Ultrasound is a portable, real-time, and radiation-free imaging method. By integrating Intracortical US imaging into multipurpose US scanners, this technique could become a widely accessible tool for evaluating cortical bone health.

\clearpage
\section{Introduction}

The nine million fragility fractures occurring globally each year \cite{cooper_iof_2017} represent a significant medical challenge. The prediction of fracture risk traditionally relies on clinical factors and areal bone mineral density (aBMD) measurements with dual energy X-ray absorptiometry (DXA). However, many individuals who are at high risk of fracture are not identified using aBMD \cite{siris_bone_2004, briot_frax_2013}. Additional bone factors not captured by DXA contribute to bone strength, such as bone microstructure and cortical bone properties. 
Cortical bone is of special interest as it plays a major role in bone mechanical stability and about 80\% of the fragility fractures involve regions comprising large amounts of cortical bone in the appendicular skeleton \cite{bala_role_2015}. Also, it has been suggested that bone loss in old age is mainly cortical and is due to the reduction of cortical thickness and the increase of porosity (pore volume fraction) \cite{zebaze_intracortical_2010}. High-resolution peripheral computed tomography (HR-pQCT) can assess cortical thickness and volumetric BMD, indirectly reflecting cortical porosity \cite{ostertag_cortical_2014}. But HR-pQCT is only available in a few hospitals and is mainly used for clinical research. 

Diagnostic techniques based on ultrasound (US) are very attractive due to the absence of ionizing radiation and their ability to provide real-time measurement with mobile equipment at a significantly lower cost than other medical imaging modalities \cite{hans_quantitative_2022}. 
Over the past few decades, quantitative US (QUS) methods have been developed to assess cortical bone employing specialized US devices to capture specific modes of US propagation in bone. These include waves transmitted through the wrist \cite{sai_novel_2010} or phalanges \cite{sakata_assessing_2004}, guided waves in the cortex along the axis of the tibia or radius \cite{minonzio_bone_2018} or echoes from cortex boundaries \cite{karjalainen_ultrasonic_2008}. These devices provide US parameters reflecting cortical thickness and intrinsic bone properties indicative of tissue mechanical quality, such as US bulk wave speeds in various anatomical directions \cite{foiret_combined_2014}.
One approach provides a surrogate measure of porosity \cite{minonzio_bone_2018,schneider_vivo_2019} based on the fact that apparent acoustical properties at the US wavelength scale in bone (approximately 1-3~mm) are strongly related to intracortical porosity \cite{grimal_quantitative_2019,peralta_bulk_2021}. 
These QUS techniques provide valuable insights into bone health, as cortical thickness and porosity are key parameters in assessing fracture risk\cite{augat_role_2006,bala_role_2015}. However, these techniques did not provide real-time anatomical images, which limits their use and creates reproducibility issues. Moreover, failure cases are not well understood due to the lack of imaging.

Ultrasonic evaluation of bone properties assisted with imaging has been suggested as a tool for assessing bone health \cite{conversano_novel_2015,armbrecht_pore-size_2021}. However, the reconstruction of the inside of bones in conventional US images shows poor contrasts and is strongly artefacted due to high US attenuation and significant acoustical impedance mismatch between bone and surrounding soft tissues \cite{bianchi_ultrasound_2007}. Renaud et al. \cite{renaud_vivo_2018} and Nguyen Minh et al. \cite{nguyen_minh_estimation_2020} have demonstrated that dedicated US sequences and image reconstruction methods can be used together with conventional array probes to overcome these barriers and enable, for instance, to measure the thickness of cortical bone. US tomography can also in principle be used to image the interior of cortical bone but such approaches have not yet been implemented \invivo~for human bone \cite{bernard_ultrasonic_2017,li_ex-vivo_2021}.
Specifically, the intracortical US imaging approach by Renaud et al. extends B-mode imaging into the bone by employing a delay-and-sum image reconstruction algorithm including correction of refraction at the bone-soft tissue interfaces \cite{renaud_vivo_2018}. This technique provides anatomical images of the cortex (periosteal and endosteal surfaces) and the medullary cavity, along with estimations of US wave speeds and their anisotropy \cite{renaud_measuring_2020}. This approach has shown promising results in a few healthy, mostly young adults, with measured cortical thickness aligning well with reference values obtained through HR-pQCT and US wave speeds in good agreement with literature. 
It remains uncertain whether the images obtained using this method in older or osteoporotic subjects would be of sufficient quality for a reliable analysis. 
Unbalanced intracortical remodeling in aging and osteoporosis leads to microstructure degradation, characterized by large pores (clustered remodeling cavities) \cite{bell_regional_1999,andreasen_understanding_2018}, a radial gradient of tissue porosity from the periosteal to the endosteal surfaces, and a trabecularization the inner cortex forming a transitional zone ressembling trabecular bone \cite{zebaze_intracortical_2010}. 
In cortical bone, porosity increases with age, from about 5\% at 30 years old to 15\% at 80 years old in females \cite{bousson_distribution_2001}, accompanied by an increase in pore diameter \cite{cooper_age-dependent_2007}. The altered intracortical microstructure in older and osteoporotic bones may significantly impact US images. 
A numerical study has shown that strong diffuse scattering by numerous and moderately large pores leads to increased attenuation and speckle, hindering the visibility of the endosteal surface \cite{dia_influence_2023}. Furthermore, individual very large pores have a large contribution to scattering \cite{iori_estimation_2020}, posing a major challenge for US imaging  of the cortex in older, osteopenic, and osteoporotic individuals. 

The aim of this study was to critically assess the performance of bone intracortical US imaging in bones with altered microstructure, characterized by signs of unbalanced remodeling, as commonly observed in aging and osteoporotic individuals. We aimed to identify the factors influencing the quality of US images of the cortex and, specifically, of the reconstruction of the endosteal surface which enables to measure cortical thickness. An ancillary objective was to validate the estimation of US wave speed in human bones representative of an osteoporotic condition, and to discuss the relationship between this wave speed and porosity.
For these purposes, we conducted an {\exvivo} study with bones from elderly human donors as surrogates for osteoporotic bones. US image reconstructions were compared to high-resolution X-ray micro-computed tomography (HR-$\upmu$CT) images serving as reference. US images were reconstructed using a delay-and-sum algorithm including a correction of refraction at the bone-soft tissue interface, similar to that used {\invivo} in young, healthy individuals \cite{renaud_vivo_2018}.

\section{Methods}

\subsection{Samples}
\begin{figure}[htb!]
    \centering
    \includegraphics[width=.6\textwidth]{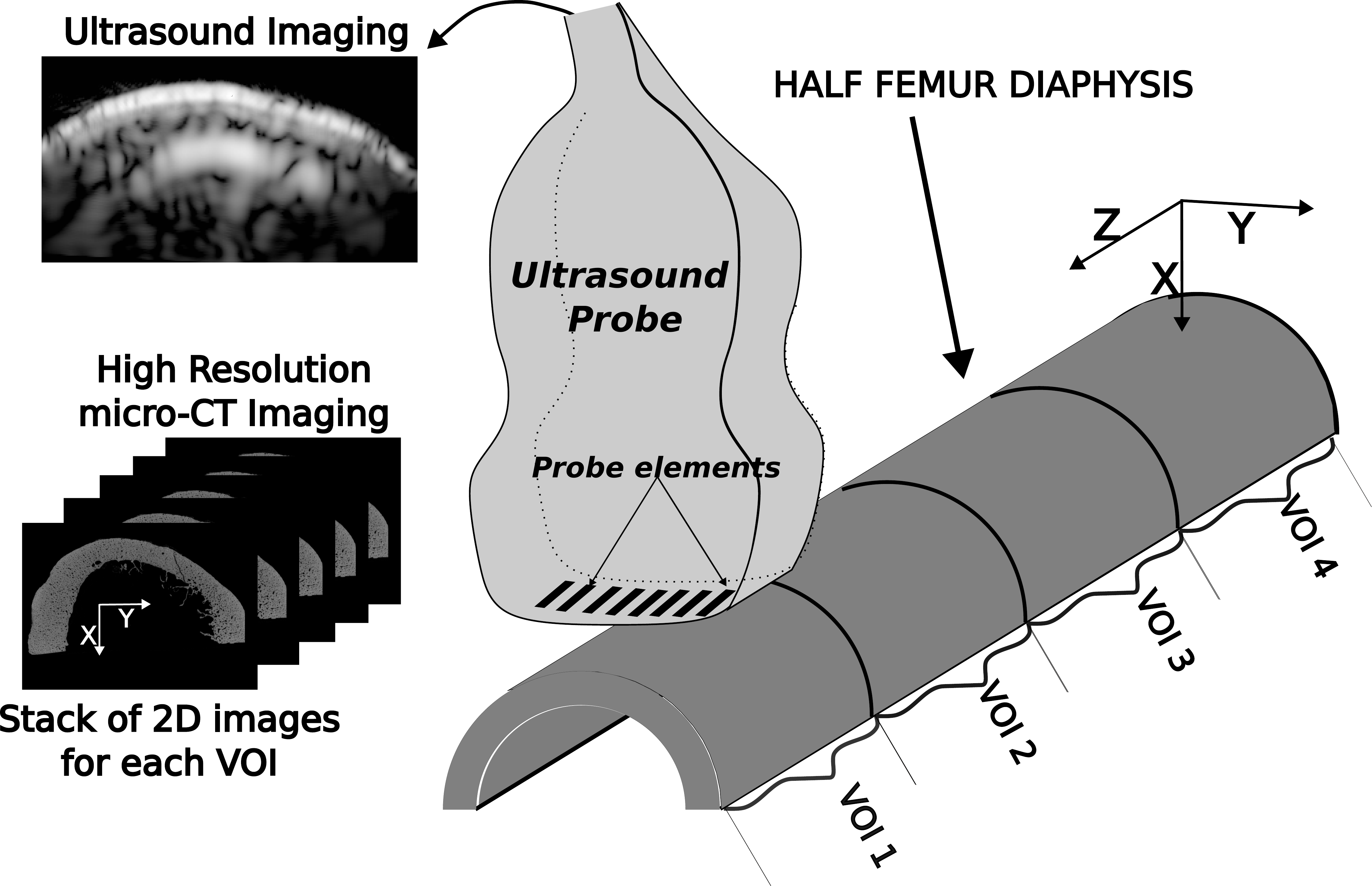}
    \caption{Experimental protocol. A half-cylinder femur sample was immersed in physiological saline solution and scanned in a plane perpendicular to the bone axis using a conventional ultrasound array transducer operating at a central frequency of 2.5 MHz. Due to the probe's elevation, four Volumes of Interest (VOI) were defined. Ten US acquisitions were performed for each region, with slight lateral ($y$-direction) or rotational adjustments of the probe between repetitions to vary the imaging area. For each VOI, a stack of 2D slices was obtained using high-resolution X-ray micro-computed tomography (voxel size: 8.8~$\upmu$m), which served as reference for image comparison.}
    \label{method:protocol}
\end{figure}

Bone samples were obtained from the femurs of five female donors. Ethical approval for sample collection was granted by the Human Ethics Committee of the Centre du don des Corps at the University Paris Descartes (Paris, France).
Informed written consent was provided by the tissue donors or their legal guardians, in accordance with the legal requirements outlined in the French Code of Public Health.
The fresh bone material, stored at -20°C, was sawed, and soft tissues were totally removed. For each femur, a half-cylinder measuring 7 cm in length was extracted from the region between the proximal third and the mid-diaphysis. The samples were subsequently fixed in alcoholic formalin-free fixative F13 (Morphisto, Germany) for 48 hours and rinsed in distilled water for 6 hours using an ultrasonic bath. Four regions of interest, each approximately 1.5 cm in length, were delineated in each sample for subsequent measurements (Figure~\ref{method:protocol}).
In the following we refer to the samples using a number. The age of the donors were 78, 78, 80, 66,  and 86 years old for samples 1 to 5, respectively.

\subsection{High-resolution X-ray micro-computed tomography (HR-\textmu CT) imaging and definition of volumes of interest} 
The samples were scanned using HR-\textmu CT using a Skyscan 1176 (SkyScan-Brücker, Kontich, Belgium) equipped with a scintillator coupled to a CCD Princeton camera with a pixel size of 12.53~$\upmu$m and a dynamic range of 16 bits. Scanning was performed at the maximum potential of 90~kV and 278~$\upmu$A. Total angular rotation and rotation step were set to 197.10° and 0.3°, respectively. Because of the large length of the samples, the HR-\textmu CT acquisition for one sample was done for four slightly overlapping regions. The total duration of the HR-\textmu CT acquisitions was about 6 hours per sample. Image reconstruction with a voxel size of 8.79~\textmu m was performed with the Feldkamp filtered back-projection algorithm (NRecon,  version 1.7.4.6, SkyScan-Brücker, Kontich, Belgium). For each sample, appropriate beam hardening and ring artifact corrections were applied. 

Four volumes of interest (VOI) were defined within the three-dimensional (3D) HR-\textmu CT image of each sample.
The thickness (direction of the bone axis) and width  of the VOIs were 15 and 20~mm, respectively.
The height of the VOIs  varied between 8 and 10 mm, depending on the cortical thickness and diameter of each sample. Figure~\ref{method:example_xray_slice} shows a two-dimensional (2D) slice in the transverse plane extracted from each VOI for all the samples.
The VOI's dimensions approximately correspond to the thickness (probe elevation direction),
lateral dimension (probe aperture) 
and depth of the US image.

         \afterpage{\begin{landscape}
        \begin{figure}[htb!]
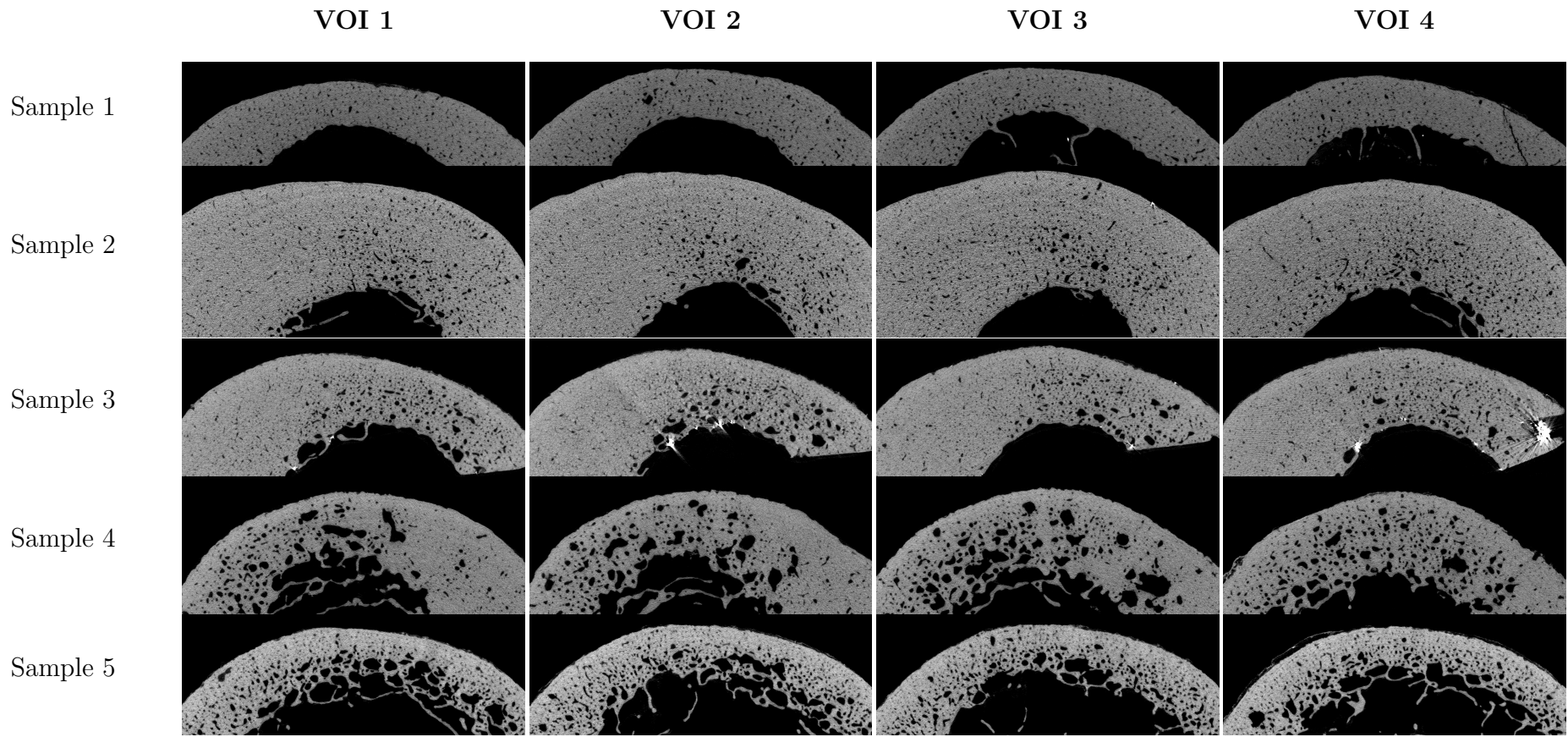

            \setlength\tabcolsep{1pt}
            \renewcommand{\arraystretch}{0.1} 
            \begin{tabularx}{\linewidth}{m{.11\linewidth}m{.225\linewidth}m{.225\linewidth}m{.225\linewidth}m{.225\linewidth}}
                & \centerline{\bf VOI 1} &\centerline{\bf VOI 2}&\centerline{\bf VOI 3}&\centerline{\bf VOI 4}\\
                    Sample 1                            
                &\includepageasasubfigure{1}
                &\includepageasasubfigure{2}
                &\includepageasasubfigure{3}
                &\includepageasasubfigure{4}\\
                    Sample 2                            
                &\includepageasasubfigure{13}
                &\includepageasasubfigure{14}
                &\includepageasasubfigure{15}
                &\includepageasasubfigure{16}\\
                    Sample 3                            
                &\includepageasasubfigure{5}
                &\includepageasasubfigure{6}
                &\includepageasasubfigure{7}
                &\includepageasasubfigure{8}\\
                    Sample 4                            
                &\includepageasasubfigure{9}
                &\includepageasasubfigure{10}
                &\includepageasasubfigure{11}
                &\includepageasasubfigure{12}\\
                    Sample 5                            
                &\includepageasasubfigure{17}
                &\includepageasasubfigure{18}
                &\includepageasasubfigure{19}
                &\includepageasasubfigure{20}\\
                \end{tabularx}
                \caption{Reference X-ray images of all samples. An example slice is shown for each Volume Of Interest (VOI).}
                \label{method:example_xray_slice}
        \end{figure}
        \end{landscape}}

\subsection{X-ray micro-computed tomography image processing: microstructure and cortical thickness} 
\label{reference_cortical_thickness}

We processed the micro-CT images to derive parameters that represent pore distribution and to measure cortical thickness which served as references for the assessment of US imaging. All processing steps 
were performed using CTAn software (CTAn; SkyScan-Brücker, Kontich, Belgium) except  where otherwise indicated. 

The initial step was to apply a Gaussian filter (round kernel of radius 2 or 3, depending on the sample) to remove acquisition noise. Then, the VOI were binarized using 3D Otsu method \cite{behrooz_automated_2017}. 
Given that this study focuses on the cortical bone tissue, we removed, in the endosteal region, the bone remnants resulting from a trabecularization \cite{zebaze_intracortical_2010} of the cortex (samples~4 and 5, see Figure~\ref{method:example_xray_slice} and Figure S1 in supplementary materials).
To do this, we used kernel filtering (either Kuwahara or Uniform kernels with a radius of 2) and morphological operations to close all pores below a specific threshold diameters, thereby generating a mask that excluded trabecular regions. The specific choice of kernel and parameters varied across samples to accommodate the high morphological variability among the samples (Figure~\ref{method:example_xray_slice}).

\begin{figure}[htb!]
        \centering
            \includegraphics[trim={0 250 0 250}, clip, width=.33\linewidth]{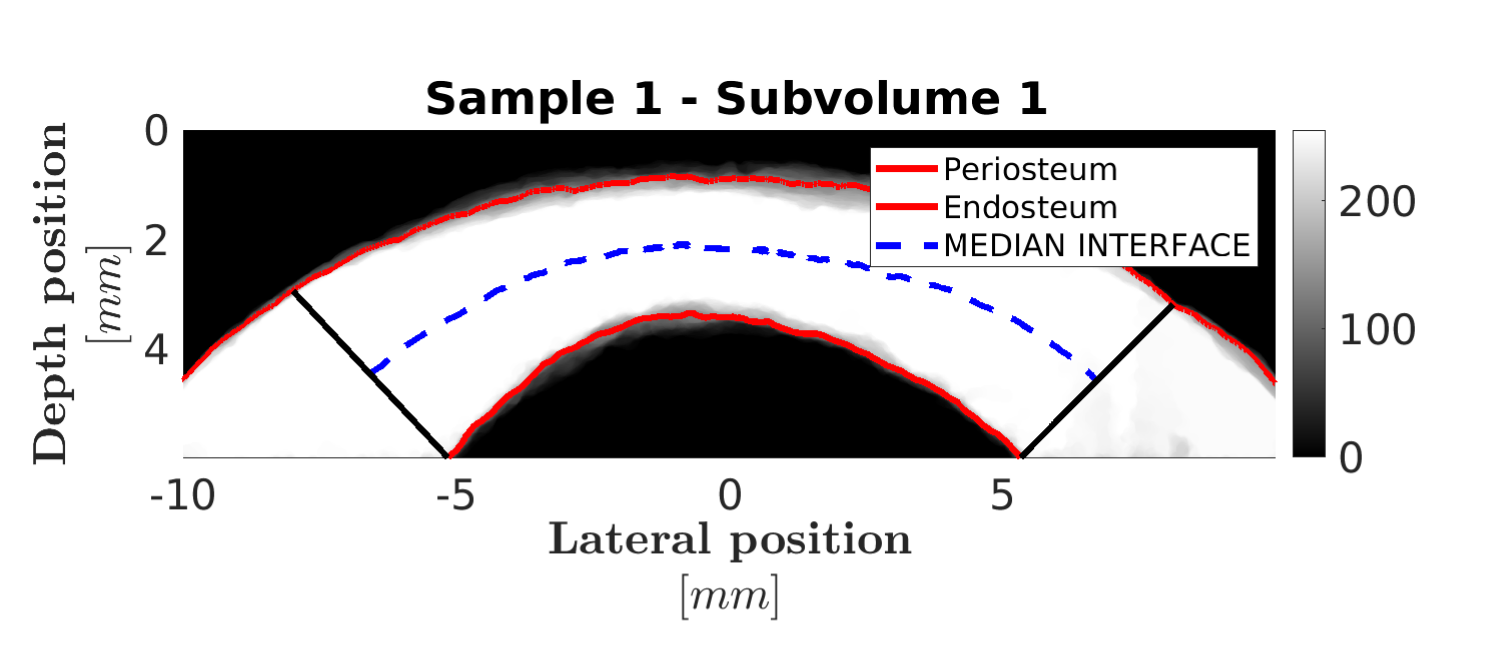}
            \includegraphics[trim={0 200 0 180}, clip, width=.33\linewidth]{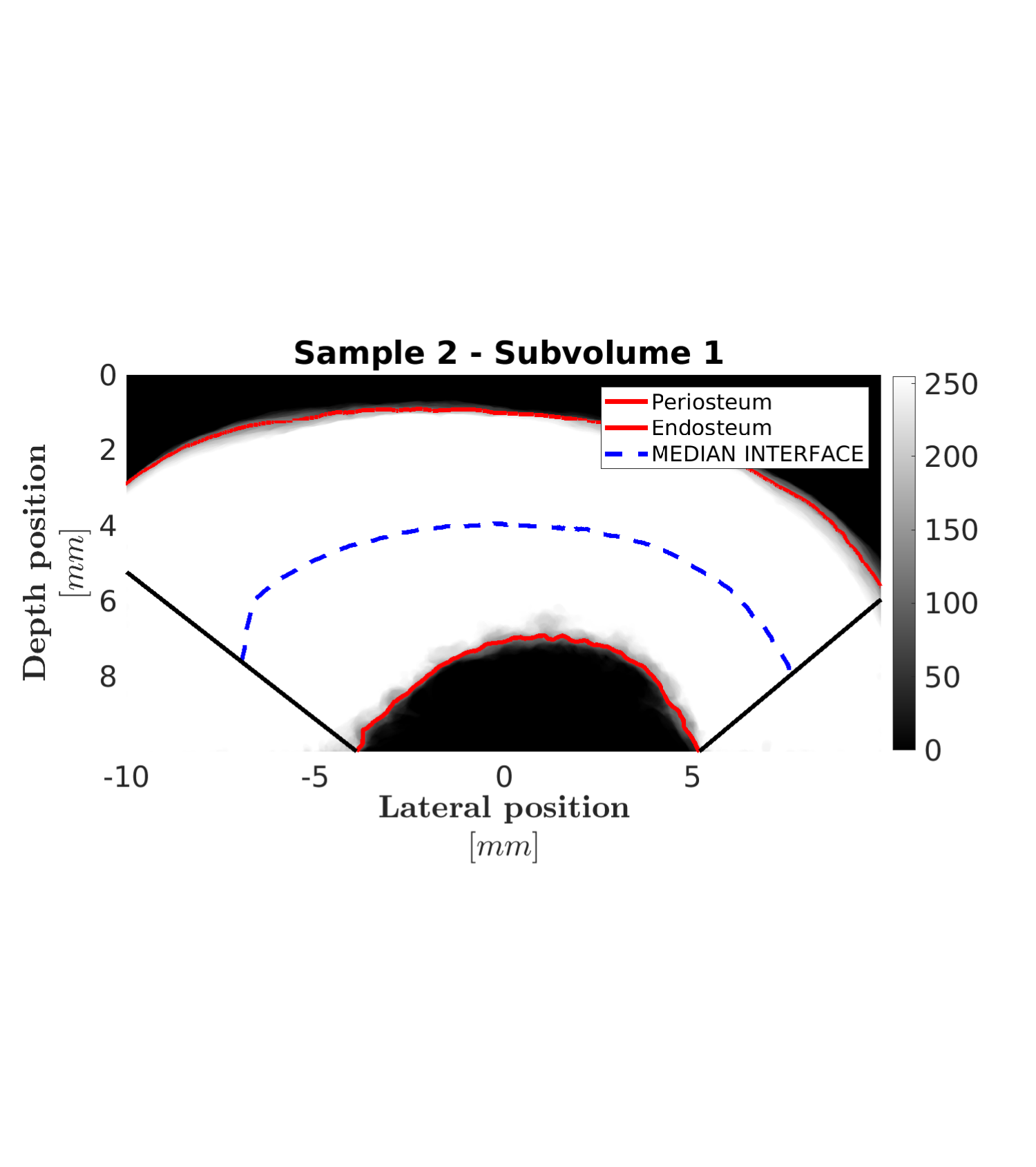}
            \includegraphics[trim={0 200 0 250}, clip, width=.33\linewidth]{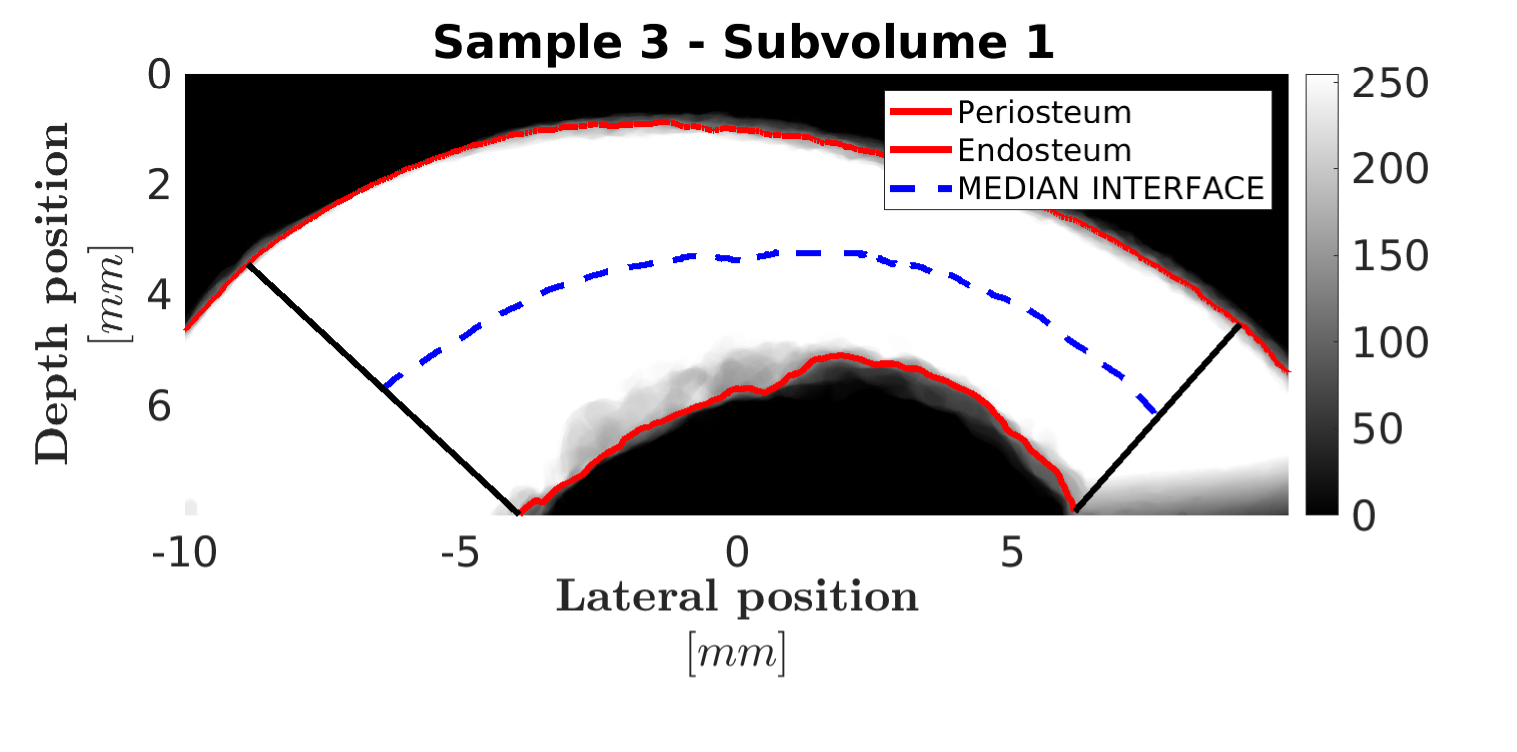}
            \includegraphics[trim={0 200 0 250}, clip, width=.33\linewidth]{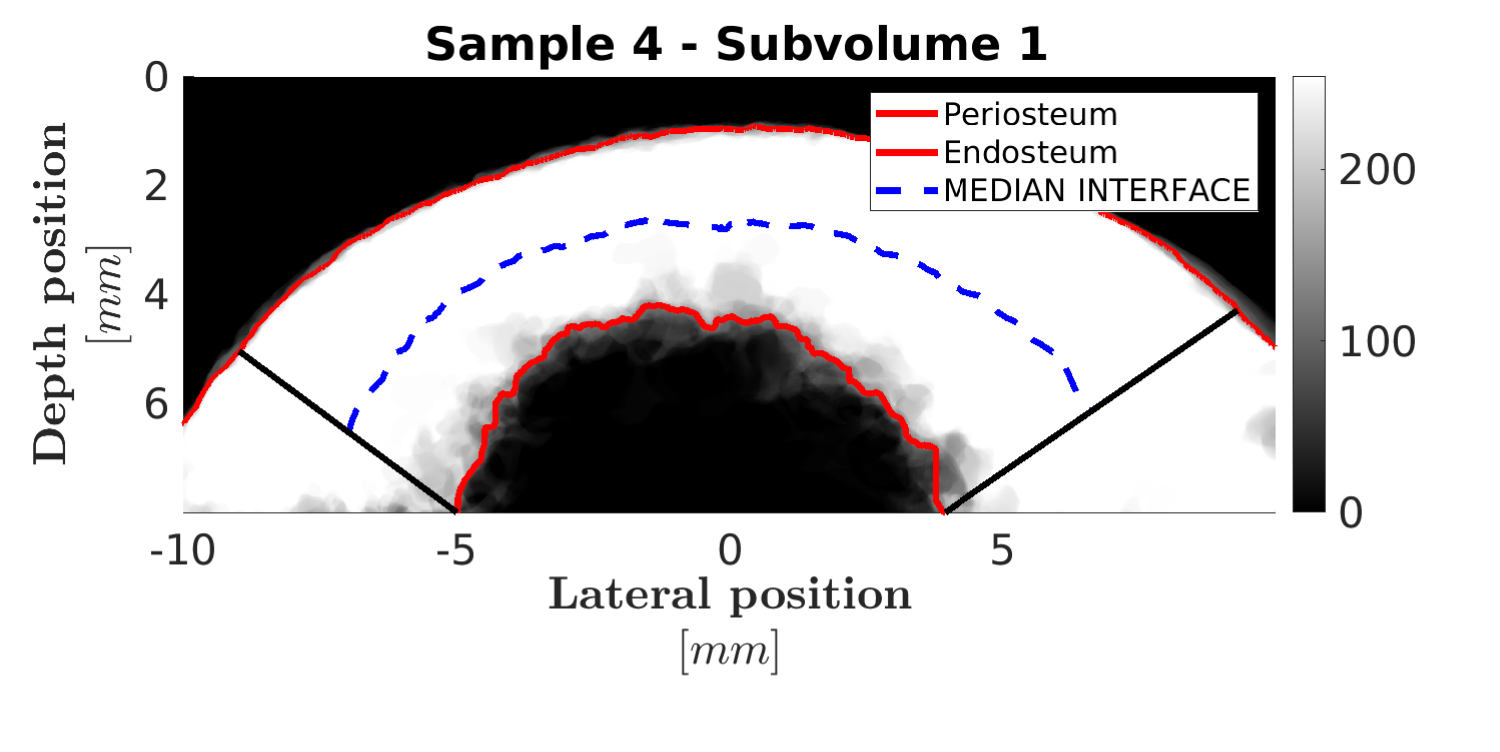}
            \includegraphics[trim={0 200 0 250}, clip, width=.33\linewidth]{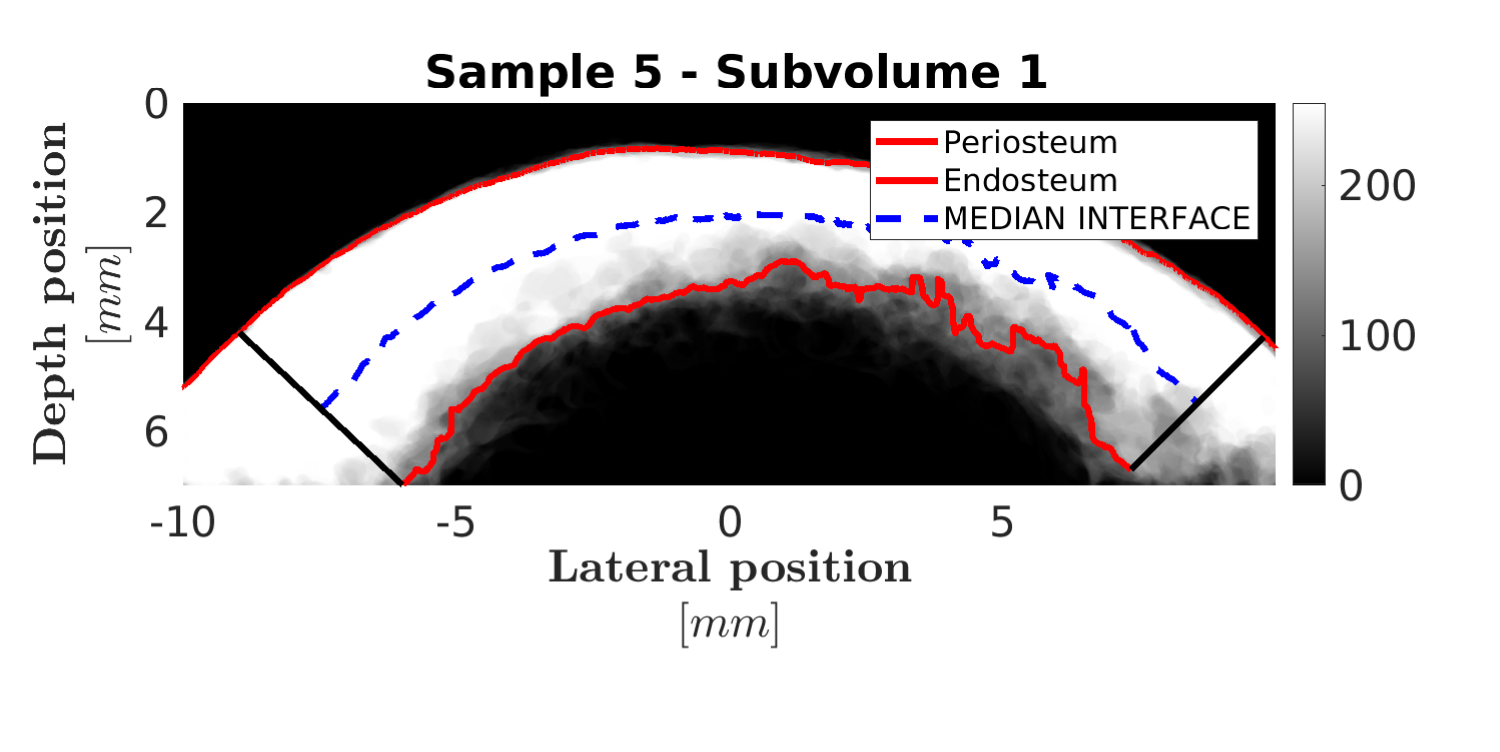}
        \caption{Illustration of the segmentation of the outer (periosteal) and inner (endosteal) surfaces. After segmentation and pore closing, the images in one VOI stack are averaged and segmented (red line). The blue dashed-line is the 'median' line at mid-distance of the periosteal and  endosteal surfaces. }
        \label{methods:average_masks}
    \end{figure}

An average cortical thickness was measured for each volume of interest (VOI), yielding four thickness values per sample,  using Matlab 2023b (Mathworks Inc., Natick, MA, USA). The procedure was as follows: (i)~The pores in the binarized 2D images were closed; (ii)~the resulting images in a VOI were averaged to produce a single 2D image representative of the VOI (Figure~\ref{methods:average_masks}) ; (iii)~the average image was binarized using Otsu method ; (iv)~the contours of the outer surface (periosteal) and inner surface (endosteal) were detected (Figure~\ref{methods:average_masks}) and  approximated by parabolas; and (v)~a 'median' line was determined at mid-distance from the two parabolas. 
Finally, cortical thickness was calculated for each region of interest as the ratio between the bone surface to the length of the median line. 
The average cortical thickness derived through this method will be compared to the cortical thickness measured with US.

We then calculated in each VOI the parameters representing the distribution of pores from the 3D binarized image stack of each VOI, using  the 3D analysis methods of CTAn software. The following parameters were obtained:
\begin{itemize}
    \item the 3D cortical porosity (Ct.Por), representing the percentage of pore volume relative to the total volume of the VOI;
    \item  the cortical pore density (Po.Dn, in units of pores/mm$^3$) was calculated as the number of pores divided by the total volume of the VOI;
    \item the statistical distribution of pore diameters was determined in each 2D image of one VOI stack using Matlab 2023b (Mathworks Inc., Natick, MA, USA). The diameter of each pore was calculated as the diameter of a disk of the same area. The pore distribution in the VOI was then obtained by pooling pore diameter distributions from each 2D slice. The distribution of pore diameters was finally characterized by the median value (Po.Dm), the 9$^{th}$ diameter decile (Dm.DC-9), the average diameter of large pores (Lg.Po.Dm) (all pores with a diameter greater than Dm.DC-9)  and the inter-decile range (Dm.IDR). These parameters were the same as those used in a previous study \cite{dia_influence_2023}.
\end{itemize}

\subsection{Ultrasound imaging} 

Due to the large sample sizes, a custom degassing protocol was developed to remove any air bubbles potentially trapped within the bone.
The setup was placed within a vacuum chamber connected to a pump. The bone sample was initially suspended in air and placed in contact with a plate connected to the vibrating diaphragm of an electrodynamic speaker driver. A water bath was also placed within the vacuum chamber. The degassing process involved two steps:  first, after air was pumped out, the sample underwent vibrations at 10 Hz for 40 minutes; second, it was then immersed in the water bath for an additional 20 minutes.
Both steps were conducted under vacuum (without opening the chamber), utilizing a magnet to initially suspend the sample in air and subsequently release it into the water bath. Then, throughout the US measurements, the samples remained immersed in water. 

US imaging was conducted using a fully programmable US system (Vantage, Verasonics Inc., Redmond, WA, USA). A phased array US transducer with 96 elements, operating at a central frequency of 2.5 MHz (P4-1 ATL/Philips, Bothell, WA, USA) with a pitch of 0.295~mm and an elevation aperture of 14~mm (US image thickness), was employed. The emitted pulse had a 80\% relative frequency bandwidth at -6 dB. 

The US probe was positioned approximately 4 mm above the outer bone surface at the center of each region of interest (Figure~\ref{method:protocol}).  Ten acquisitions were performed for each region, with slight lateral ($y$-direction) or rotational adjustments of the probe between repetitions to vary the imaging area. After each repositioning, probe placement was fine-tuned relying on the real-time visualization of intra-cortical bone images, as in \invivo~ acquisitions, to ensure clear visualization of the periosteal and endosteal surfaces.

We used a synthetic aperture imaging sequence with single-element transmissions (\cite{karaman_synthetic_1995, jensen_synthetic_2006}): each element of the array was activated sequentially, followed by the recording of the backscattered signals with all the elements.

\subsection{Ultrasound image reconstruction and processing}

We employed a delay-and-sum (DAS) algorithm optimized specifically for intra-osseous bone imaging, incorporating a correction for refraction at tissue interfaces and the adjustment of wave speed in water and bone for each measurement. 
A similar implementation of the method was previously used for \invivo~ measurements \cite{renaud_vivo_2018, salles_revealing_2021} and simulations \cite{dia_influence_2023}. 
For comprehensive details, readers may refer to \cite{renaud_single-sided_2022}.
The method is briefly described below. A Hanning window was applied to the receiver sub-aperture, and a constant f-number in receive of 0.5 with a receive angle of 0$^{\circ}$ was used throughout the image \cite{perrot_so_2021}. The 96 received datasets recorded for each transmission in the synthetic aperture sequence were reconstructed individually to obtain the so-called low-resolution images. These images were then coherently summed to obtain the so-called high-resolution image with optimal contrast \cite{jensen_synthetic_2006}. The reconstruction of each image was done sequentially in layers corresponding to water (between the probe and the periosteal surface) and the cortex. 
The method proceeded as follows: (i)~the wave speed in water was estimated from the head wave at the probe lens-water interface, yielding values between 1480 and 1492 m/s \cite{renaud_measuring_2020}. (ii)~an image was reconstructed with DAS using wave speed in water, from which the periosteal surface was segmented using Dijkstra's algorithm \cite{hong_medical_2012} and subsequently fitted with a parabola. 
(iii)-the image of the cortical bone region was then reconstructed using the travel-times of longitudinal waves in bone (shear waves were disregarded). For each array receive element and image pixel, Fermat's principle \cite{waltham_two-point_1988} was employed to calculate travel-times across a three-layer medium including the lens of the probe (front layer made of silicone rubber), water, and cortical bone. This accounted for refraction at the interface between the lens and water and between water and the periosteal surface. 
To determine the wave speed in cortical bone, we used an autofocus approach \cite{treeby_automatic_2011}. This approach identifies the US wave speed yielding the sharpest image by maximizing global energy and normalized pixel intensity variance. The selected wave speed serves as the best estimate of the true wave speed. Following \cite{granke_change_2011,dia_influence_2023}, the algorithm was supplied with a wave speed range from 2500 to 3500 m/s, in 25 m/s increments. This optimal image reconstruction, with wave speeds adjusted to each specific measurement, was conducted offline. For real-time imaging, the image reconstruction was done using nominal wave speeds values in water and cortical bone. (iv)~the final high-resolution image is generated by coherently summing the 96 low-resolution images.

Cortical thickness of the samples was measured from US images by segmenting the periosteal and endosteal surfaces using Dijkstra's algorithm. Following segmentation, surfaces were fitted with parabolas, and cortical thickness was calculated using the same method described for HR-\textmu CT imaging (see paragraph~\ref{reference_cortical_thickness}).

\subsection{Alignment of US and HR-\textmu CT images for comparative analysis}

To enable qualitative comparison between US and averaged 2D X-ray images, X-ray images were superimposed onto their US counterparts. Optimal rigid body rotation and translation matrices were determined to minimize the least squares errors between the raw segmented periosteal surfaces from each modality \cite{arun_least-squares_1987}. The final image overlays were created using Inkscape software (Inkscape project, Inkscape version 0.92.5).

\subsection{Reference data for wave speed in cortical bone}

We used data from a previous study \cite{cai_anisotropic_2019} to assess the accuracy of the wave speed derived from US imaging with the autofocus approach. Briefly, 52 cortical bone specimens were harvested from the femur diaphysis of 29 human cadavers (16 females and 13 males, aged 77.8 $\pm$ 11.4 years old).
The nominal dimensions of the specimens were $3\times 4 \times 5$~mm$^3$. 
Resonant ultrasound spectroscopy (RUS) was used to measure the elastic coefficients assuming that cortical bone is a transversely isotropic material. For the present study, we calculated the longitudinal wave speed in the transverse plane (perpendicular to the axis of the diaphysis) using the elastic coefficients and the mass density reported in \cite{cai_anisotropic_2019}. The porosity of these samples was determined from HR-\textmu CT images obtained with synchrotron radiation micro-computed tomography with a voxel size of 6.5~$\mathrm{\upmu}$m.

\section{Results}

\subsection{Bone thickness and microstructure from HR-\textmu CT data}
The 20 VOIs (4 VOIs in each of the 5 samples) showed a great diversity in terms of cortical thickness, porosity, and distribution of pores (Figure~
\ref{method:example_xray_slice}). 
Cortical thickness estimated from micro-CT images ranged from 2.5 to 6.3~mm (Table~\ref{table_chap_4:pore_stat_samples_combined}).
The microstructure parameters are summarized in Table~\ref{table_chap_4:pore_stat_samples_combined}. Samples 1 and 2 had the lowest porosity (5–6.4~\% and 5.3–6.7~\%, respectively) and no large pores (Lg.Po.Dm $<$ 175 $\upmu$m). Sample 3 showed moderate porosity (7–12.3~\%) with moderately large pores (Lg.Po.Dm in the range 224–228~$\upmu$m). Samples 4 and 5 presented the highest porosity (12.2–16~\% and 16.4–16.6~\%, respectively) with 10\% of the pores with diameter above 300 $\upmu$m.
Sample~4 displayed the largest pores (Lg.Po.Dm in the ranged 334–423 $\upmu$m), which can be clearly seen in Figure~\ref{method:example_xray_slice}.
Porosity values and the porosity gradient from the periosteal to the endosteal surface seen in some samples are consistent with the literature data for elderly donors \cite{bousson_distribution_2001}.

\afterpage{
\begin{landscape}
    \begin{table}[htb!]
    \centering
    \setlength\tabcolsep{2pt}
    \renewcommand{\arraystretch}{1.5} 
    \begin{adjustbox}{width=\linewidth}page
    \begin{tabular}{|c|cccc|cccc|cccc|cccc|cccc|}
    \toprule
     & \multicolumn{4}{c|}{Sample 1} & \multicolumn{4}{c|}{Sample 2} & \multicolumn{4}{c|}{Sample 3} & \multicolumn{4}{c|}{Sample 4} & \multicolumn{4}{c|}{Sample 5} \\
    \midrule
    ROI &  1 &  2 &  3 &  4 &  1 &  2 &  3 &  4 &  1 &  2 &  3 &  4 &  1 &  2 &  3 &  4 &  1 &  2 &  3 &  4 \\
    \midrule
    Ct.Th (mm) & 2.5 & 2.7 & 3.1 & 3.1 & 6.3 & 5.8 & 6.0 & 5.8 & 4.5 & 4.6 & 4.7 & 4.7 & 3.6 & 3.6 & 3.8 & 3.4 & 2.9 & 2.7 & 2.8 & 2.7\\
    Ct.Por (\%)         & 5.0  & 6.5  & 6.2  & 6.4  & 5.3  & 5.6  & 6.2  & 6.7  & 11.5 & 12.3 & 10.93 & 7.0  & 12.2 & 14.4 & 16.5 & 16.0 & 16.6 & 16.4 & 16.5 & 16.4 \\
    Po.Dn (mm$^{-3}$) & 10.7 & 11.7 & 12.4 & 14.4 & 9.7  & 9.4  & 9.9  & 10.6 & 10.8 & 11.9 & 14.3 & 17.1 & 8.0  & 7.1  & 6.8  & 7.9  & 12.5 & 13.0 & 12.8 & 13.1 \\
    Po.Dm ($\upmu$m)  & 56 & 57 & 55 & 52 & 55 & 56 & 57 & 55 & 57 & 53 & 53 & 50 & 50 & 59 & 65 & 64 & 49 & 50 & 52 & 52 \\
    Dm.DC-9 ($\upmu$m)   & 115& 127& 120& 111& 124& 126& 131& 136& 161 & 155 & 140& 134& 190& 210& 225& 218& 182& 187& 191& 190 \\
    Lg.Po.Dm ($\upmu$m)  & 155& 175& 165& 156& 159& 163& 168& 175& 268& 265& 227& 224& 334& 383& 423& 383& 307& 302& 303& 300 \\
    Dm.IDR ($\upmu$m)  & 91 & 101& 94 & 87 & 100  & 102& 107& 111& 134& 128& 114& 110& 166& 186& 201& 194& 157& 162& 166& 165 \\
    \bottomrule
    \end{tabular}
    \end{adjustbox}
    \caption{Reference cortical thickness (Ct.Th) and parameters of the cortical bone microstructure for all ROIs and samples.  Ct.Por : cortical porosity; Po.Dn cortical pore density;  Po.Dm, Dm.DC-9, Lg.Po.Dm, Dm.IDR characterize the distribution of pore diameters, repectively : median value, 9$^{th}$ decile, average diameter of large pores (diameter greater than Dm.DC-9), and inter-decile range.}
    \label{table_chap_4:pore_stat_samples_combined}
    \end{table}
\end{landscape}
}

\subsection{Wave speed}
Estimation of wave speed with the autofocus method failed for sample~4 because of its high heterogeneity and the presence of numerous large pores. Therefore, sample 4 was excluded from further analysis. 
The estimated wave speed for the other samples is reported in Table~\ref{table:median_speed}. 
Across all VOIs, the median wave speed ranged from 3022~m/s to 3374~m/s. Sample 5 exhibited the lowest and sample 2 has the highest wave speed. 
The relative range of variations of wave speeds across repetitions ranged between 4.0 and 7.1~\%  reflecting both the intrinsic reproducibility of the technique and the heterogeneity of the samples (between each repetition, the probe was purposely placed in a slightly different position from one acquisition to another). 

\begin{figure}[htb!]
\centering
\includegraphics[width=.75\textwidth]{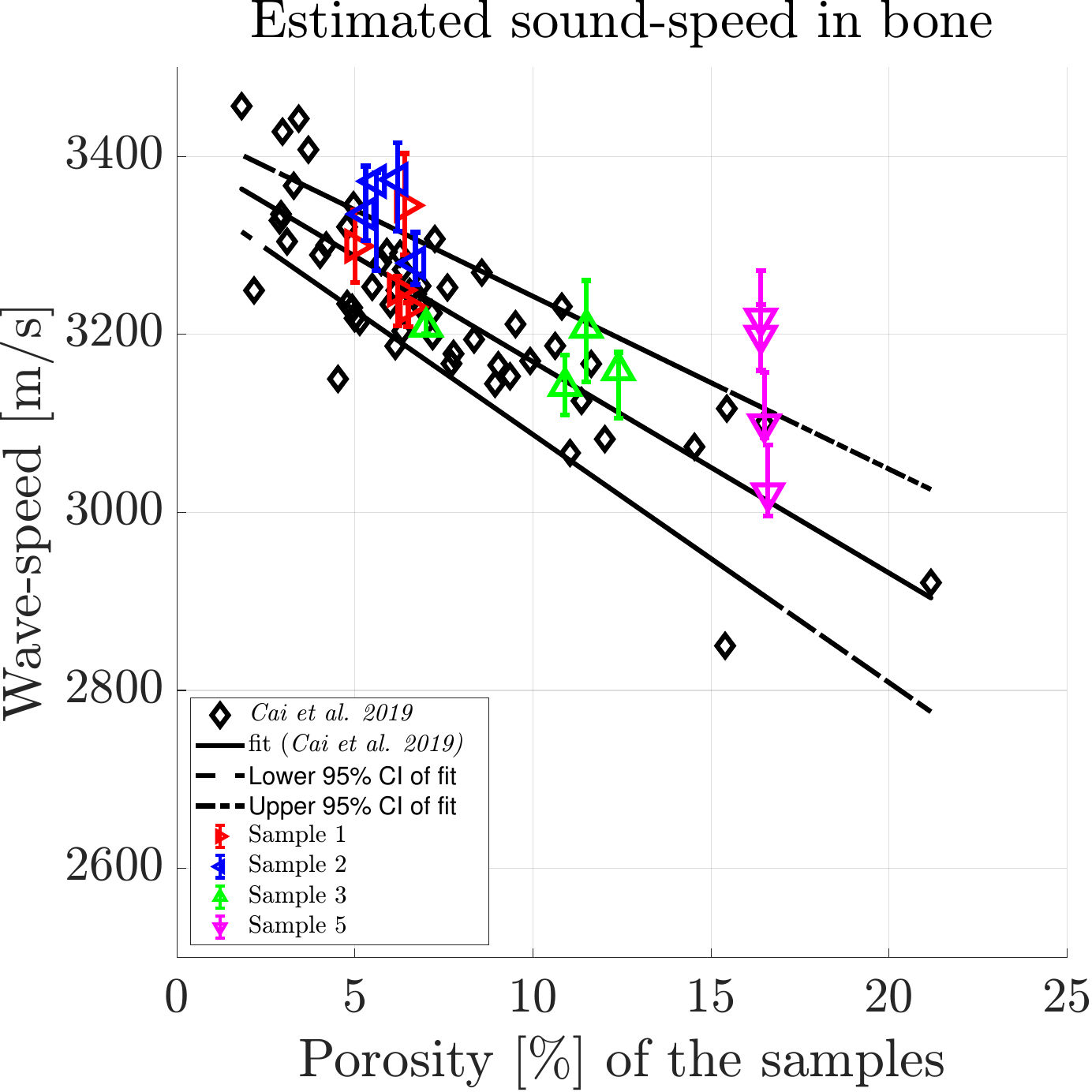}
\caption{Longitudinal wave speed in cortical bone vs. porosity for all VOIs and samples. Triangle symbols and vertical bars indicate the median value and range (min value and max value) across the 10 repetitions for each VOI, respectively. Reference data from \cite{cai_anisotropic_2019} is given for comparison: diamond symbols indicate the wave speed and porosity values for 52 cortical bone samples of millimeter size. The black lines are the linear regression of the reference data and the corresponding 95~\% confidence interval.}
\label{fig_chap_4:speed_of_sound_vs_xiran}
\end{figure}

Median wave speed was found to decrease with porosity (Figure~\ref{fig_chap_4:speed_of_sound_vs_xiran}). The wave speed values and their trend of variations with porosity matched well with reference wave speed data obtained with Resonant Ultrasound Spectroscopy (RUS) \cite{cai_anisotropic_2019} for millimetric samples of known porosity. Most of the wave speed values derived from US imaging in the present study fell within the 95~\% confidence interval of the linear regression of the reference data.
For sample~5, we observed large differences of wave speeds across VOIs (about 200~m/s) while a similar porosity of about 16\% was calculated for these VOIs. This may be due to the strong heterogeneity within the VOIs, the small thickness of this sample and the fact that the autofocus method will estimate the wave speed in a volume not exactly matching the volume in which the porosity is calculated from HR-\textmu CT images.

     \begin{table}[htb!]
	\centering
	\begin{tabular}{|c|cccc|}
		\toprule
		VOI&Sample 1&Sample 2&Sample 3&Sample 5 \\
		\midrule
		 1&3299(5.7)&3335(6.4)&3207(6.8) &3022 (5.0) \\
		 2&3231(4.5)&3372(5.8)&3159(4.0)&3199(5.4) \\
		 3&3250(6.3)&3374(6.9)&3142(6.2)&3099(6.3) \\
		 4&3345(7.1)&3280(5.6)&3208(4.7)&3218(5.2) \\
            \midrule
		Average&3281&3340&3179&3134 \\
		\bottomrule
	\end{tabular}
	\caption{US wave speed in cortical bone in m/s (median value (relative range in \%)) measured with the autofocus method for each sample and VOI. Relative range was calculated from the 10 repetitions. The average wave speed in the 4 VOIs is provided in the last row.}
	\label{table:median_speed}
        \end{table}
        
\subsection{Ultrasound images} In all US images (Figure~\ref{results:reconstructed_images} and Figures S2-5 of supplementary materials), the periosteal surface appeared as a very bright line. The brightness of the endosteal surface was found to decrease with increasing porosity and with the presence of large pores.
The endosteal surface was relatively bright in Samples~1 and~2 which had a small number of large pores and a small porosity; while the contrast of the endosteal surface was relatively weak in Samples~3 and~5 which had a higher porosity and some large pores near the endosteal surface (see the HR-\textmu CT images (panels (a), (d), (g) and (j) of Figure~\ref{results:reconstructed_images}).
Segmentation with Dijkstra's algorithm successfully delineated periosteal and endosteal surfaces for all measurement zones in all samples (see Figure~\ref{results:reconstructed_images}). 
US images were aligned on HR-\textmu CT images based on the periosteal surface segmentation. Periosteal surfaces from the US images accurately matched the periosteal surfaces segmented from the HR-\textmu CT images: for all VOIs and all measurement repetitions, the root mean square error (RMSE) between the two segmentations was smaller than 0.3~mm, which value corresponds to half a US wavelength in water at 2.5~MHz (i.e., approximately the resolution limit). 

We observed a good match between the segmented endosteal surface in the US image and the boundary of the cortex as seen in the HR-\textmu CT image. The segmented endosteal surface closely followed the shape of the bone in the samples~1 and~2 which were the less porous and less heterogeneous. It is noteworthy that, despite the large thickness and complex geometry of sample 2, the endosteal surface was very bright in the US image. 
In the more porous and heterogeneous samples~3 and~5, the segmented endosteal surface followed the gross shape of the bone but was inside the cortex as seen in the HR-\textmu CT image, above some large pores found close to the endosteal cortex boundary.

\begin{figure}[htb!]
        \centering
        \begin{subfigure}{.30\linewidth}
            \includegraphics[width=\linewidth]{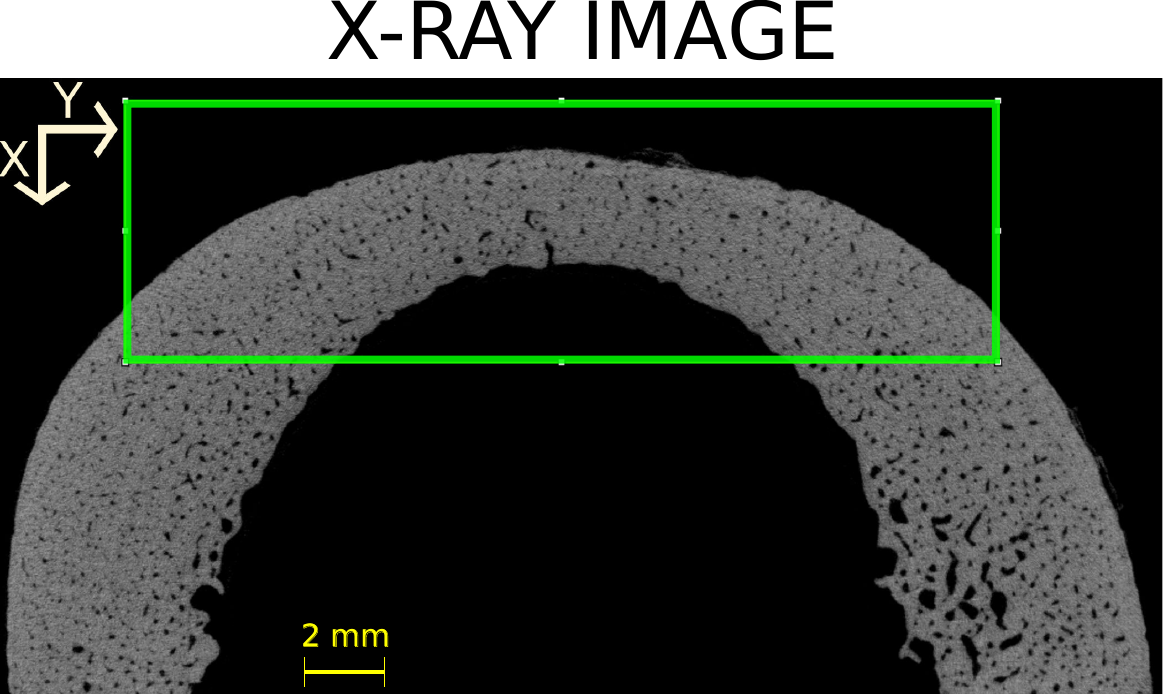}
            \caption{}
         \end{subfigure}
        \begin{subfigure}{.30\linewidth}
            \includegraphics[width=\linewidth]{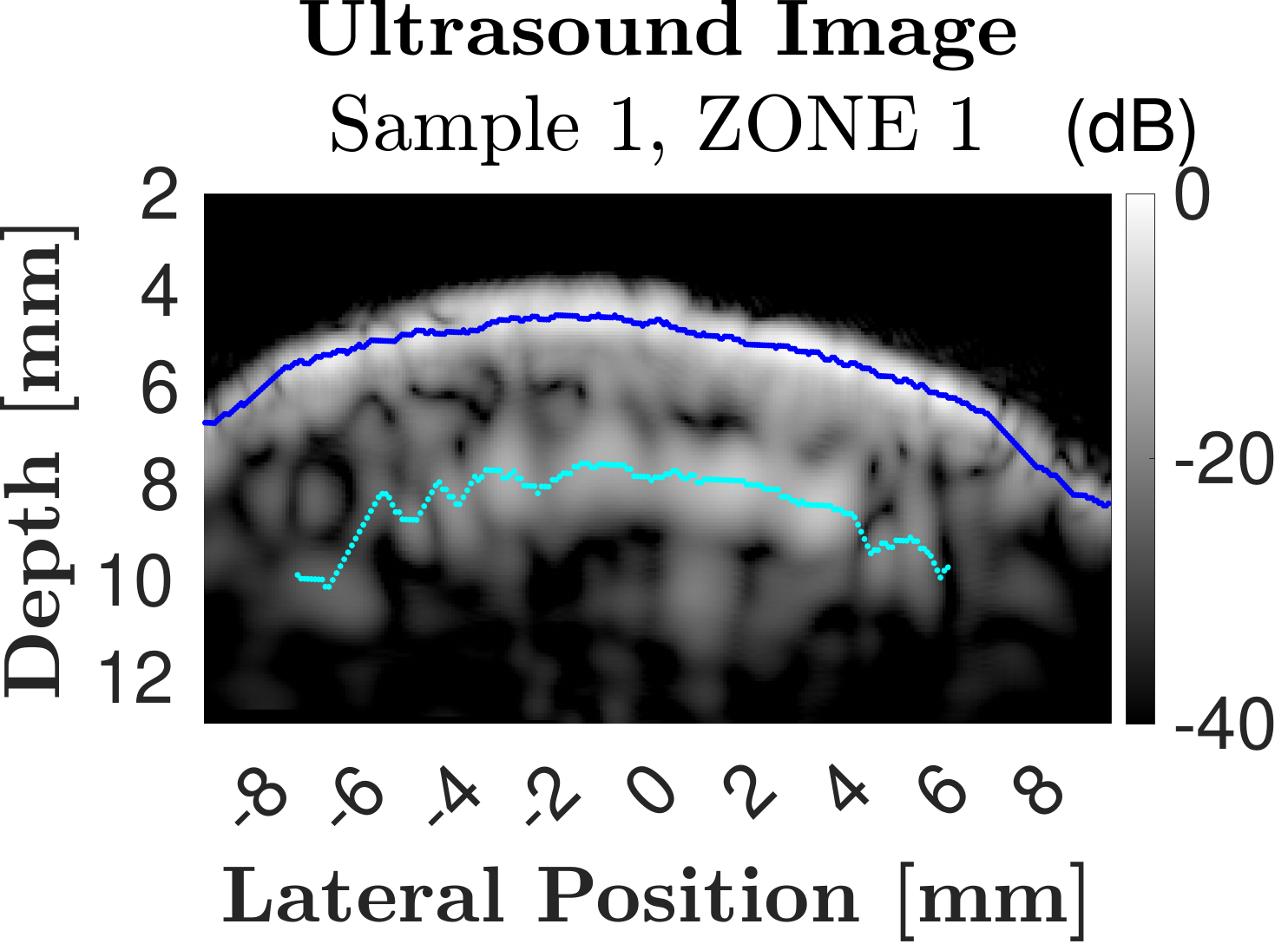}
            \caption{}
        \end{subfigure}         
        \begin{subfigure}{.30\linewidth}
                \includegraphics[width=\linewidth]{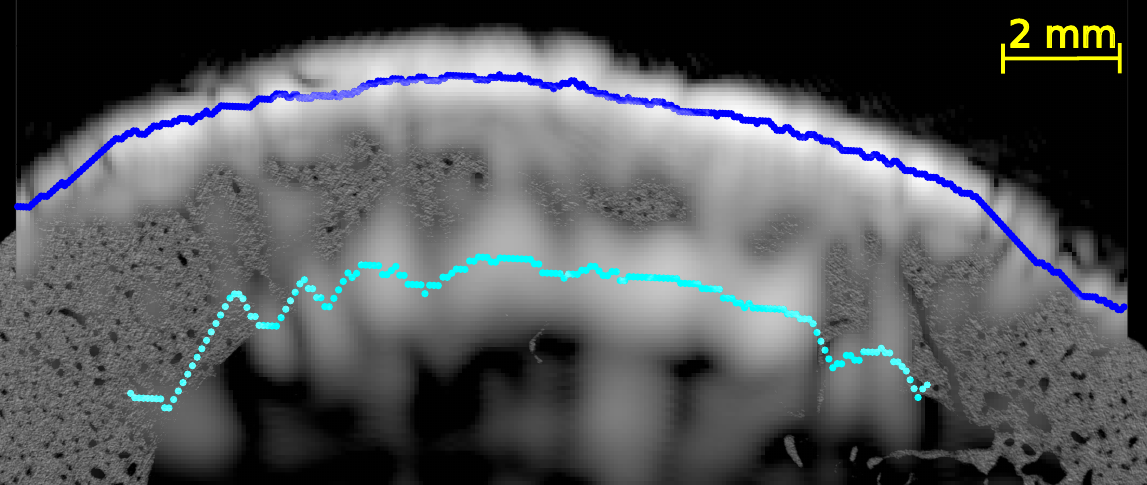}
                \caption{}
                \label{subfiga:sample-1}
            \end{subfigure}
            
        \begin{subfigure}{.30\linewidth}
            \includegraphics[width=\linewidth]{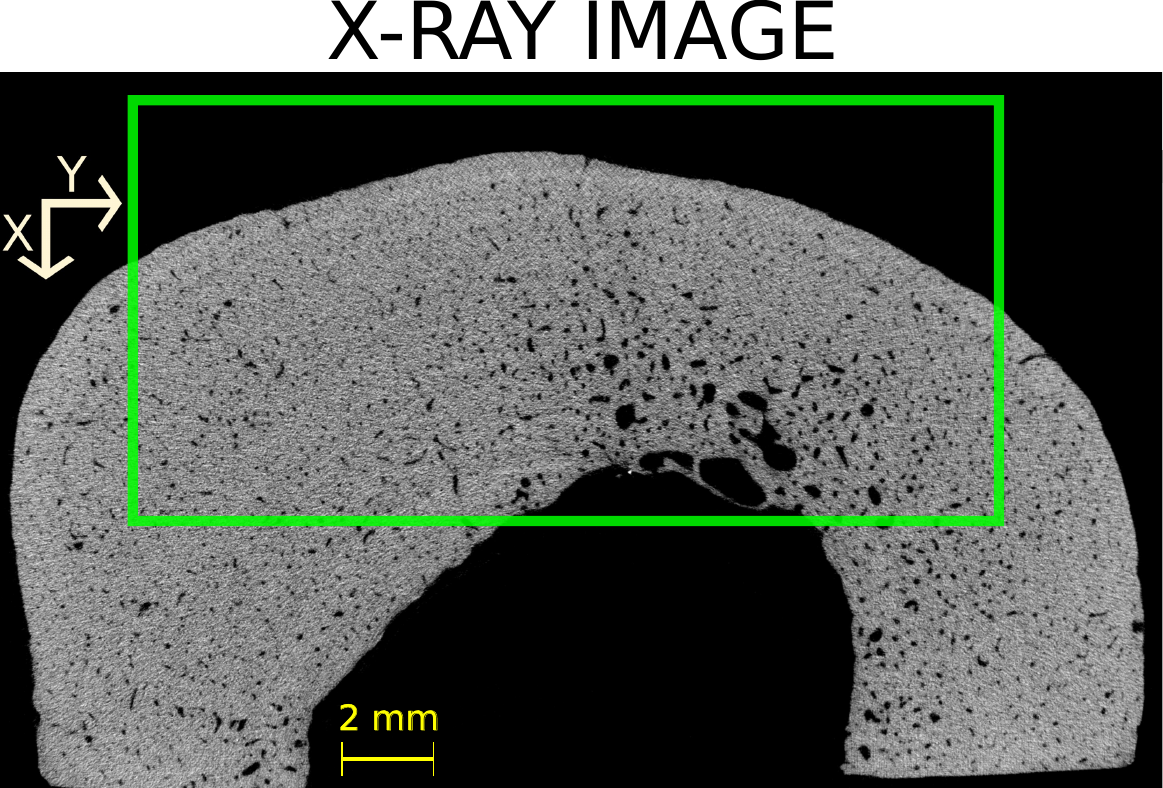}
            \caption{}
         \end{subfigure}
         \begin{subfigure}{.30\linewidth}
            \includegraphics[width=\linewidth]{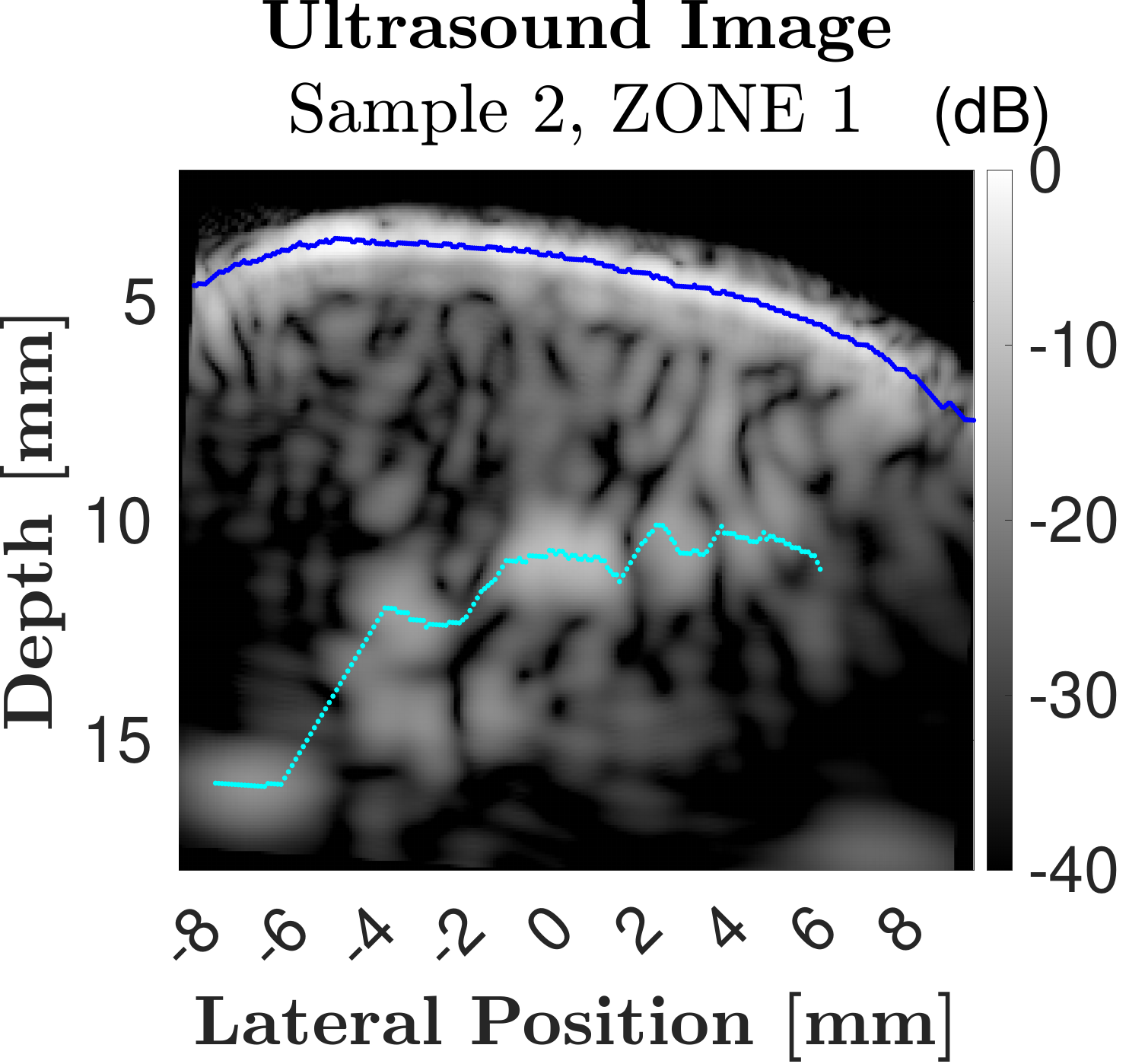}
            \caption{}
        \end{subfigure}
        \begin{subfigure}{.30\linewidth}
                \includegraphics[width=\linewidth]{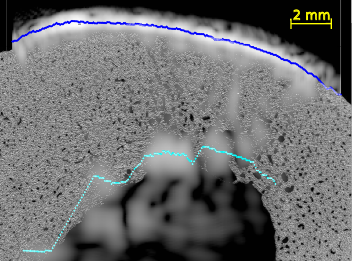}
                \caption{}
                \label{subfigb:sample-2}
            \end{subfigure}
            
        \begin{subfigure}{.30\linewidth}
            \includegraphics[width=\linewidth]{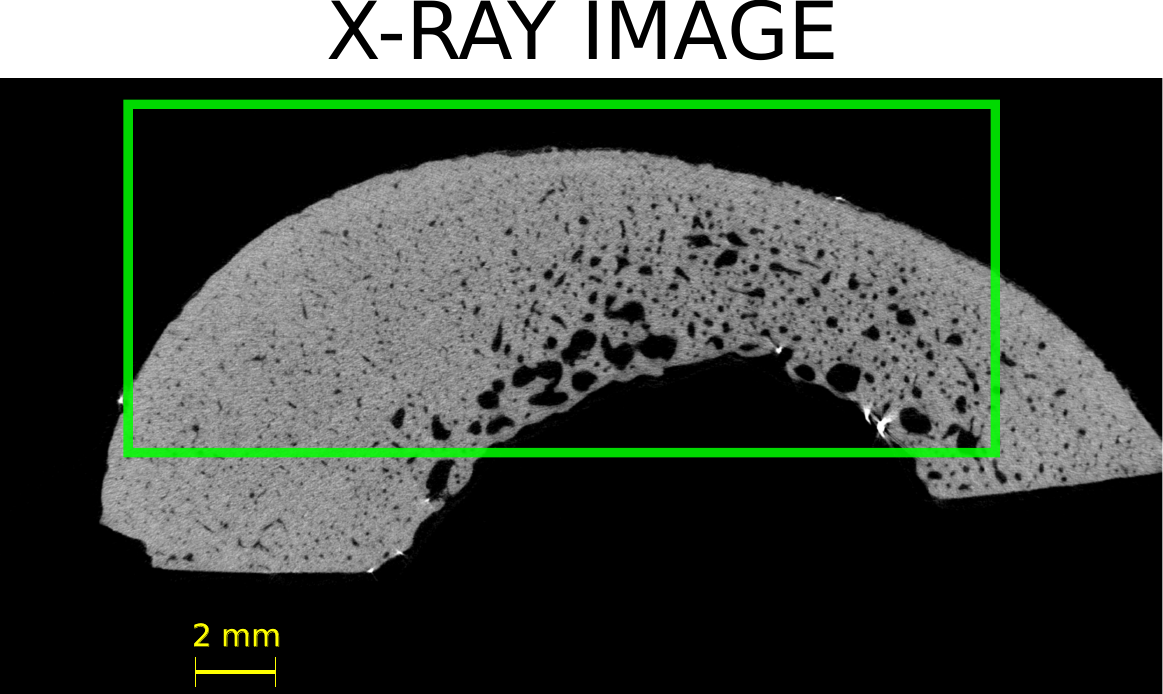}
            \caption{}
         \end{subfigure}
         \begin{subfigure}{.30\linewidth}
            \includegraphics[width=\linewidth]{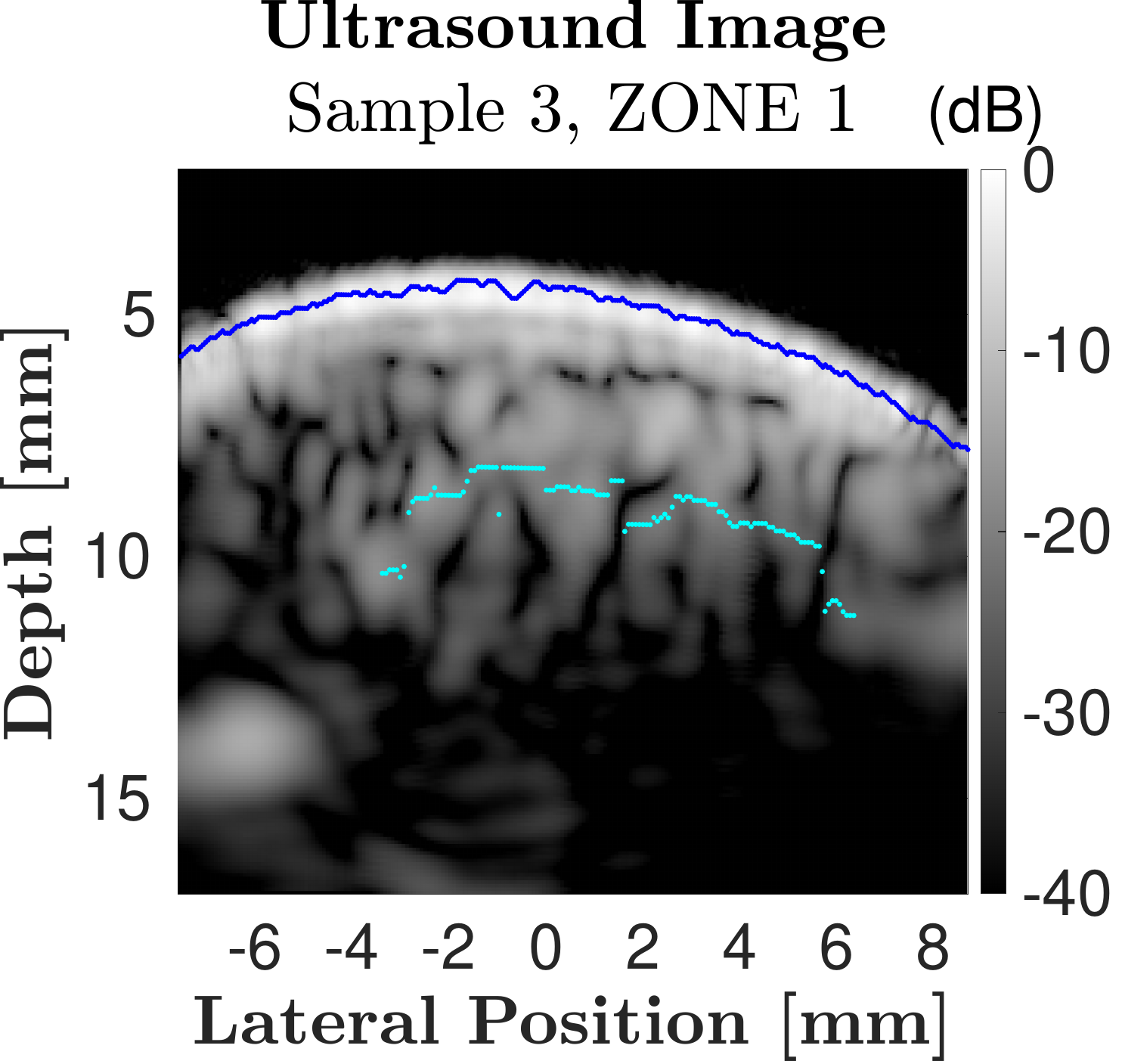}
            \caption{}
        \end{subfigure}
        \begin{subfigure}{.30\linewidth}
                \includegraphics[width=\linewidth]{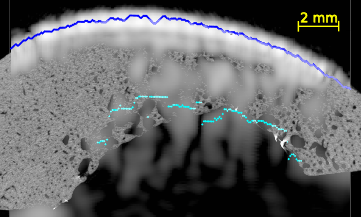}
                \caption{}
                \label{subfigc:sample-3}
            \end{subfigure}
            
        \begin{subfigure}{.30\linewidth}
            \includegraphics[width=\linewidth]{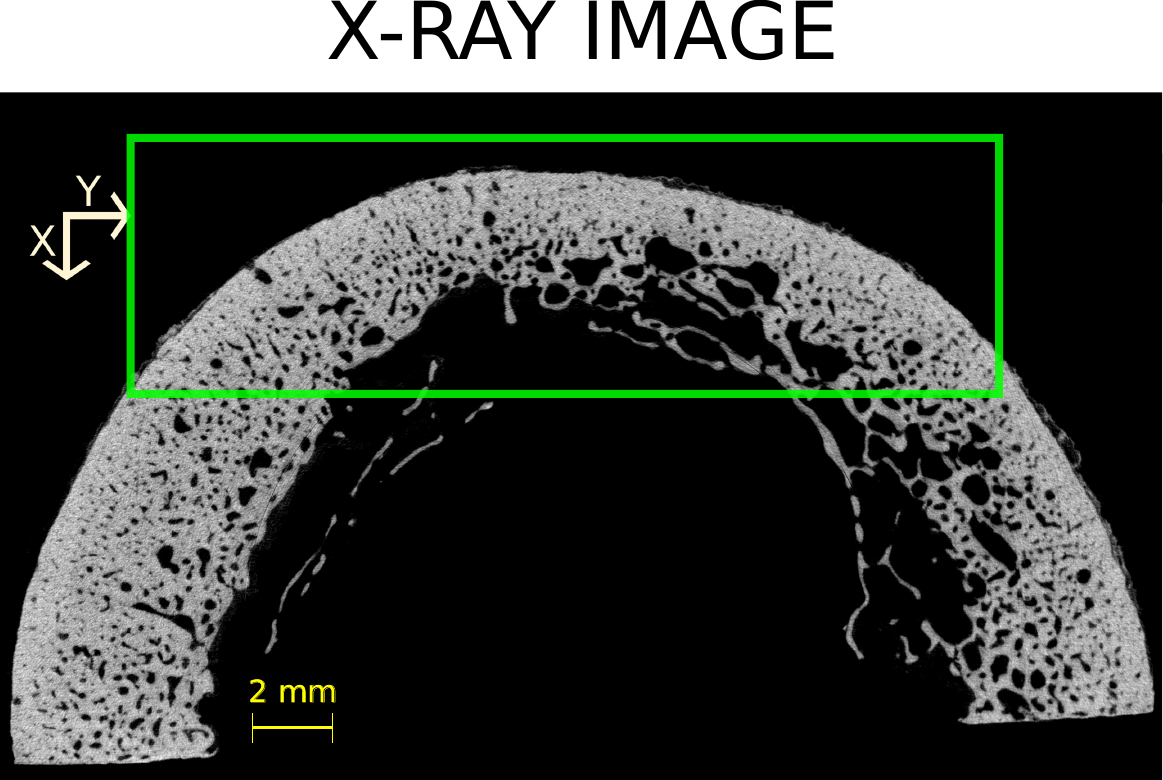}
            \caption{}
         \end{subfigure}         
        \begin{subfigure}{.30\linewidth}
            \includegraphics[width=\linewidth]{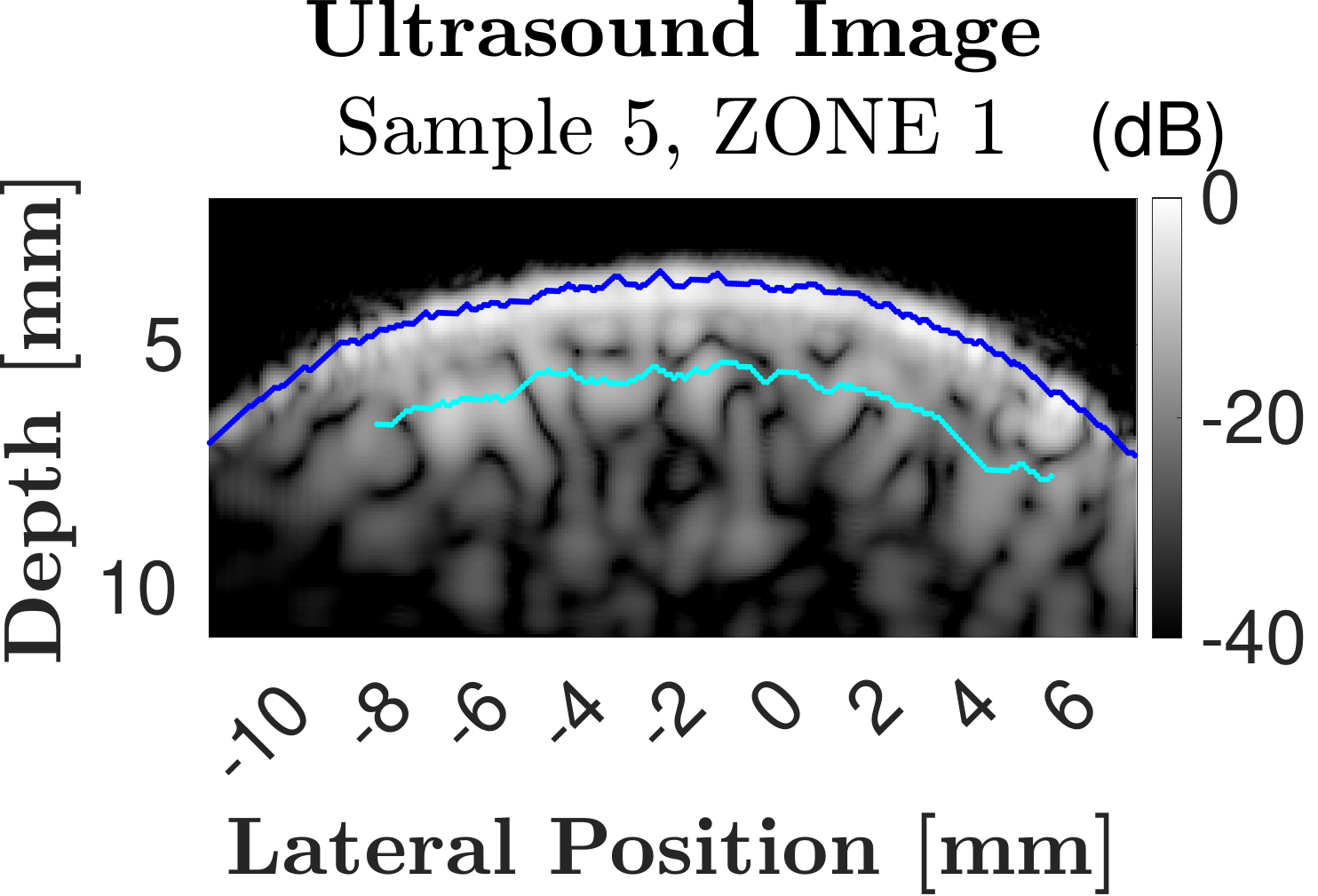}
            \caption{}
        \end{subfigure}
        \begin{subfigure}{.30\linewidth}
                \includegraphics[width=\linewidth]{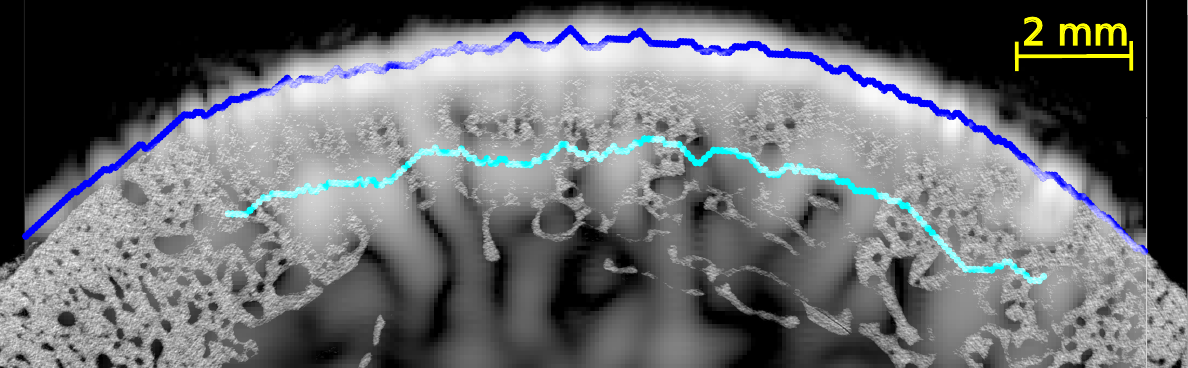}
                \caption{}
                \label{subfigd:sample-5}
            \end{subfigure}
        \caption{Comparison of micro-CT and ultrasound images. One image, i.e., from one of the repetitions, and for one of the VOIs, was selected for each sample. The first column displays X-ray images; the ROI in the green rectangle corresponds approximately to the area probed with ultrasound. The second column shows ultrasound images displaying raw segmentation (from Dijkstra' algorithm) of the cortex (periosteal and endosteal interfaces). The last column shows the superimposition of ultrasound and X-ray images and the raw segmentation of the cortex from ultrasound images.
        The US reconstructions and comparison to HR-$\upmu$CT images for all measurement zones and all samples is provided in supplemenatry material in Figures S2-5.}
        \label{results:reconstructed_images}
    \end{figure}

\subsection{Ultrasound measurement of cortical thickness.} 
\begin{figure}[htb!]
\centering
\includegraphics[width=0.6\linewidth]{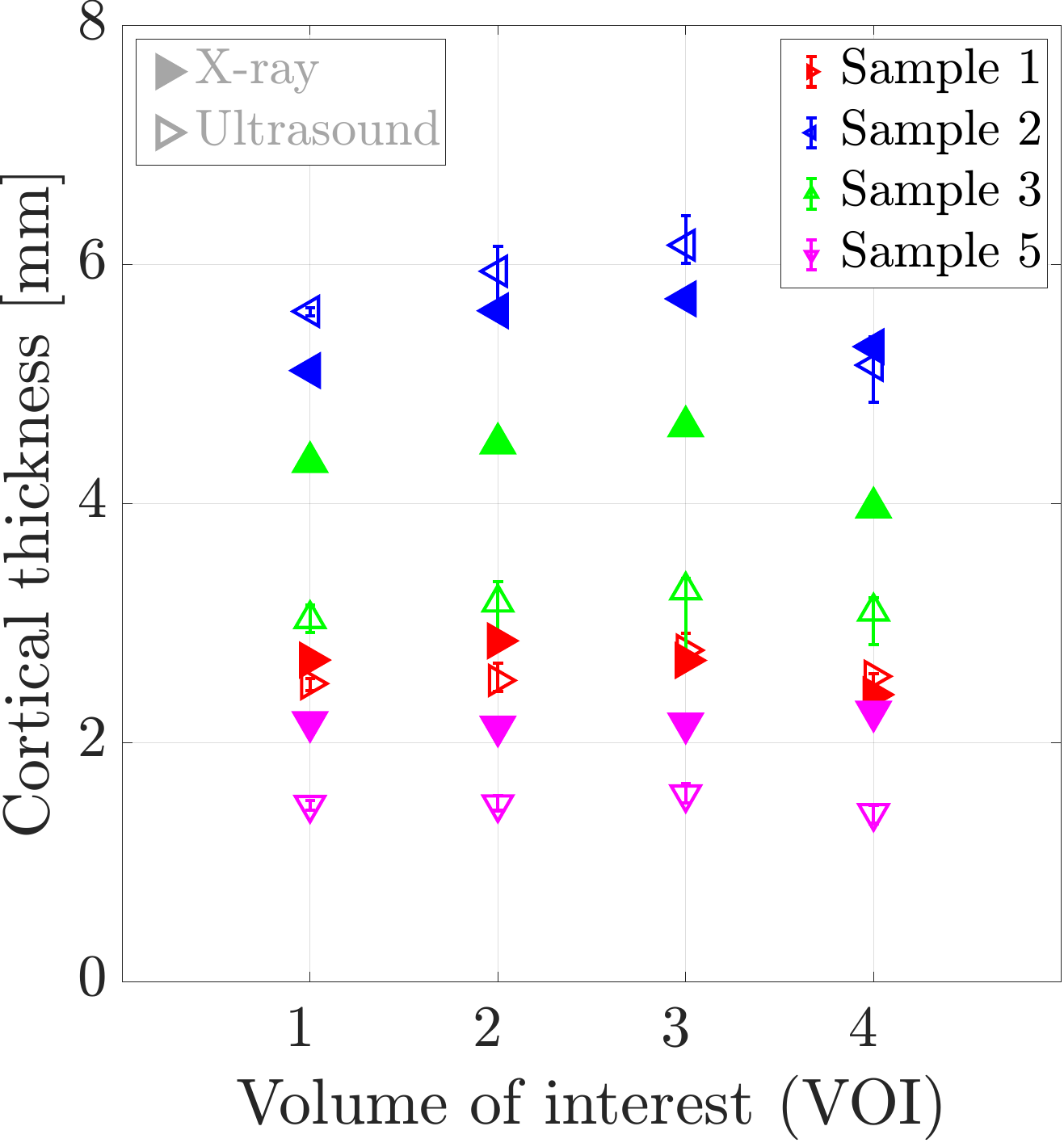}
\caption{Estimated cortical thickness using US imaging (empty symbols) for each sample and each VOI compared to reference cortical thickness measured from HR-\textmu CT images (filled symbols). The symbols and the bars represent the median value and the inter-quartile range across repetitions, respectively.}
\label{fig_chap_4:us_cortical_thickness}
\end{figure}

US imaging could accurately determine the cortical thickness for samples~1 and~2 (across the 4 VOIs, the relative error, compared to the HR-\textmu CT reference, was between 2 and 16~\% and between 3 and 11~\%,  respectively) and less accurately for samples~3 and~5 (across the 4 VOIs, the relative error was between 31 and 34~\% and between 45 and 49~\%,  respectively)
(Figure~\ref{fig_chap_4:us_cortical_thickness}).
Except for the thickest sample (sample~2, which had a more complex geometry), US underestimated the thickness obtained from HR-\textmu CT images. 

\section{Discussion}
This study represents the first critical evaluation of the potential for US to image the bone cortex in elderly individuals, with the aim to diagnose bone health. We adopted the imaging approach developed by Renaud et al. \cite{renaud_vivo_2018}, which utilizes a 2.5~MHz phased array probe for data acquisition and employs a delay-and-sum reconstruction algorithm that accounts for refraction effects at the bone-soft tissue interface. This method extends conventional B-mode imaging to bone tissue, providing anatomical images of the cortex and estimating US wave speed \cite{renaud_measuring_2020}. Prior research has validated the feasibility of this imaging approach \invivo~in healthy, mostly young, volunteers \cite{renaud_vivo_2018, salles_revealing_2021}, and cortical thickness measured with US imaging was validated for two subjects by comparing it with that measured with HR-pQCT \cite{renaud_vivo_2018}. 
This study goes much further in validating the technique by providing, on the one hand, a precise comparison of the geometric reconstruction of the cortex obtained with ultrasound imaging against a HR-$\upmu$CT reference, and on the other hand, a critical assessment of the measured wave-speed as it reflects porosity.
This was done for 20 imaging regions (four measurement sites in each of the five bones) in bones of elderly donors  which exhibit significant variability in cortical thickness and microstructure, with some showing signs of unbalanced remodeling.
The results of this \exvivo~study suggest that the US imaging method has the potential to assess bone health in elderly or osteoporotic individuals by assessing cortical thickness and cortical bone tissue mechanical competence, as reflected by wave-speed. 

\subsection{High-resolution micro-CT images}
\Exvivo~studies are an essential step to validate an imaging technique avoiding some limitations of \invivo~testing. 
In \exvivo~settings, the geometry and  microstructure of the samples can be accurately characterized using HR-$\upmu$CT. For this purpose, in this study, we used the highest HR-$\upmu$CT resolution available ex-vivo for our voluminous samples (voxel size of 8.8~$\upmu$m). This resolution enabled the quantification of pore size, which is known to increase with the progression of osteoporosis \cite{andreasen_understanding_2018} and is a known factor of US image quality \cite{dia_influence_2023}. 
The bones from elderly human donors, used in this study exhibited a range of microstructural characteristics typically found at different stages of osteoporosis and showing signs of unbalanced remodeling.
Additionally, HR-$\upmu$CT images provided a reliable reference for the geometry of the cortex and the cortical thickness. The bones used in this study exhibited a large range
of cortical thicknesses (2.5 to 6.3~mm).

\subsection{US image reconstructions and cortical thickness measurement}
We evaluated the quality of US anatomical reconstructions and wave speed measurements across sixteen VOIs from bones of four individuals. Indeed, the bone of a fifth individual (sample~4) was too porous and heterogeneous to be effectively imaged through US. Our results indicate that the cortex boundaries and cortical thickness can be accurately determined for homogeneous and moderately porous VOIs, even for a thick cortex, as observed in samples~1 and~2 (mean thickness 2.9 and 5.9~mm and porosity between 5.0 and 12.3\%, respectively). In these cases, the overlaying of US and HR-$\upmu$CT images showed that the segmentation of the surfaces in US images with the Dijkstra's method closely aligns with the actual boundary,  achieving thickness measurements with mean relative errors of 9~\% and 4~\%, respectively.
In regions with higher porosity and larger pore diameters (samples~3 and~5 with porosity between 10.9 and 16.6\% and Lg.Po.Dm between 224 and 307~$\upmu$m, respectively), the brightness of the endosteal surface is significantly reduced (from 7~dB to -5~dB). Nevertheless, Dijkstra’s algorithm was able to delineate endosteal boundaries that generally lied within the cortex, above the region with very large pores nearest to the endosteal surface.
Consequently, in these cases, US-based cortical thickness measurements underestimated the reference (mean relative errors of 32~\% and 47~\%). 
However, for these highly remodeled bones with large resorption cavities leading to a trabecularization of the endosteal region, the delineation of the endosteal boundary as the end of the cortex and the beginning of the medullary canal is ambiguous, making it challenging to obtain a meaningful estimate of cortical thickness in the HR-$\upmu$CT images \cite{zebaze_intracortical_2010}.
Another limiting factor in comparing cortical thicknesses determined from HR-$\upmu$CT and from US images lies in the slight differences in the regions of interest used for each VOI. The region of interest used for HR-$\upmu$CT images (Figure~\ref{methods:average_masks}) were slightly larger than that used for US images. In the latter, as shown in Figure ~\ref{results:reconstructed_images},  the surfaces could only be reconstructed for a limited portion of the cortex due to the finite US probe aperture. 

It is noteworthy that the inner cortex surface reconstructed with US was consistently within the bone, indicating that it effectively detects the most compact region corresponding to the tissue above the large pores in the endosteal region. 

In highly porous samples exhibiting large pores, the reconstructed endosteal surface may show a reduced brightness and a lack of continuity.
Indeed, previous simulations demonstrated that increased pore size has a strong detrimental effect on endosteal interface brightness \cite{dia_influence_2023}, primarily due to the enhanced scattering by large pores. 
This \exvivo~ study corroborates these simulations results by providing an experimental proof that pore size is a major factor influencing intracortical US image quality. 

In certain cases, such as sample~4 in this study,  the US imaging method may fail to reveal the inner cortex surface on the reconstructed image. In an \invivo~ application of the method, such failure would strongly suggest that the bone has undergone excessive resorption and has very large pores. Consequently, the inability to visualize the endosteal surface could serve as a marker of an advanced alteration of the cortical bone tissue.

\subsection{US wave-speed estimated with the autofocus approach}

The wave speed within the bone cortex is known to be strongly influenced by cortical porosity \cite{granke_change_2011}. Given that the samples used in this study exhibited a substantial porosity heterogeneity, with values ranging from 5 to 17~\% (Table~\ref{table_chap_4:pore_stat_samples_combined}), we anticipated a range of variation of wave speed of about 300~m/s \cite{peralta_bulk_2021} accross the samples and VOIs.
Wave speed values measured in this study align with previous studies, following the expected decreasing trend with increasing porosity \cite{peralta_bulk_2021, eneh_effect_2016}.
The similarity between wave speed values measured in the 12 VOIs of samples 1, 2, and 3, and those from the reference data for the same porosities (Figure~\ref{fig_chap_4:speed_of_sound_vs_xiran}), suggest that the autofocus method provides accurate estimations of wave speed in cortical bone.
As far as we are aware, the reference dataset, obtained from femoral bones of a separate cohort of elderly donors, is the most accurate set of wave speeds and porosity values available for human cortical bone material. Wave speed was assessed using resonant ultrasound spectroscopy, an accurate technique for the characterization of small samples \cite{bernard_accurate_2013},  while porosity was quantified by HR-$\upmu$CT imaging from a synchrotron source, providing optimal contrast and resolution. Furthermore, the wave speed and porosity data were derived from millimetric samples, which were only slightly larger than a representative volume element of bone material \cite{grimal_determination_2011}, ensuring an optimal scale for determining a relationship between local tissue properties and porosity.

In contrast, in the autofocus method, a large region encompassing all the cortex within the US image of a VOI is used to derive a single wave speed value.  As can be seen in Figure~\ref{method:example_xray_slice}, the imaged regions may exhibit highly heterogeneous pore distributions, resulting in local wave speed variations. The error bars in Figure~\ref{fig_chap_4:speed_of_sound_vs_xiran} for each VOI partly reflect the slight variations in the determined wave speeds across the repeated measurements due to this microstructural heterogeneity and the variation of probe orientation across repetitions.  
For each measurement, the wave speed value determined with the autofocus method may be considered as an average wave speed for the corresponding cortical region.
The influence of the material heterogeneity within the imaged region on wave speed determination remains unclear and warrants further investigation. 

In US imaging, any error in the assumed wave speed used for image reconstruction process results in a misplacement of the reconstructed surface positions (shortly, surface position is obtained from the product of an echo time with wave speed). This is why in our imaging approach, the wave speed is determined in bone and soft tissues for each measurement in order to accurately reconstruct bone geometry \cite{renaud_vivo_2018}.
In the present study we found that the endosteal surfaces reconstructed from US images closely aligned with the actual surfaces position observed in HR-$\upmu$CT images (Figure~\ref{results:reconstructed_images}). This provides further evidence that the wave speed determined by the autofocus method closely approximated the true wave speed. 

Overall, our results suggest that US imaging combined with the autofocus method provides a reliable estimation of wave speed in cortical bone samples with moderate microstructural heterogeneity.In cases of pronounced heterogeneity, the measured wave speed may be highly sensitive to the exact positioning of the probe.

\subsection{Limitations}
This study has some limitations. Femur bones were used due to availability constraints, rather than tibia or radius which are typically the bones measured \invivo~ because these are more accessible to US. The femur’s curvature is more pronounced than that of the antero-medial tibia which is a common US measurement site \cite{hans_quantitative_2022}. Due to US refraction, a stronger curvature reduces the imaged cortex area and limits the extent of the endosteal surface that can be reconstructed. Consequently, we expect that the imaging method would perform even better in less curved bone sites. 

We used only five bones in this study. One of these  could not be measured, highlighting a significant limitation of the method when large pores are present. In the other bones, four regions were analyzed, resulting in a total of 16 independent imaged VOIs. This diverse sample set represents a broad range of cortex geometries, thicknesses and microstructural patterns, enabling a critical assessment of the US imaging technique.
The osteoporotic status of the donors, which would ideally be assessed using dual-energy X-ray absorptiometry, was unknown. Nevertheless, we can classify the samples used in this study into different groups. Samples 1 and 2 can be considered representative of healthy bones, while samples 3 and 5 show characteristics of osteoporotic bones with clear signs of unbalanced remodeling. Sample 4 shows signs of a very advanced osteoporotic condition.
Overall, we expect that our conclusions would extend to larger sample sets and more generally to other measurement sites, such as the tibia and radius.

\subsection{Conclusion}
With this study, we could show that the implemented US imaging approach, using Dijkstra segmentation, can accurately reconstruct the endosteal boundary in bones of elderly donors with signs of osteoporosis, regardless of bone thickness (up to 6.3 mm in this study) and marked curvature, except in cases with very large pores. When a transitional zone (trabecularized cortical bone region) is present, the reconstructed endosteal surface tends to be above the transitional zone. When the endosteal surface is accurately reconstructed, the method provides an accurate measurement of the cortical thickness. Additionally, we also showed that the autofocus method provides consistent values of wave speeds which agree well with tabulated values and reflect the mean porosity of the imaged region of the cortex. 

Cortical bone porosity is recognized as a fracture risk factor \cite{ahmed_measurement_2015,bala_role_2015,zebaze_denosumab_2016} as well as cortical thickness \cite{mayhew_relation_2005,nishiyama_postmenopausal_2010}.
US wave speed is strongly related to bone strength and porosity \cite{lee_tibial_1997,eneh_effect_2016, peralta_bulk_2021}. For instance, in an \exvivo~study on millimetric samples using resonant ultrasound spectroscopy to measure wave speed and mechanical testing to measure compressive strength, Peralta et al \cite{peralta_bulk_2021} found that, when porosity increased from 5~\% to 20~\%, wave speed decreased from 3300 m/s to 2900 m/s, and strength decreased from 160 to 120~MPa, which corresponds to a reduction of strength of 25~\%.  Importantly, our results suggest that US imaging associated with the autofocus method could discriminate, based on determined wave speed values, a bone with a low porosity (i.e., less than 5\%) from a bone with moderate or high porosity (i.e., larger than 10\%).

The present study paves the way for the application of intracortical US imaging  to diagnose bone health, in particular to detect increased cortical porosity and reduced cortical thickness. The precision and sensitivy of the technique still need to be investigated \invivo~ in patients. This technique could be implemented on a standard B-mode US imaging system as an additional imaging mode, and consequently could be largely available.

\clearpage
\bibliographystyle{unsrt}
\bibliography{biblio_exvivo}

\clearpage

\section*{Supplemental materials}
\begin{figure}[htb!]
    \centering
    \begin{subfigure}{.327\textwidth}
        \includegraphics[width=\textwidth]{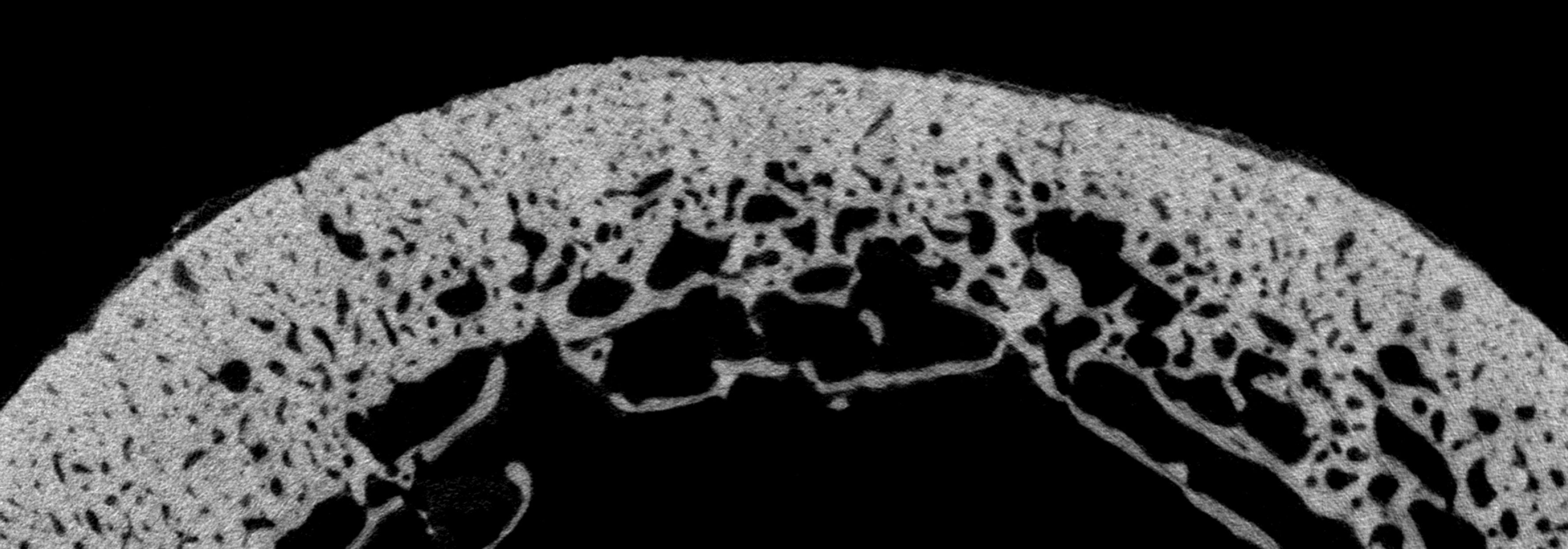}
        \caption{}
    \end{subfigure}
    \begin{subfigure}{.327\textwidth}
        \includegraphics[width=\textwidth]{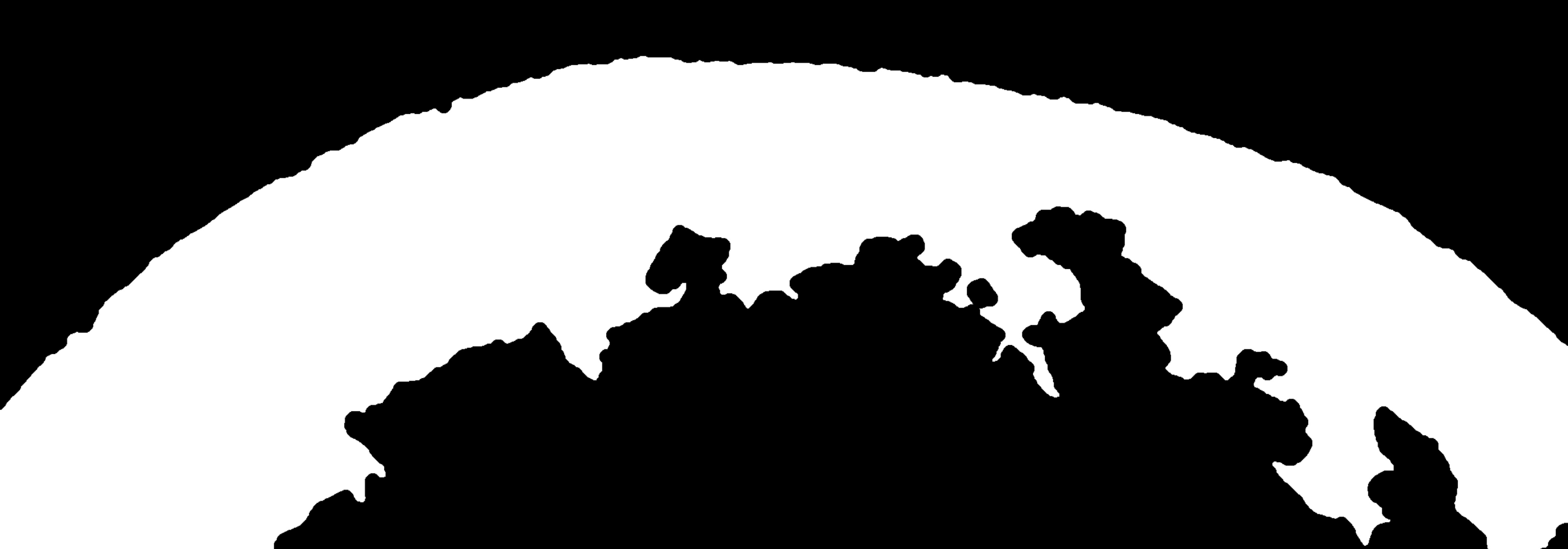}
        \caption{}
    \end{subfigure}
    \begin{subfigure}{.327\textwidth}
        \includegraphics[width=\textwidth]{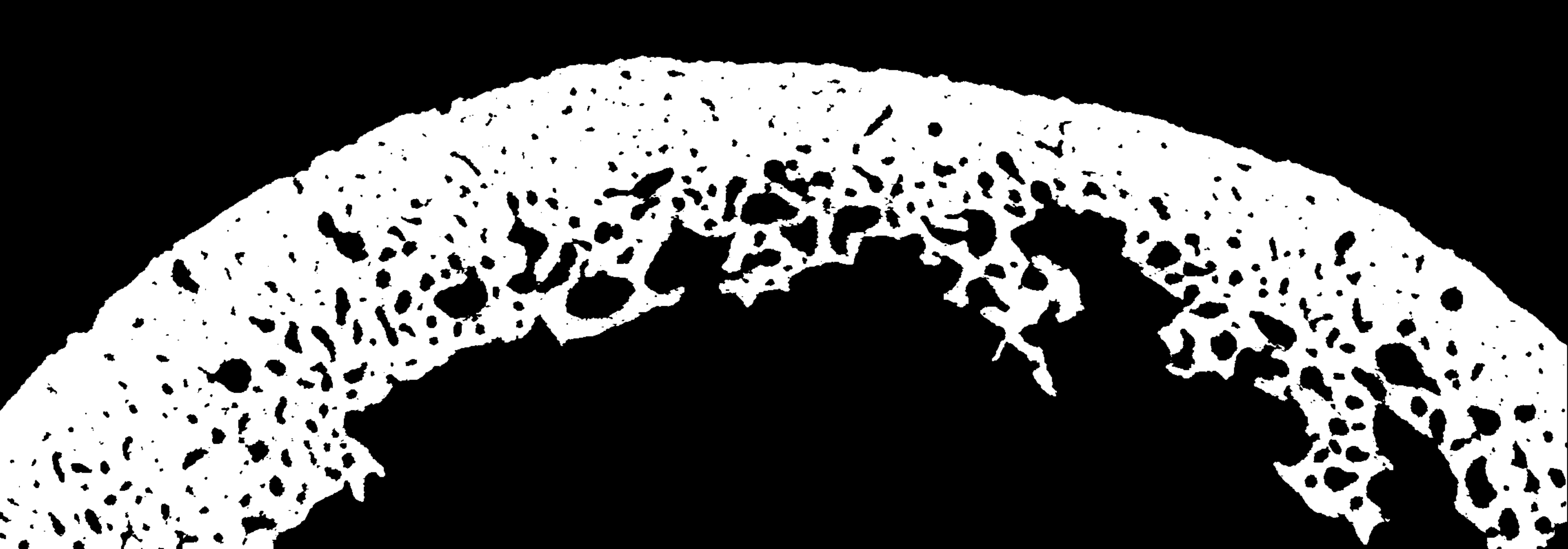}
        \caption{}
    \end{subfigure}               
    \caption{Cross-section of a bone sample illustrating the original micro-CT image (a), the generated mask excluding trabecular regions (b), and the resulting binarized image after trabecular removal (c).}
    \label{fig_chap_4:trabecular_removing}
\end{figure}

\afterpage{
\begin{landscape}
\begin{figure}[htb!]
        \centering
        \begin{subfigure}{.24\linewidth}
            \includegraphics[width=\linewidth]{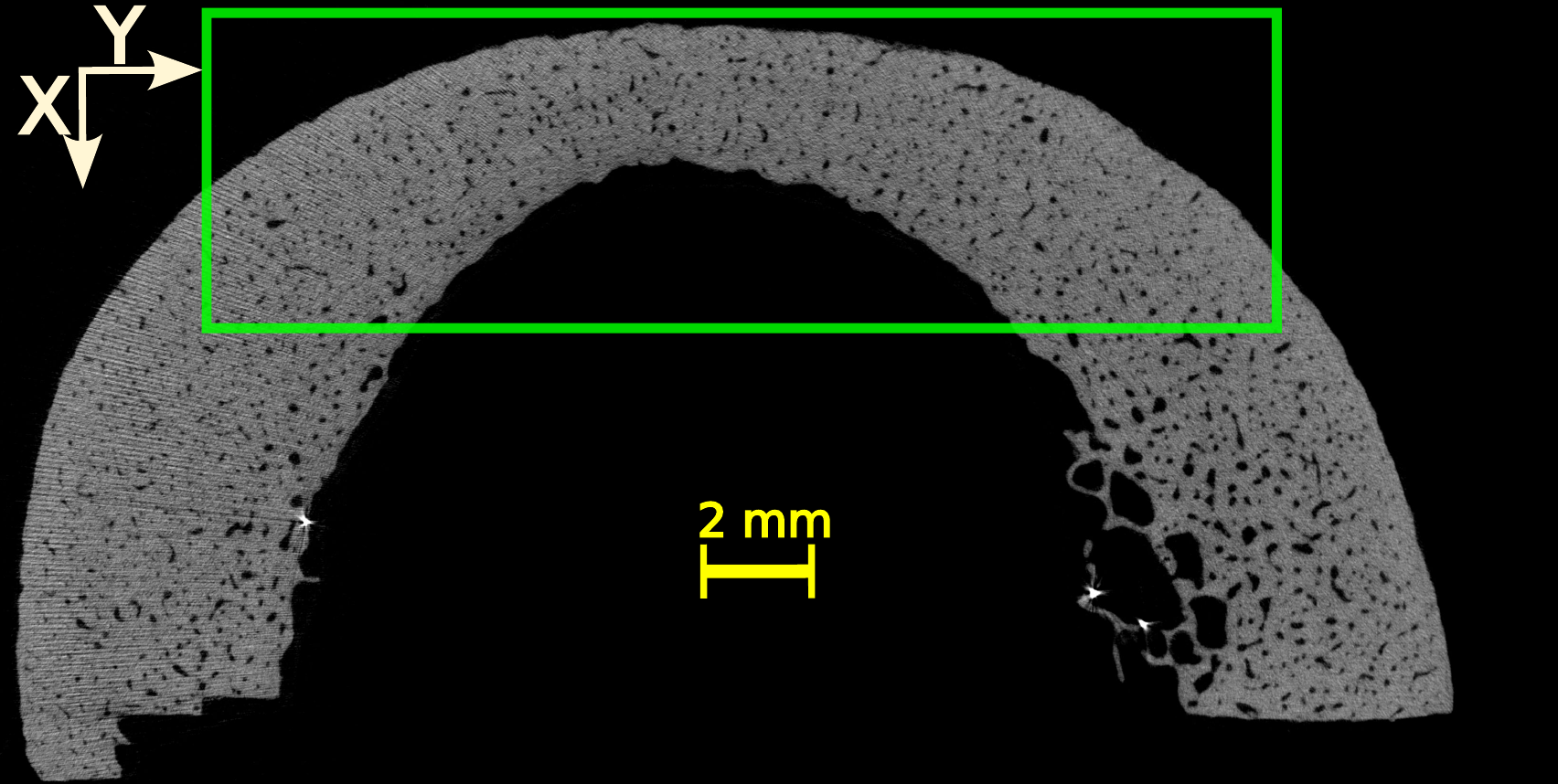}
            \caption{}
         \end{subfigure}
        \begin{subfigure}{.24\linewidth}
            \includegraphics[width=\linewidth]{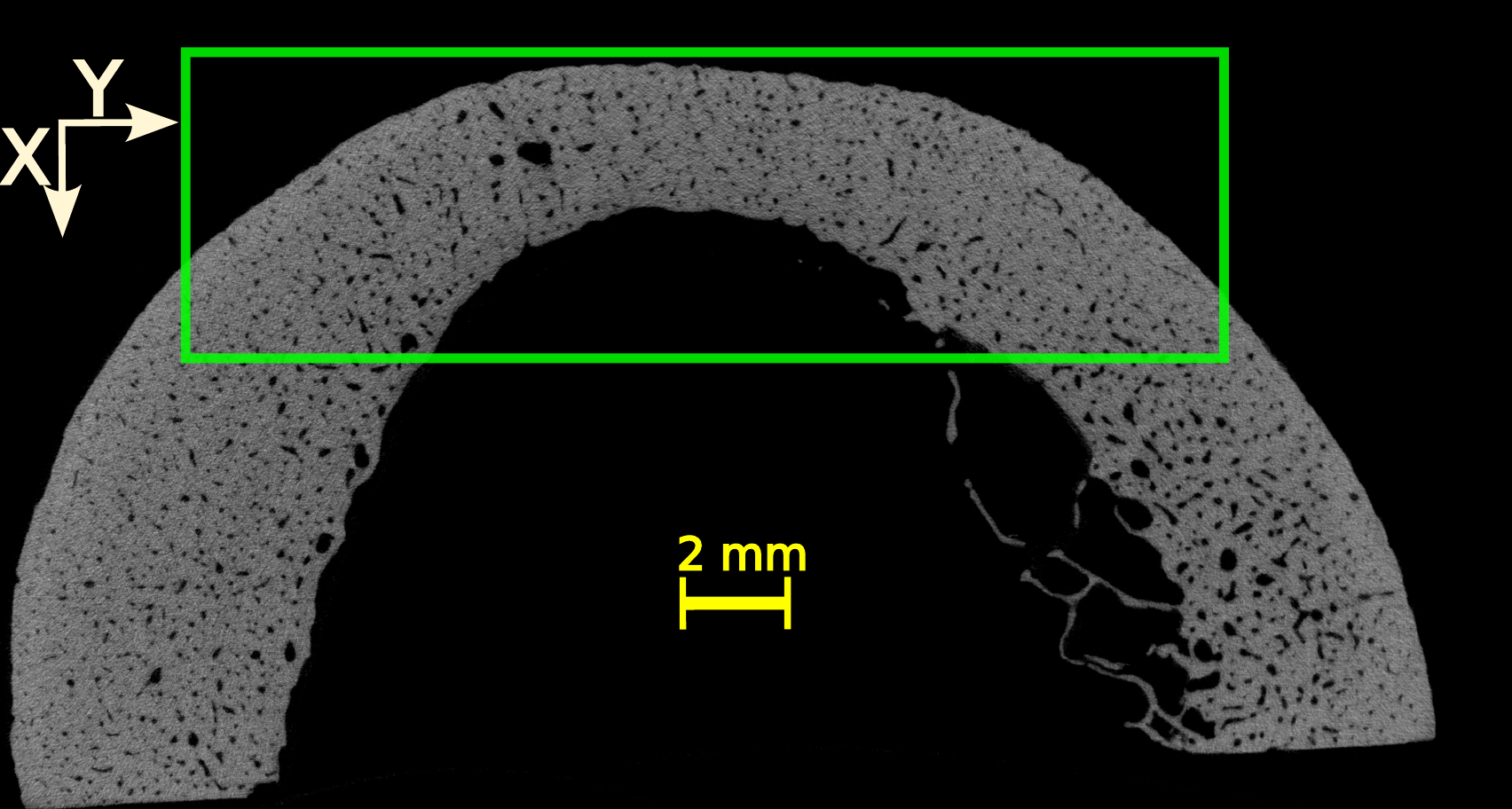}
            \caption{}
         \end{subfigure}
        \begin{subfigure}{.24\linewidth}
            \includegraphics[width=\linewidth]{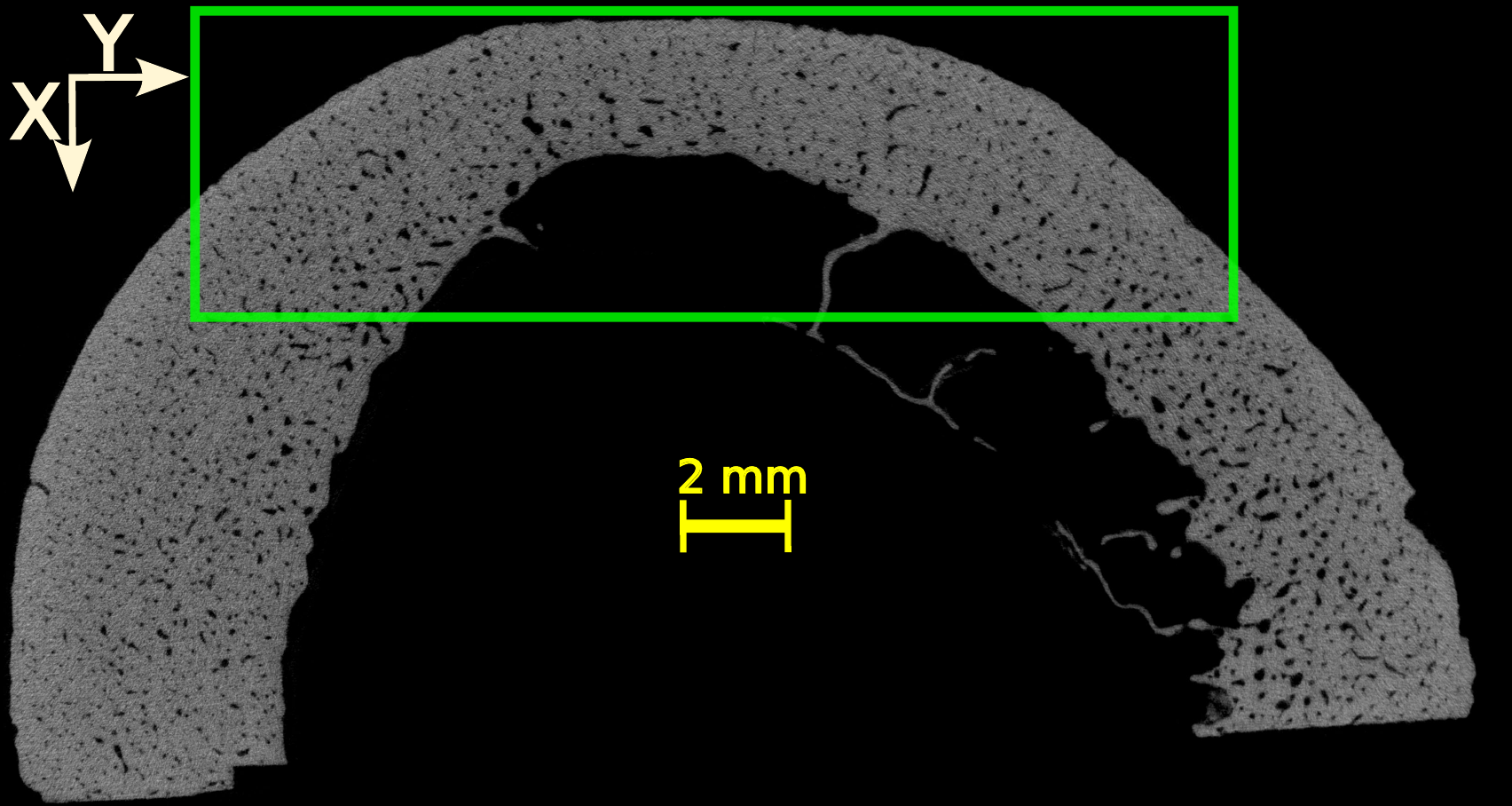}
            \caption{}
         \end{subfigure}
        \begin{subfigure}{.24\linewidth}
            \includegraphics[width=\linewidth]{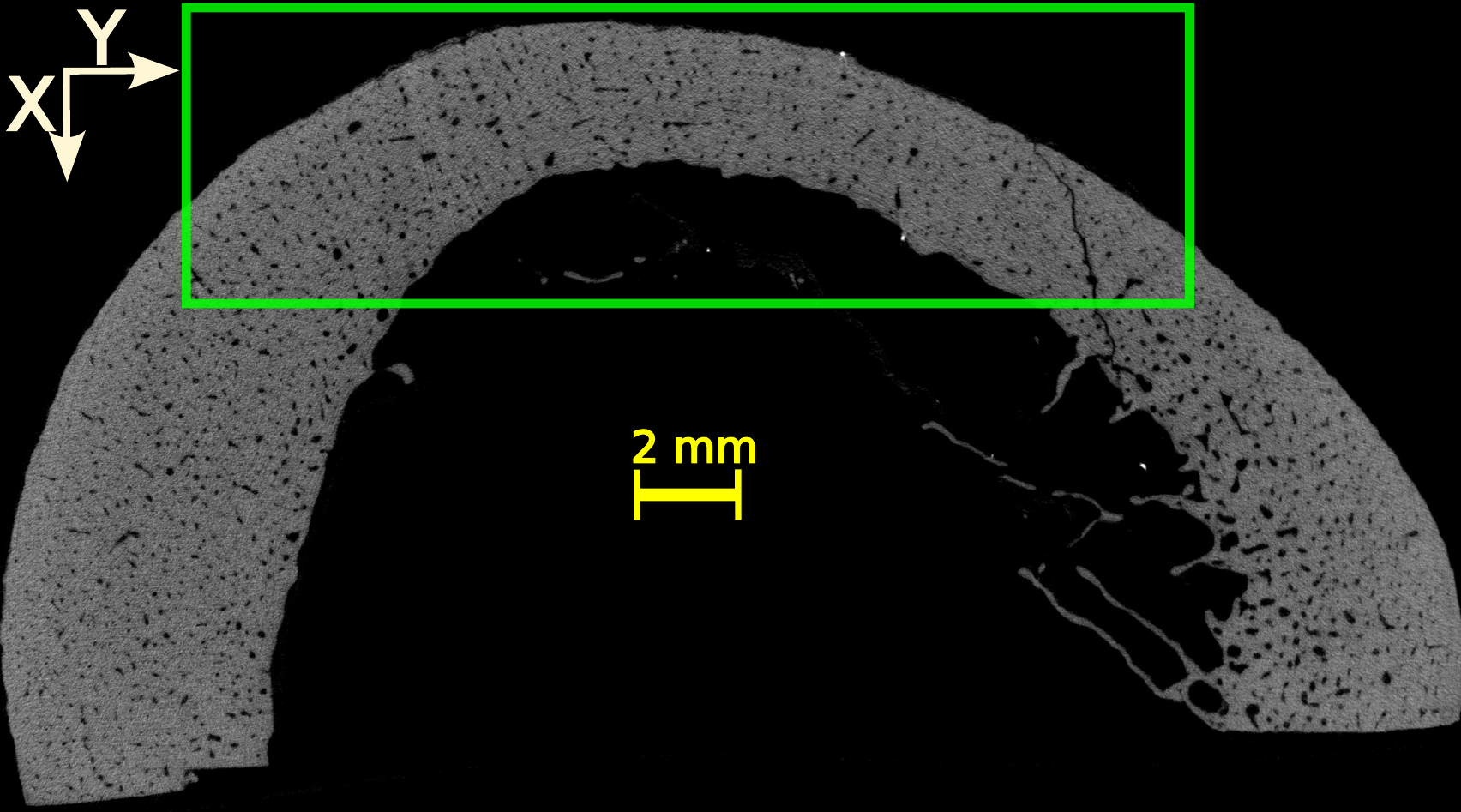}
            \caption{}
         \end{subfigure}
        \begin{subfigure}{.24\linewidth}
            \includegraphics[width=\linewidth]{images/results/rotated_HW_SA_Trans_Phantom_227G_B05_1.pdf}
            \caption{}
        \end{subfigure}   
        \begin{subfigure}{.24\linewidth}
            \includegraphics[width=\linewidth]{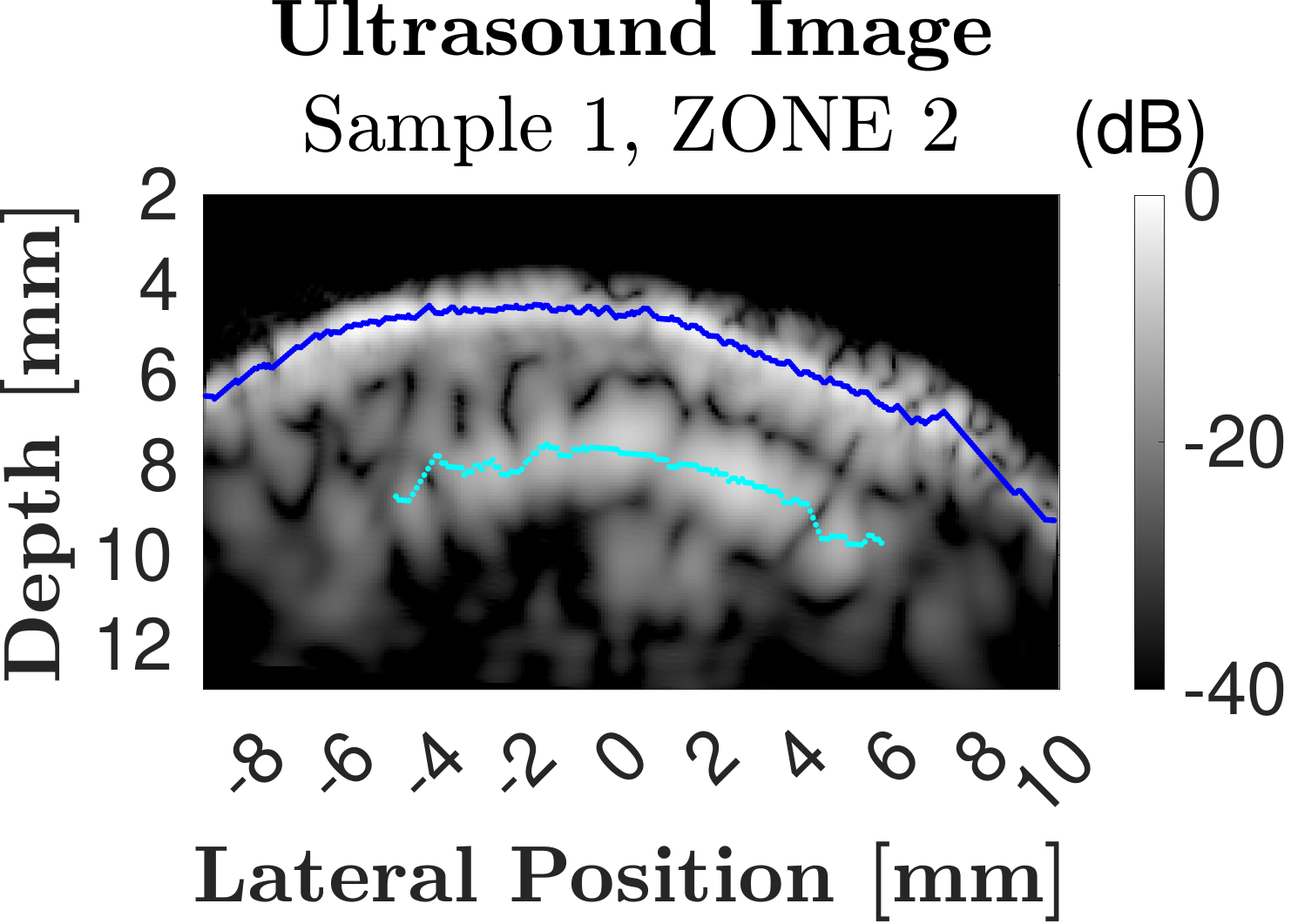}
            \caption{}
        \end{subfigure}
        \begin{subfigure}{.24\linewidth}
            \includegraphics[width=\linewidth]{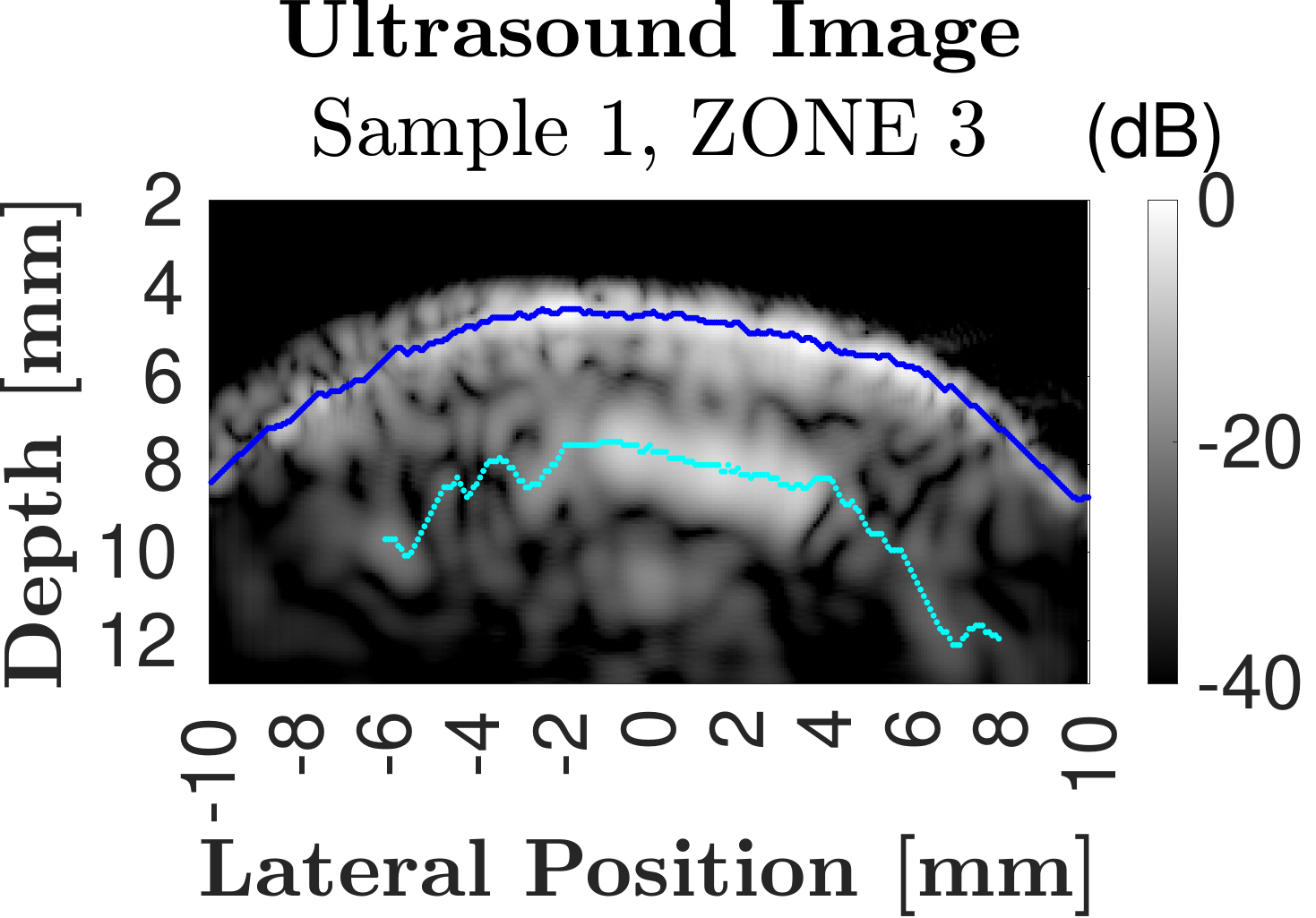}
            \caption{}
        \end{subfigure}   
        \begin{subfigure}{.24\linewidth}
            \includegraphics[width=\linewidth]{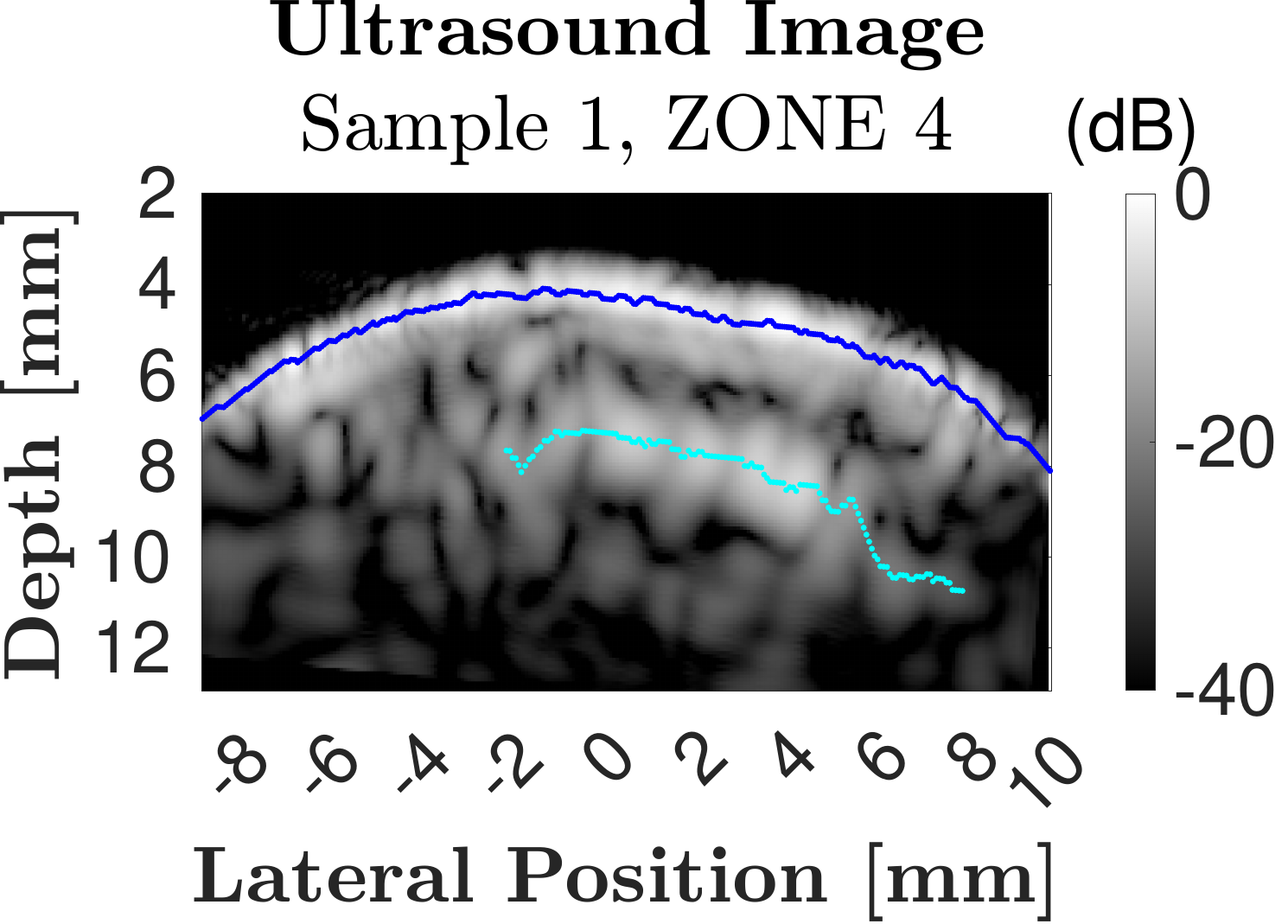}
            \caption{}
        \end{subfigure}   
        \begin{subfigure}{.24\linewidth}
            \includegraphics[width=\linewidth]{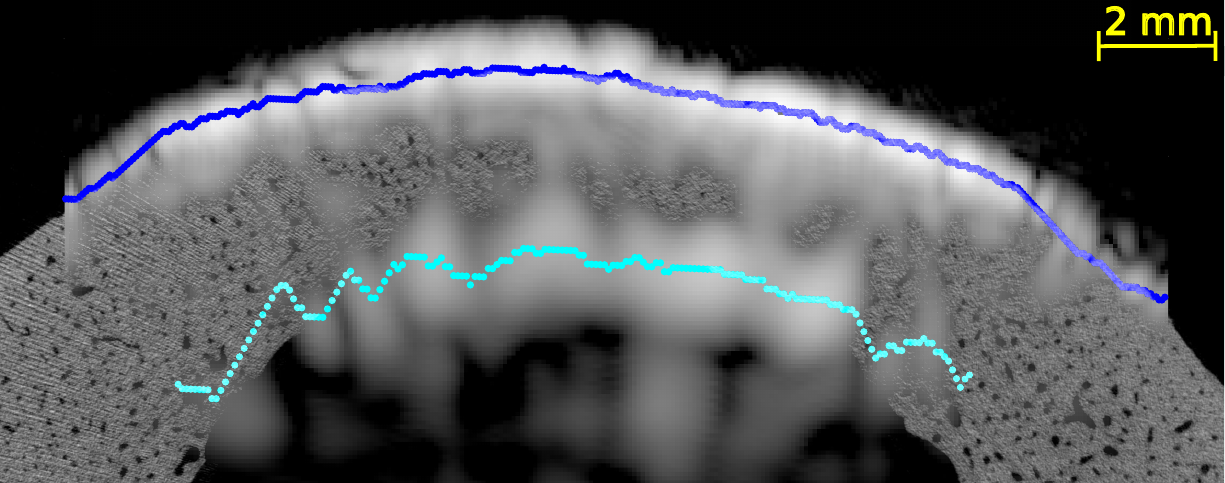}
            \caption{}
        \end{subfigure} 
        \begin{subfigure}{.24\linewidth}
            \includegraphics[width=\linewidth]{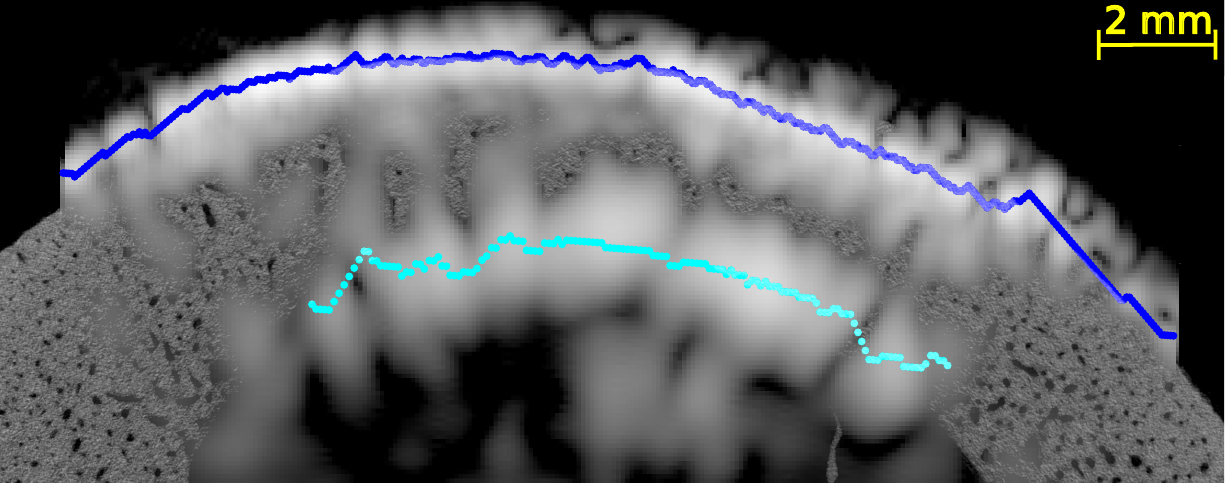}
            \caption{}
        \end{subfigure} 
        \begin{subfigure}{.24\linewidth}
            \includegraphics[width=\linewidth]{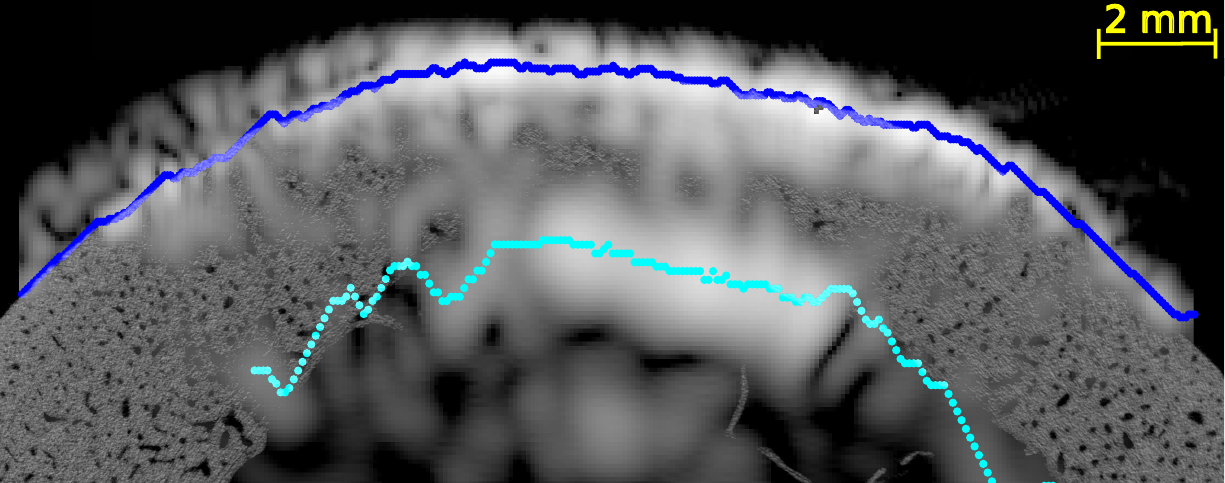}
            \caption{}
        \end{subfigure} 
        \begin{subfigure}{.24\linewidth}
            \includegraphics[width=\linewidth]{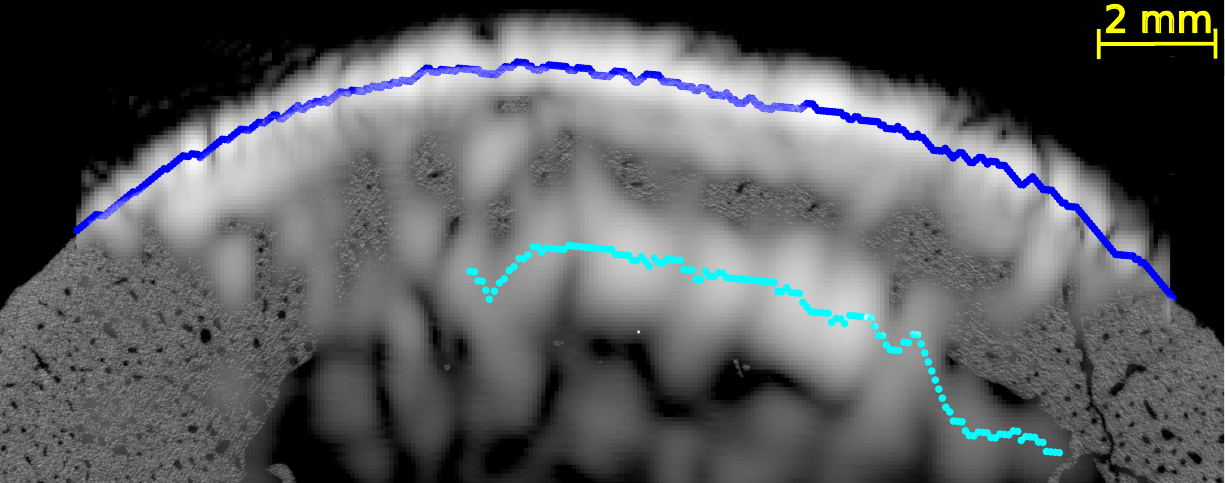}
            \caption{}
        \end{subfigure} 
        
        \caption{Comparison of micro-CT and ultrasound images for sample 1. One image, i.e., from one of the repetitions, was selected for each VOI. The first row displays X-ray images; the ROI in the green rectangle corresponds approximately to the area probed with ultrasound. The second row shows ultrasound images displaying raw segmentation (from Dijkstra' algorithm) of the cortex (periosteal and endosteal interfaces). The last row shows the superimposition of ultrasound and X-ray images and the raw segmentation of the cortex from ultrasound images.}
        \label{supplementary_materials:sample_1}
    \end{figure}
    
    \end{landscape}}

\afterpage{
\begin{landscape}
\begin{figure}[htb!]
        \centering
        \begin{subfigure}{.24\linewidth}
            \includegraphics[width=\linewidth]{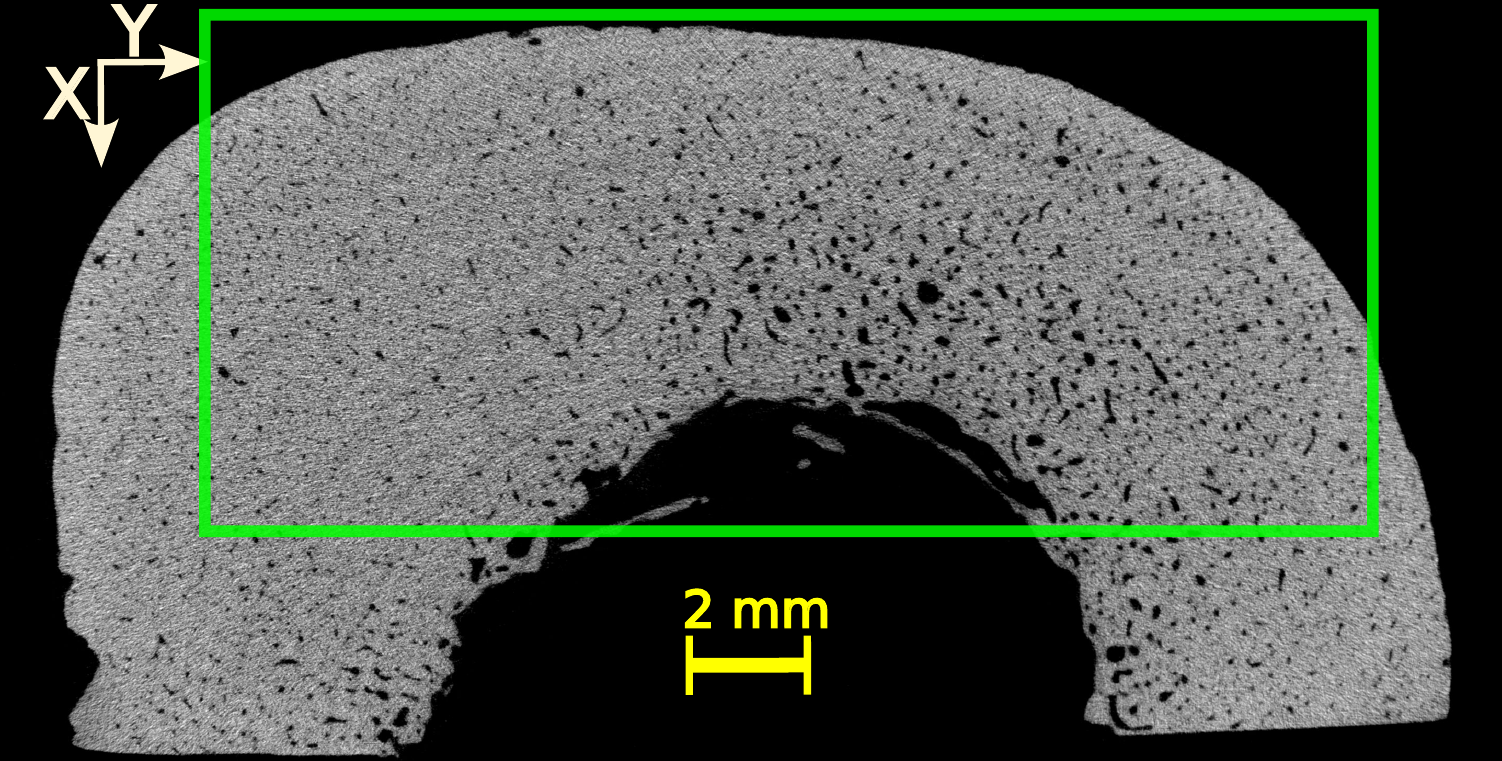}
            \caption{}
         \end{subfigure}
        \begin{subfigure}{.24\linewidth}
            \includegraphics[width=\linewidth]{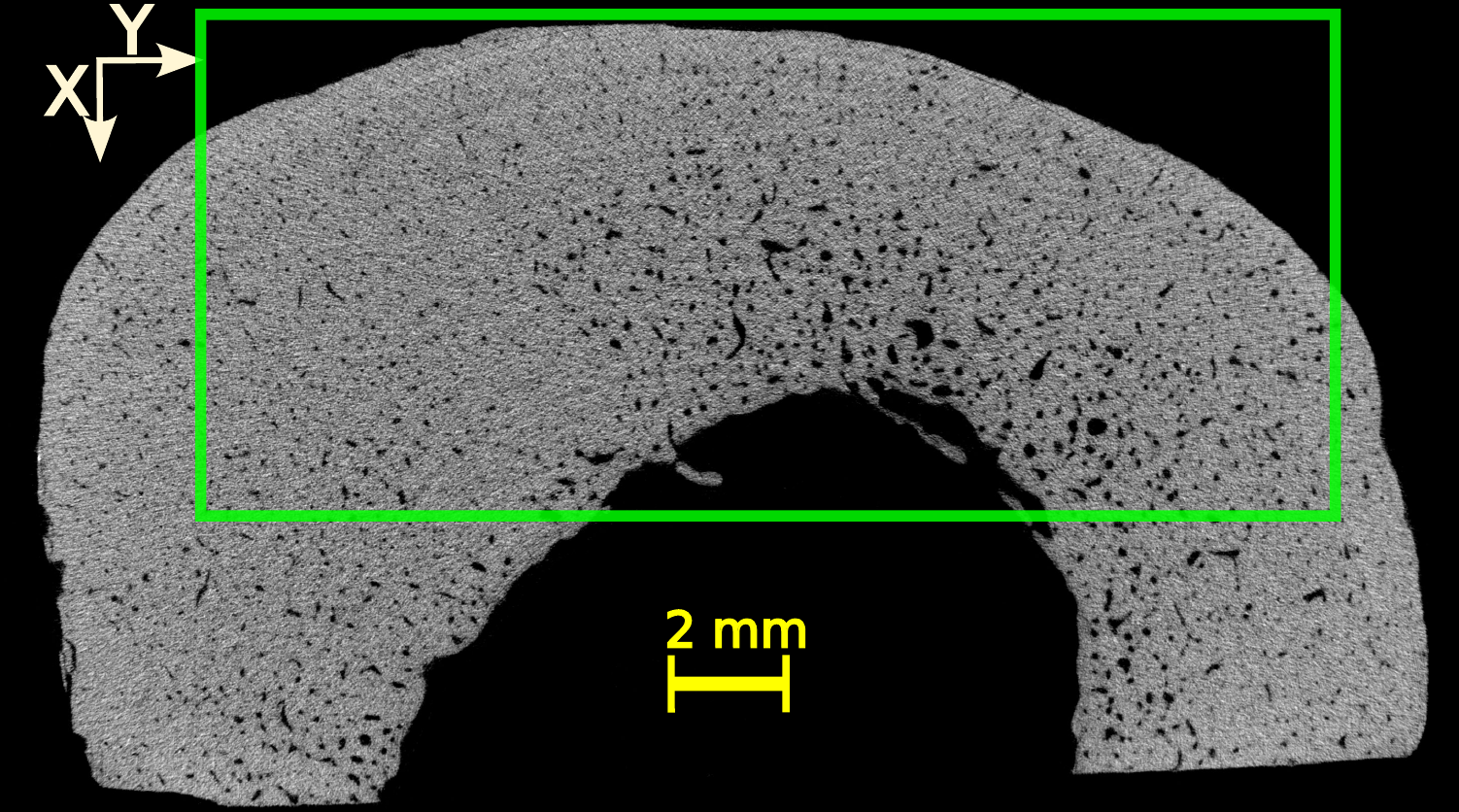}
            \caption{}
         \end{subfigure}
        \begin{subfigure}{.24\linewidth}
            \includegraphics[width=\linewidth]{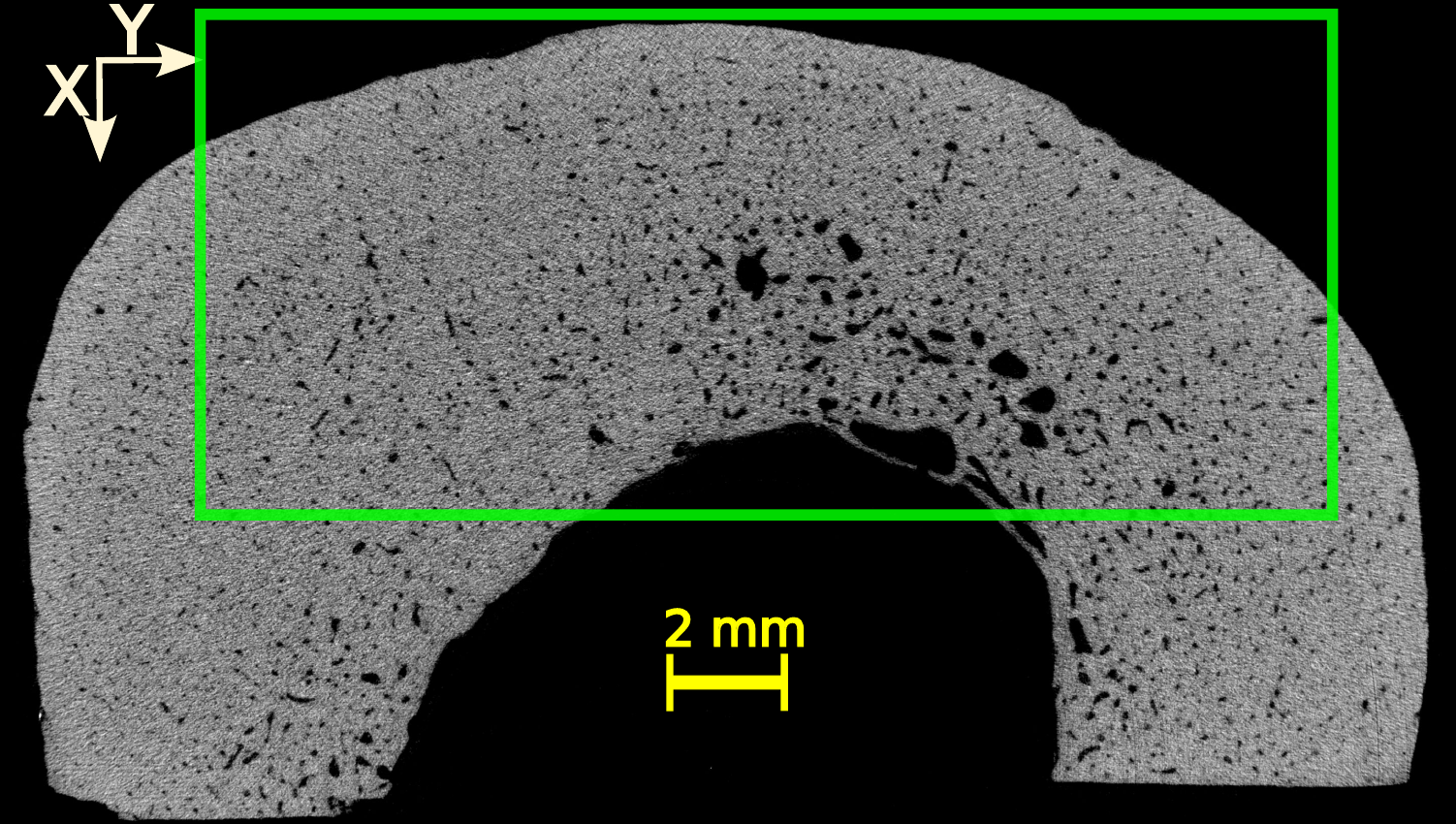}
            \caption{}
         \end{subfigure}
        \begin{subfigure}{.24\linewidth}
            \includegraphics[width=\linewidth]{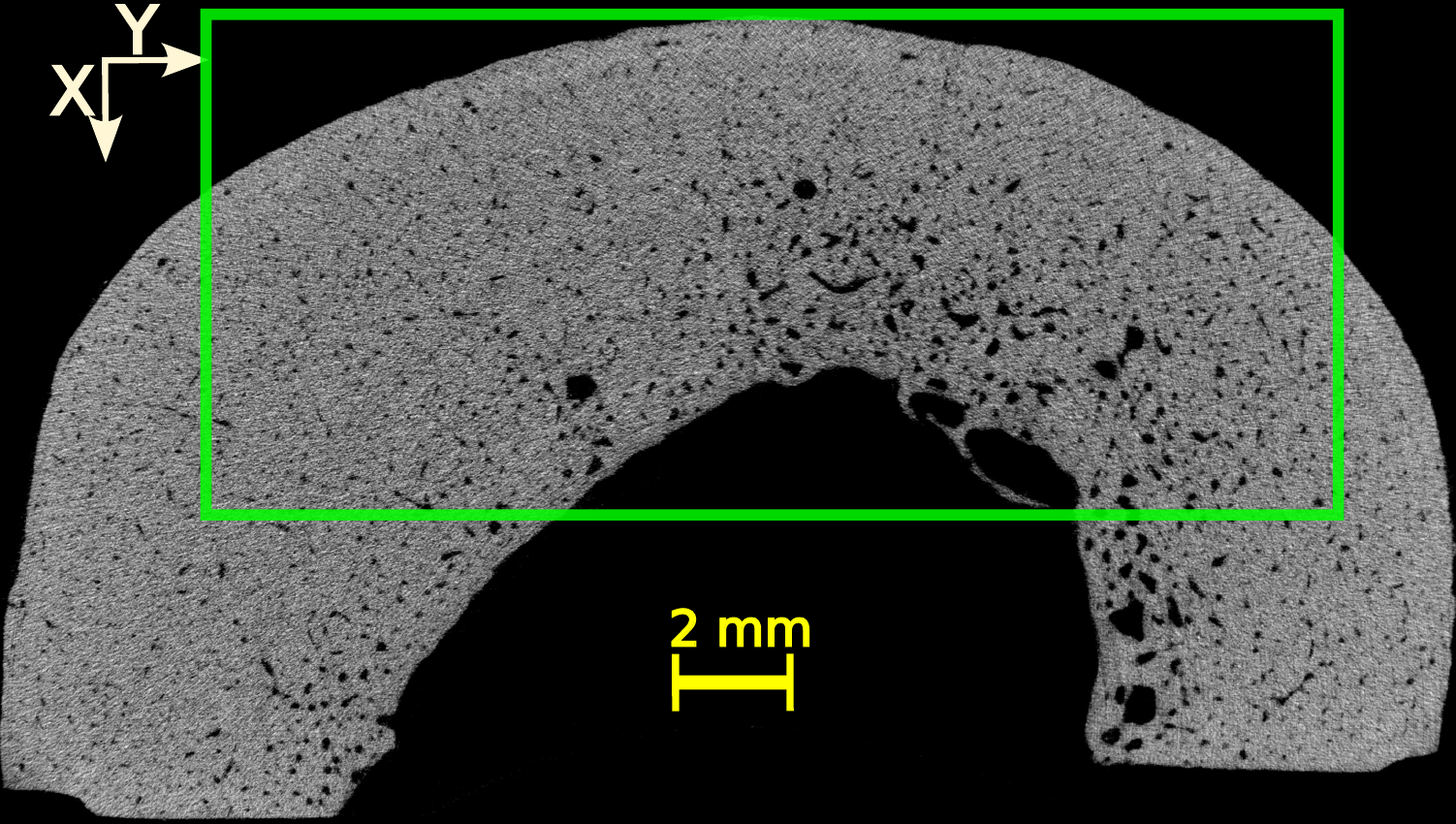}
            \caption{}
         \end{subfigure}
        \begin{subfigure}{.24\linewidth}
            \includegraphics[width=\linewidth]{images/results/rotated_HW_SA_Trans_Phantom_267G_B06_1.pdf}
            \caption{}
        \end{subfigure}   
        \begin{subfigure}{.24\linewidth}
            \includegraphics[width=\linewidth]{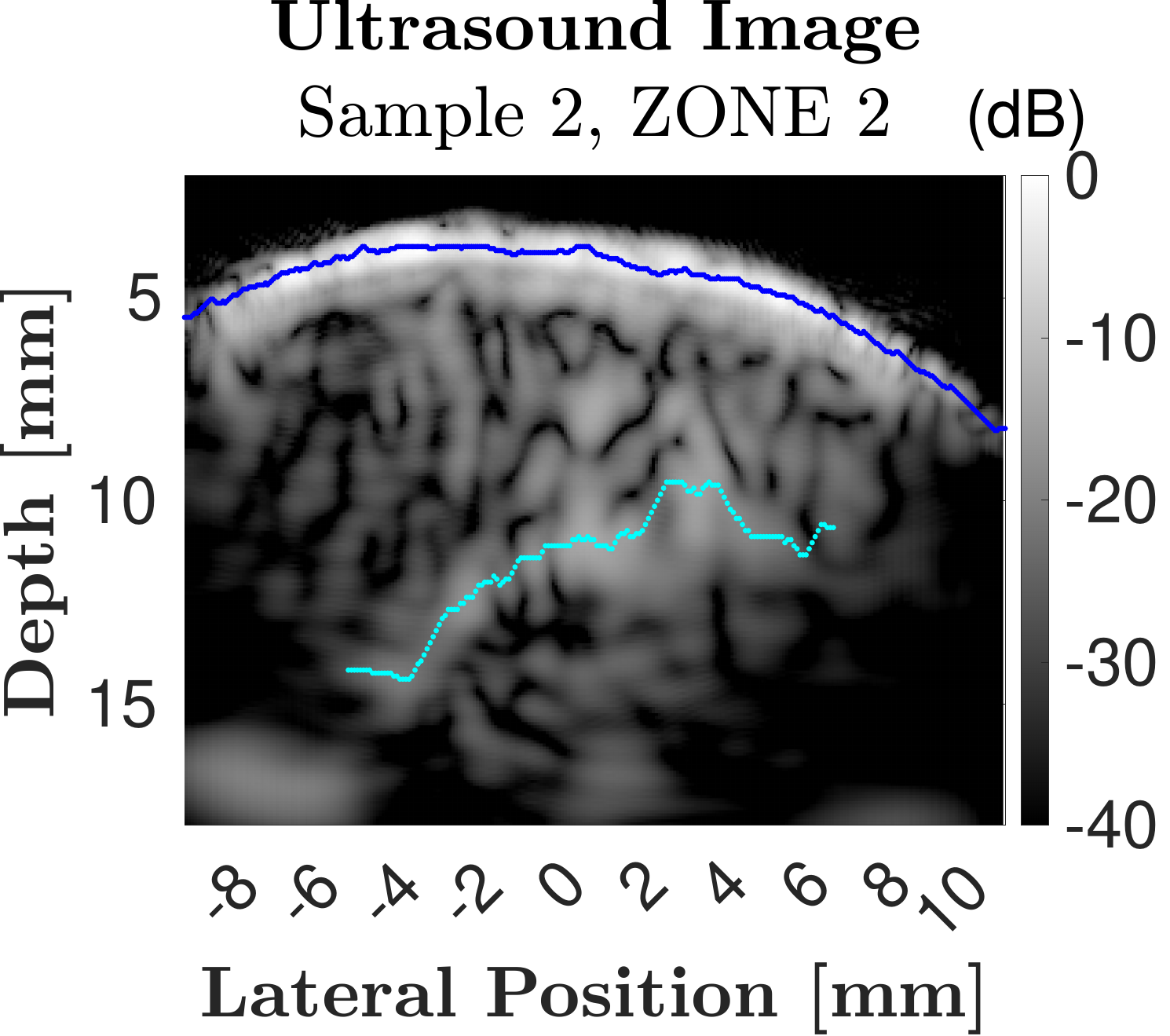}
            \caption{}
        \end{subfigure}
        \begin{subfigure}{.24\linewidth}
            \includegraphics[width=\linewidth]{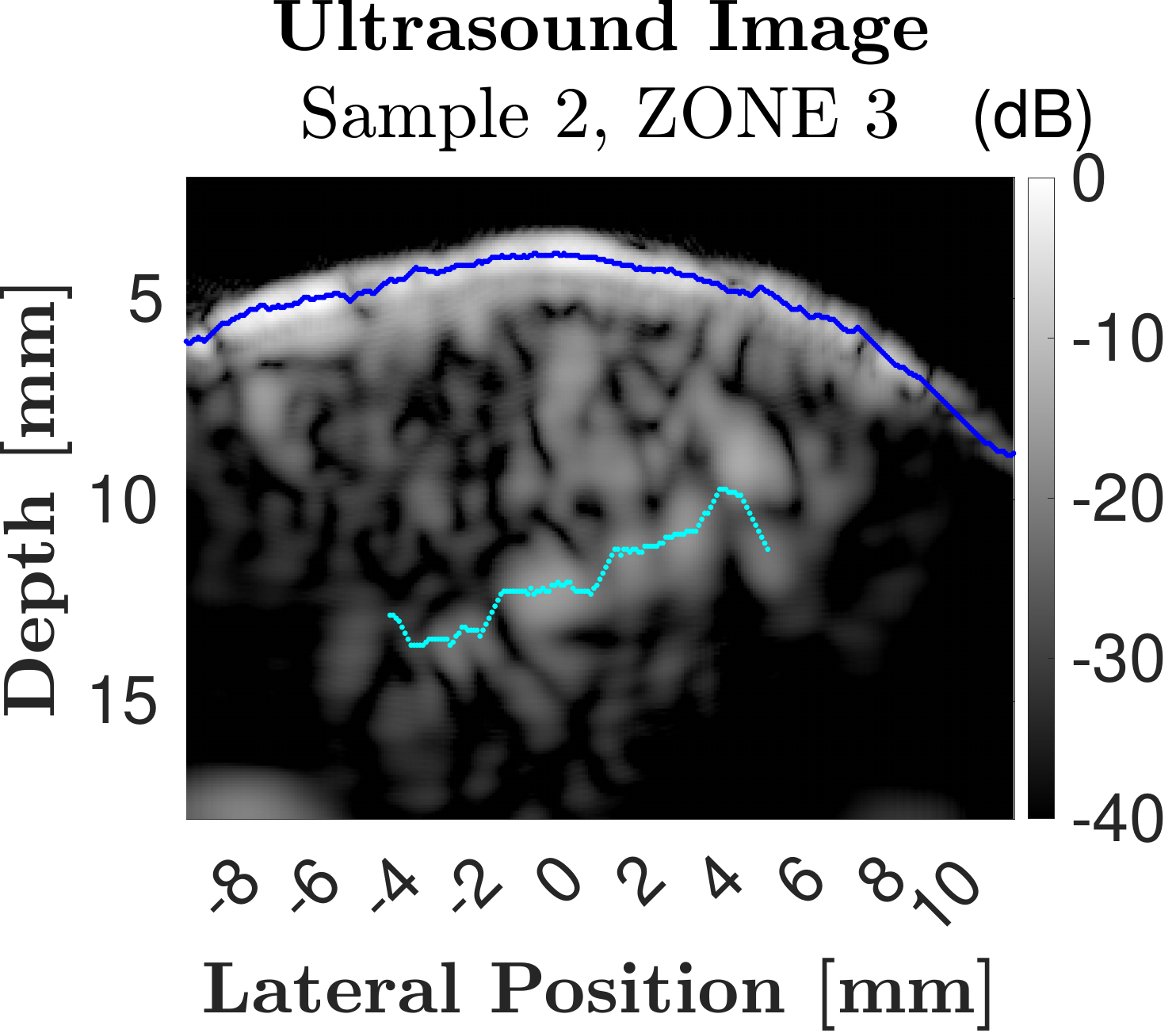}
            \caption{}
        \end{subfigure}   
        \begin{subfigure}{.24\linewidth}
            \includegraphics[width=\linewidth]{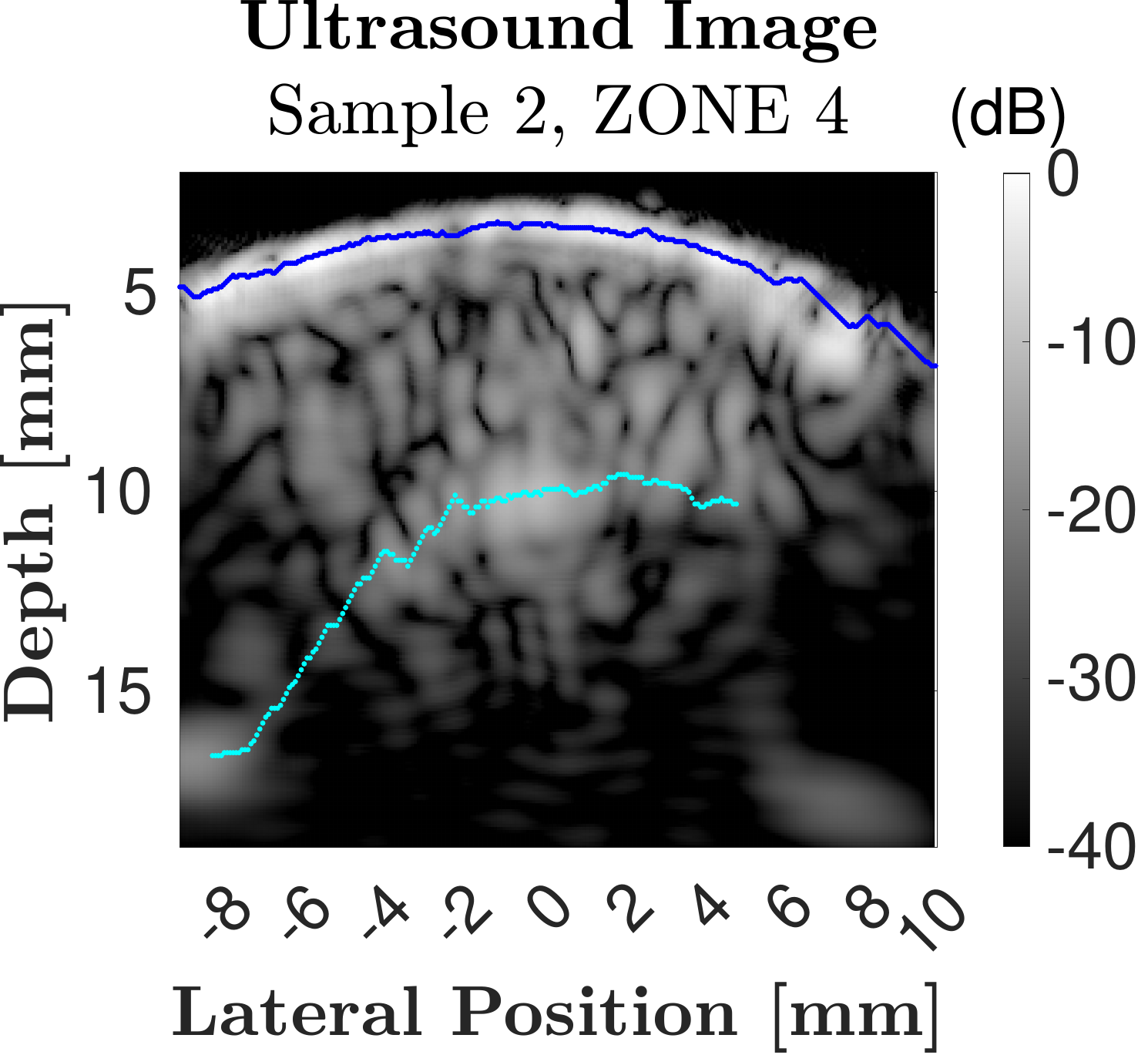}
            \caption{}
        \end{subfigure}   
        \begin{subfigure}{.24\linewidth}
            \includegraphics[width=\linewidth]{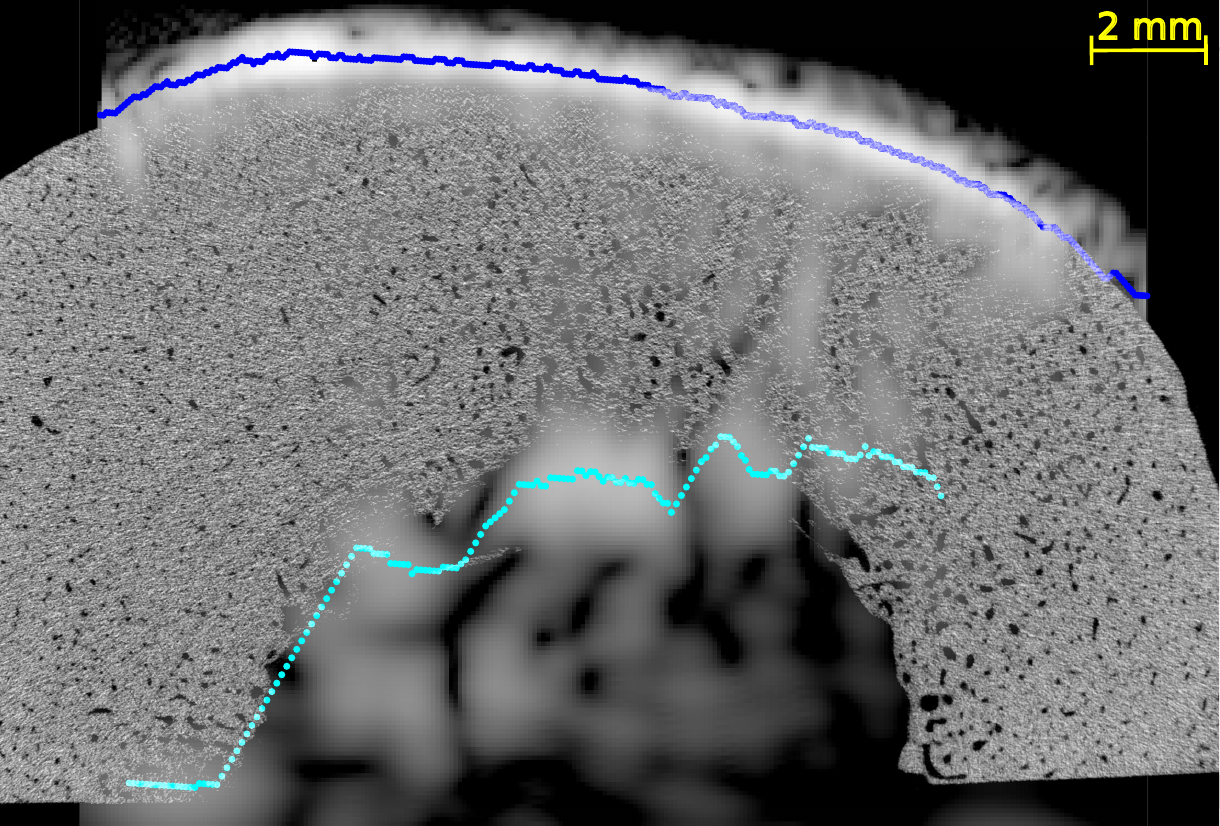}
            \caption{}
        \end{subfigure} 
        \begin{subfigure}{.24\linewidth}
            \includegraphics[width=\linewidth]{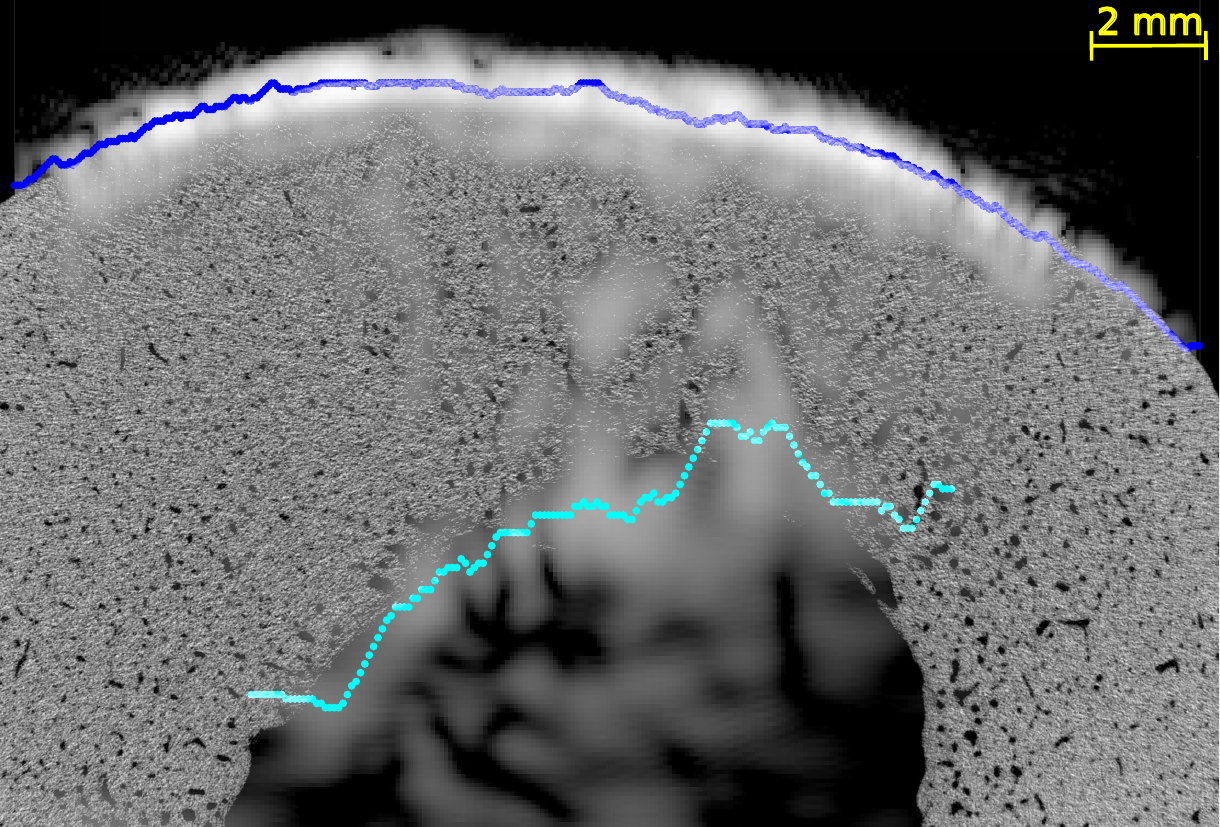}
            \caption{}
        \end{subfigure} 
        \begin{subfigure}{.24\linewidth}
            \includegraphics[width=\linewidth]{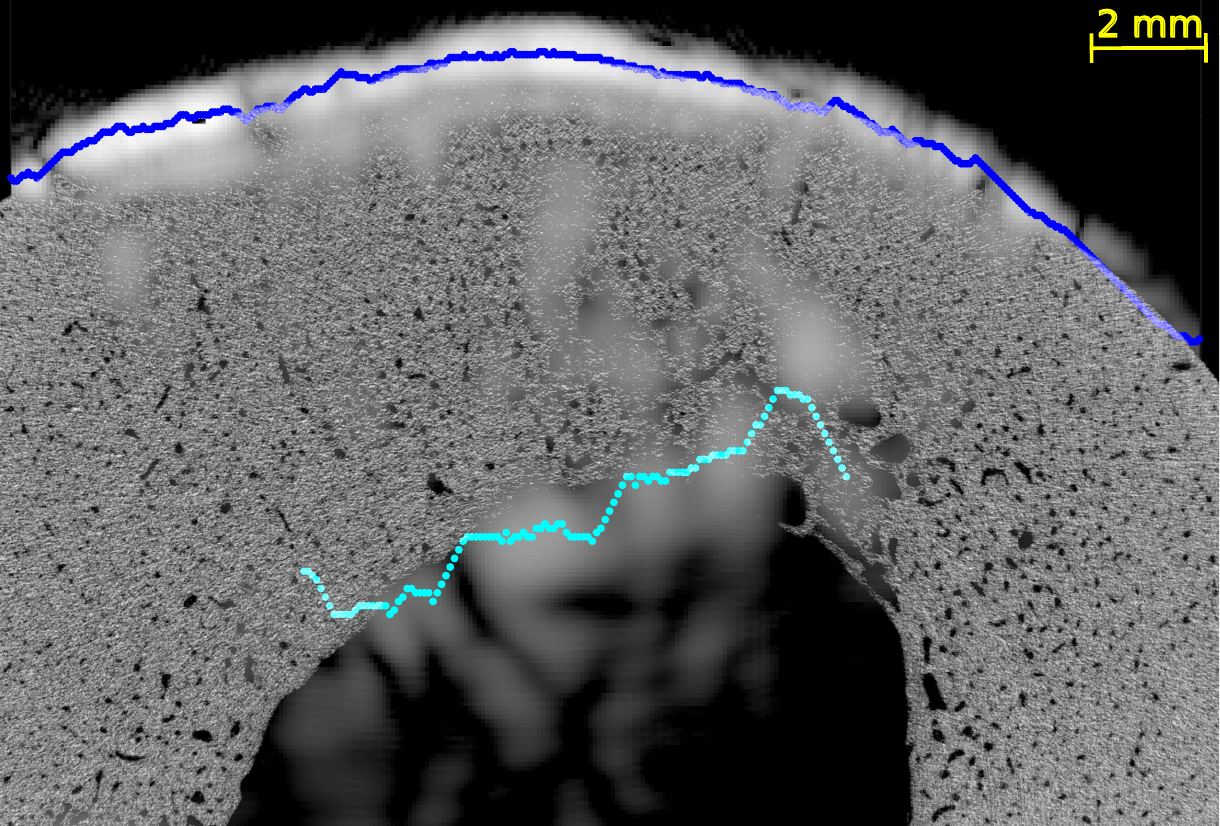}
            \caption{}
        \end{subfigure} 
        \begin{subfigure}{.24\linewidth}
            \includegraphics[width=\linewidth]{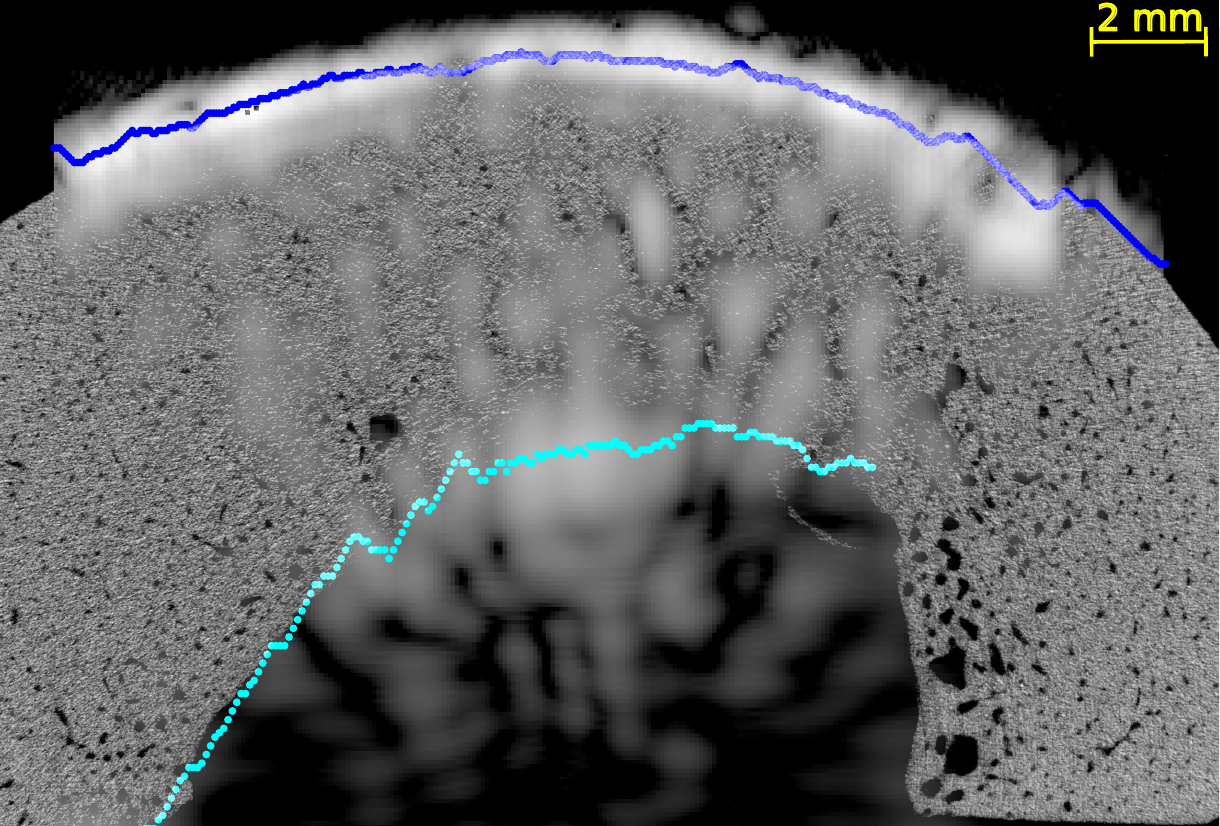}
            \caption{}
        \end{subfigure} 
        
        \caption{Same caption as Figure~\ref{supplementary_materials:sample_1}, applied to sample 2.}
        \label{supplementary_materials:sample_2}
    \end{figure}
    
    \end{landscape}
    }
{\begin{landscape}
\begin{figure}[htb!]
        \centering
        \begin{subfigure}{.24\linewidth}
            \includegraphics[width=\linewidth]{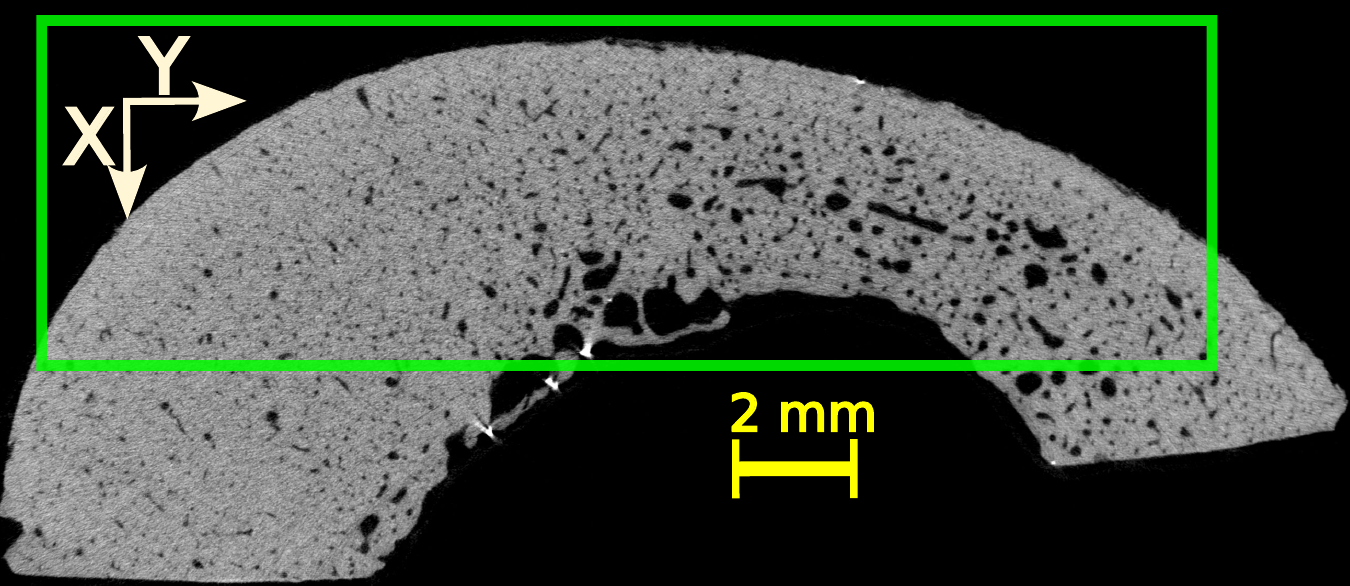}
            \caption{}
         \end{subfigure}
        \begin{subfigure}{.24\linewidth}
            \includegraphics[width=\linewidth]{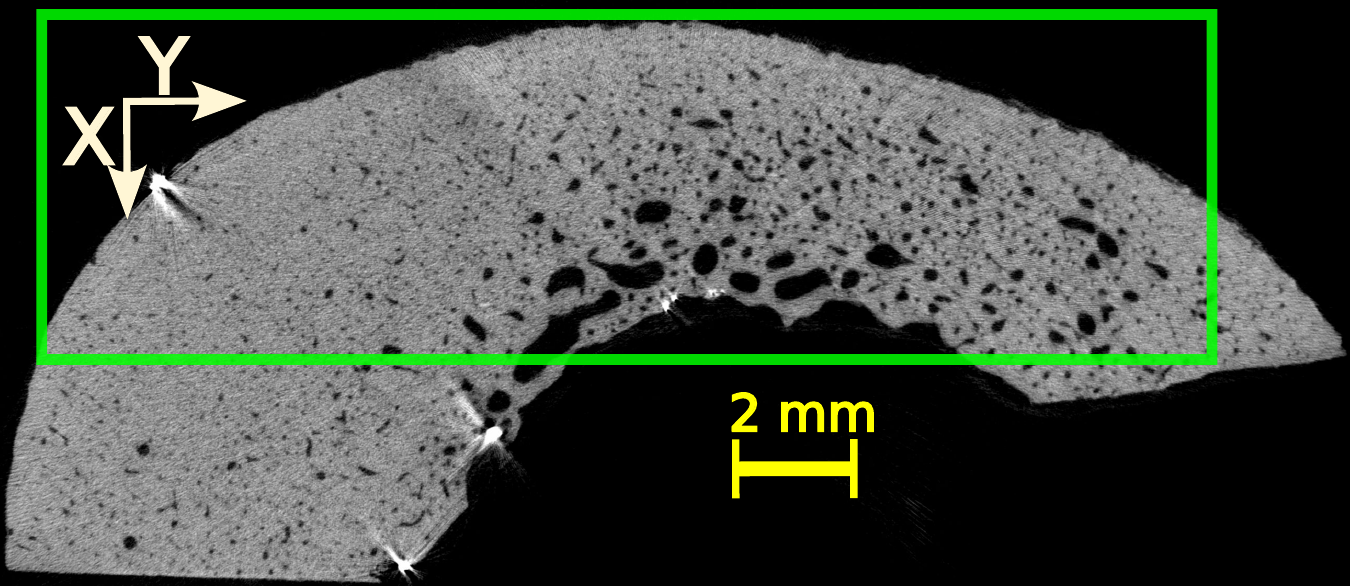}
            \caption{}
         \end{subfigure}
        \begin{subfigure}{.24\linewidth}
            \includegraphics[width=\linewidth]{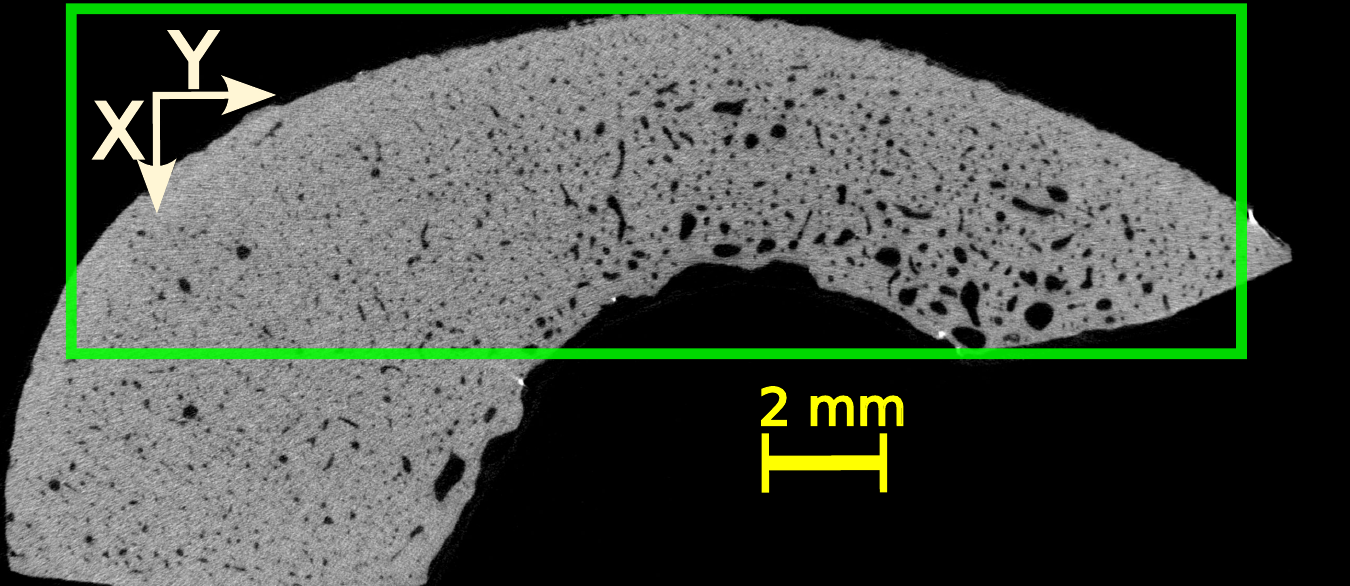}
            \caption{}
         \end{subfigure}
        \begin{subfigure}{.24\linewidth}
            \includegraphics[width=\linewidth]{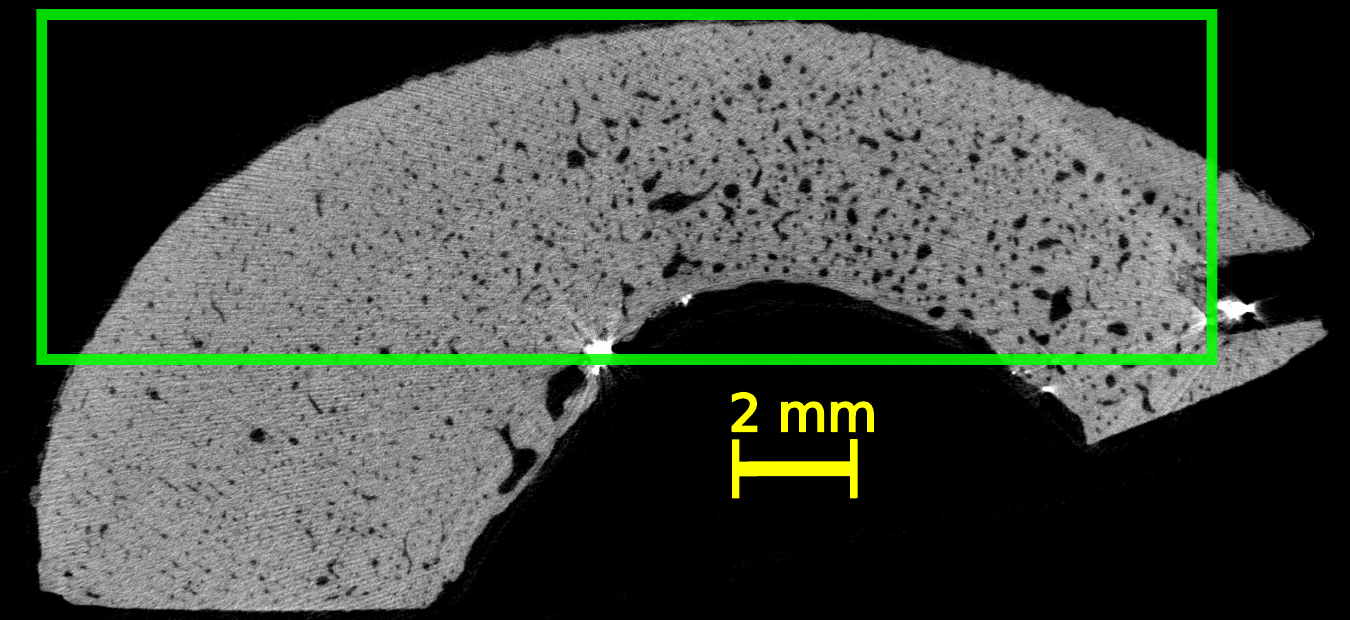}
            \caption{}
         \end{subfigure}
        \begin{subfigure}{.24\linewidth}
            \includegraphics[width=\linewidth]{images/results/rotated_HW_SA_Trans_Phantom_245D_B01_1.pdf}
            \caption{}
        \end{subfigure}   
        \begin{subfigure}{.24\linewidth}
            \includegraphics[width=\linewidth]{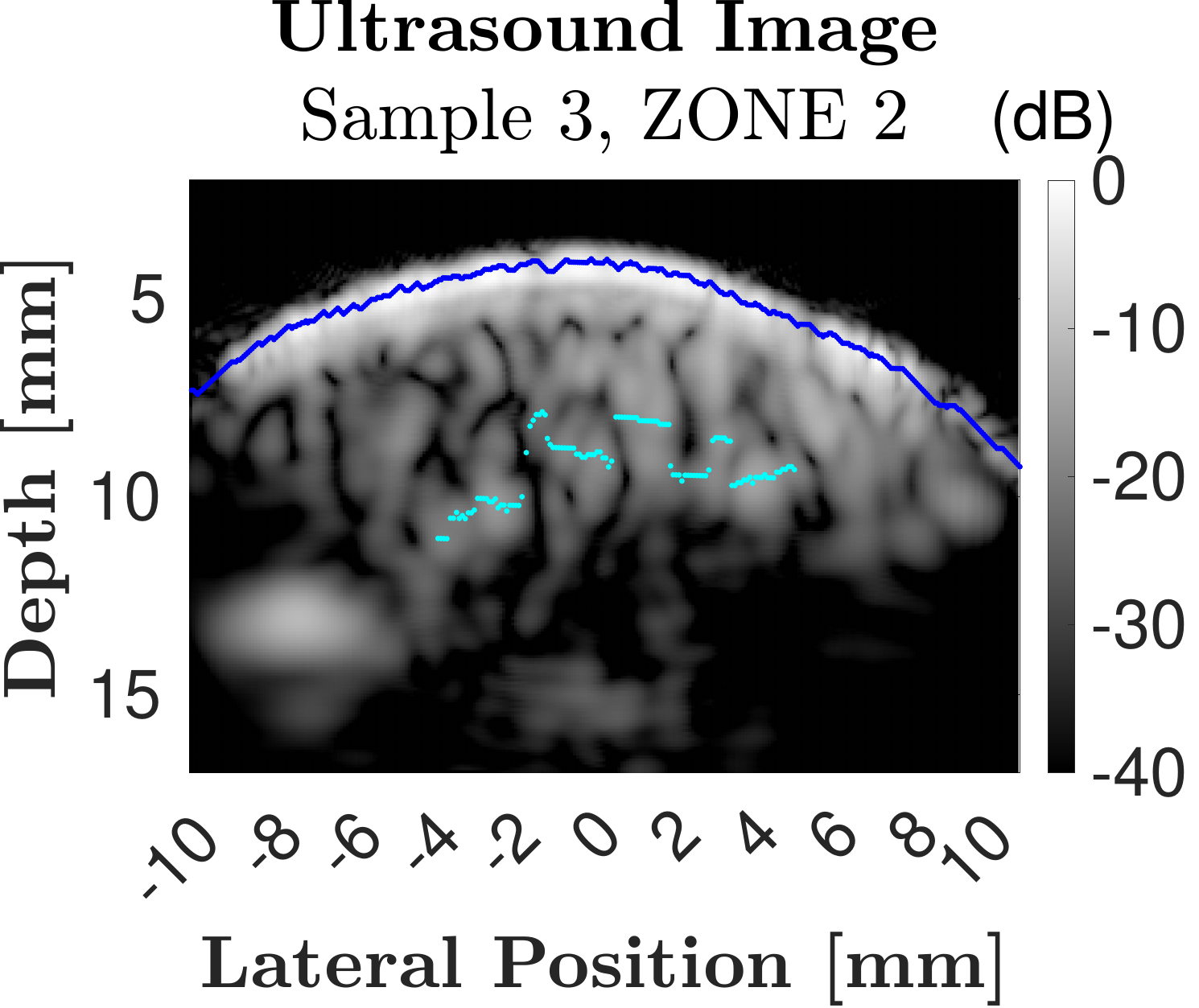}
            \caption{}
        \end{subfigure}
        \begin{subfigure}{.24\linewidth}
            \includegraphics[width=\linewidth]{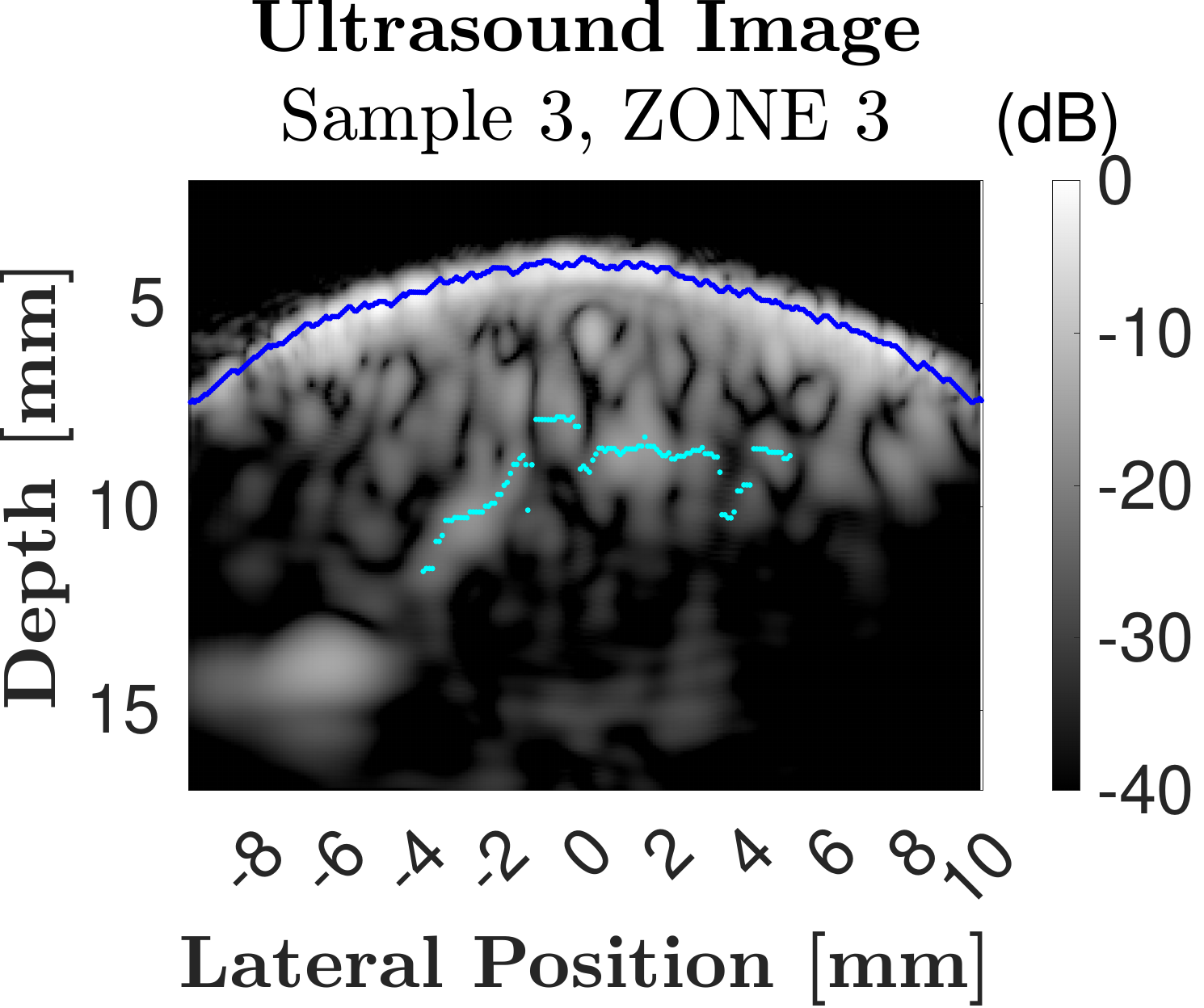}
            \caption{}
        \end{subfigure}   
        \begin{subfigure}{.24\linewidth}
            \includegraphics[width=\linewidth]{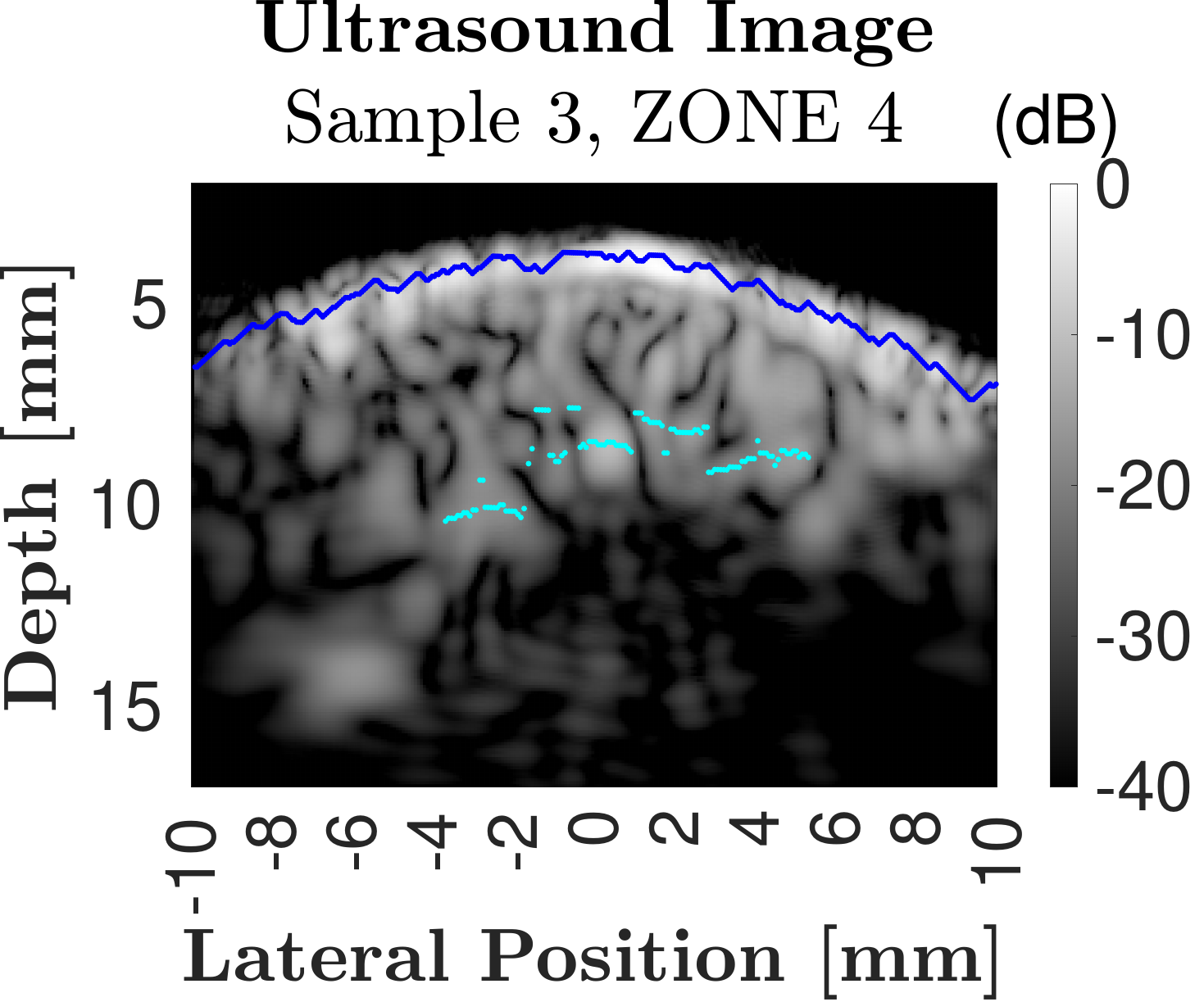}
            \caption{}
        \end{subfigure}   
        \begin{subfigure}{.24\linewidth}
            \includegraphics[width=\linewidth]{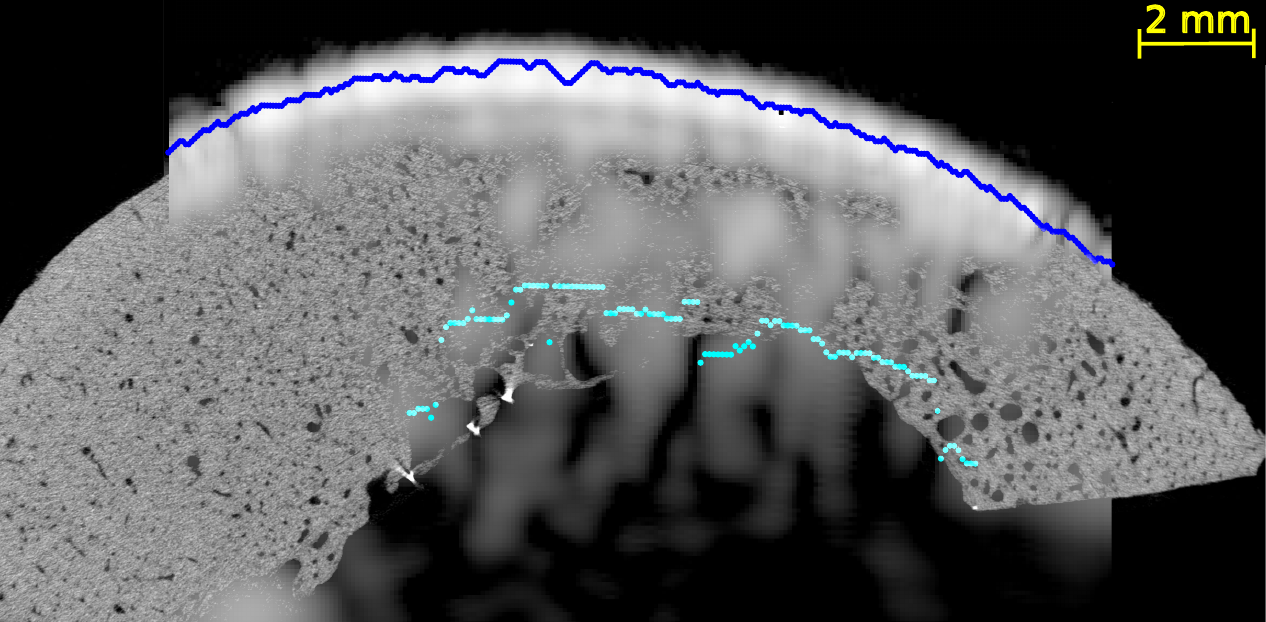}
            \caption{}
        \end{subfigure} 
        \begin{subfigure}{.24\linewidth}
            \includegraphics[width=\linewidth]{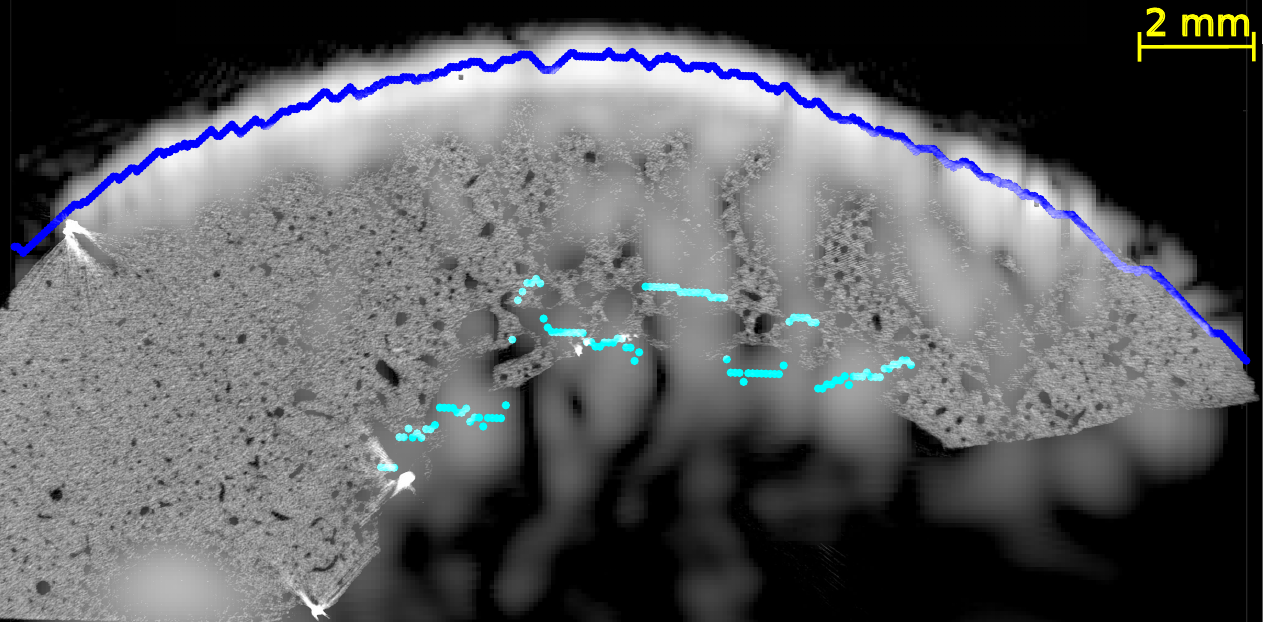}
            \caption{}
        \end{subfigure} 
        \begin{subfigure}{.24\linewidth}
            \includegraphics[width=\linewidth]{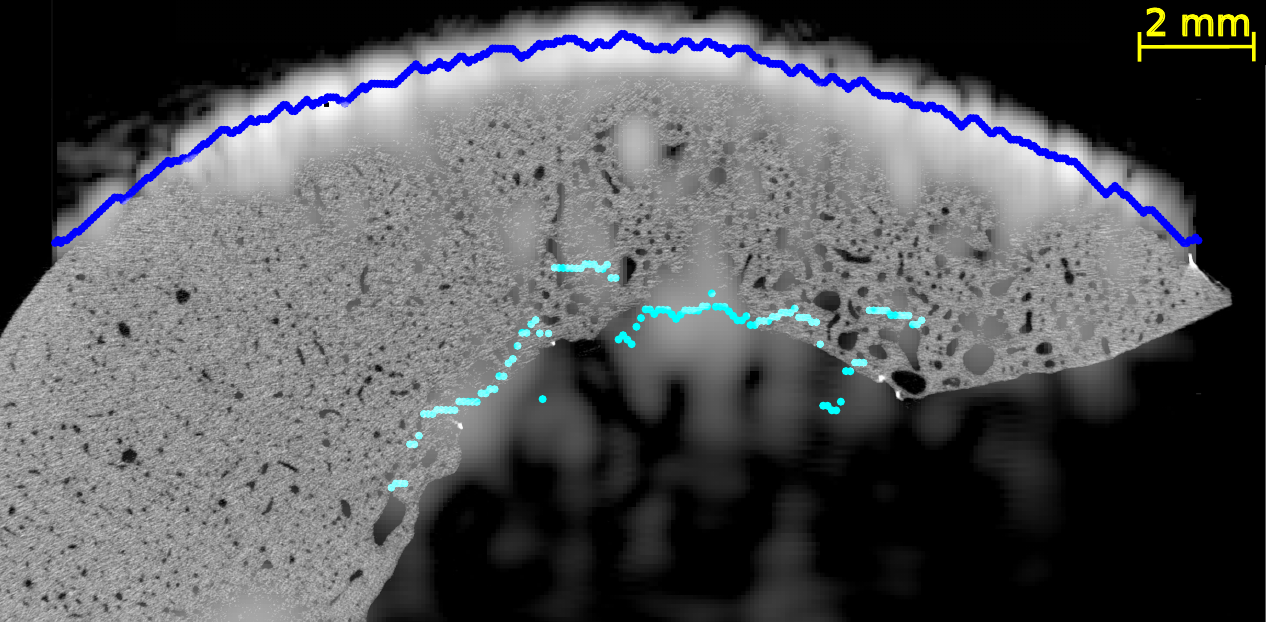}
            \caption{}
        \end{subfigure} 
        \begin{subfigure}{.24\linewidth}
            \includegraphics[width=\linewidth]{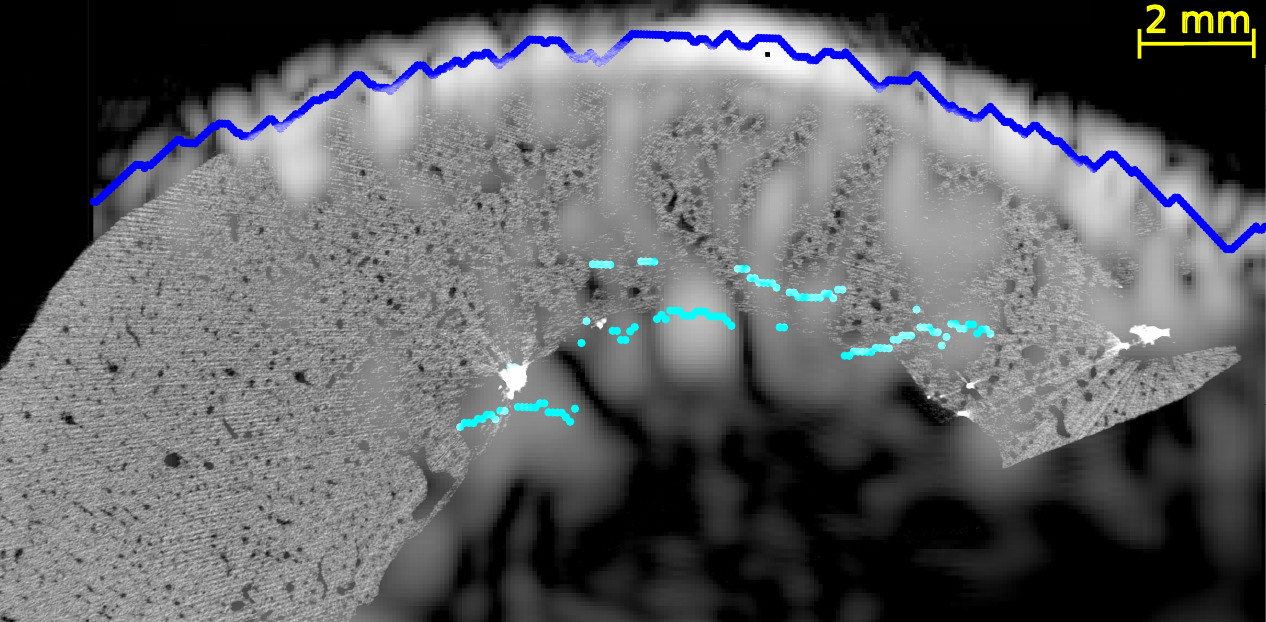}
            \caption{}
        \end{subfigure} 
        
        \caption{Same caption as Figure~\ref{supplementary_materials:sample_1}, applied to sample 3.}
        \label{supplementary_materials:sample_3}
    \end{figure}
    
    \end{landscape}
    }
{
\begin{landscape}
\begin{figure}[htb!]
        \centering
        \begin{subfigure}{.24\linewidth}
            \includegraphics[width=\linewidth]{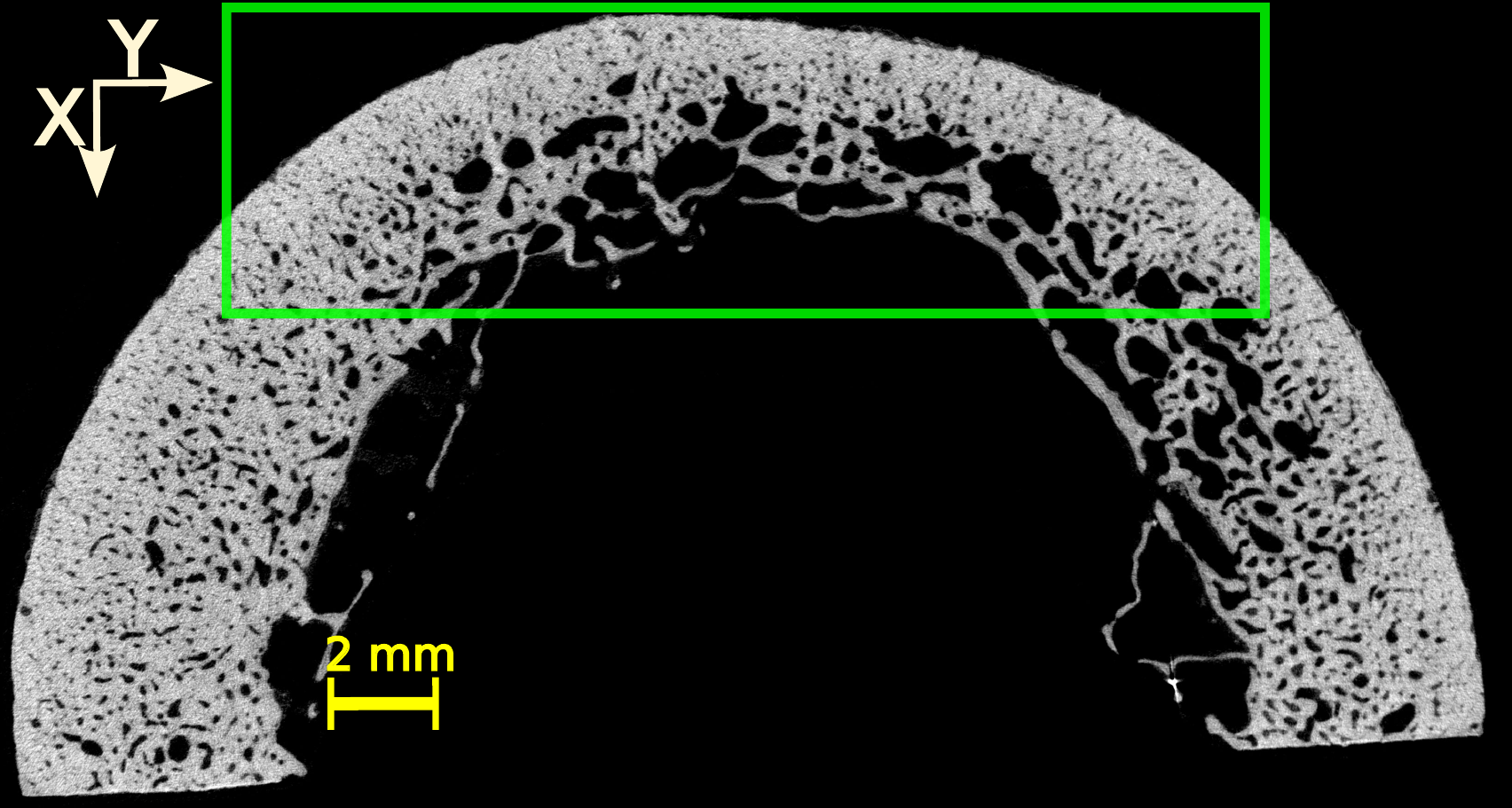}
            \caption{}
         \end{subfigure}
        \begin{subfigure}{.24\linewidth}
            \includegraphics[width=\linewidth]{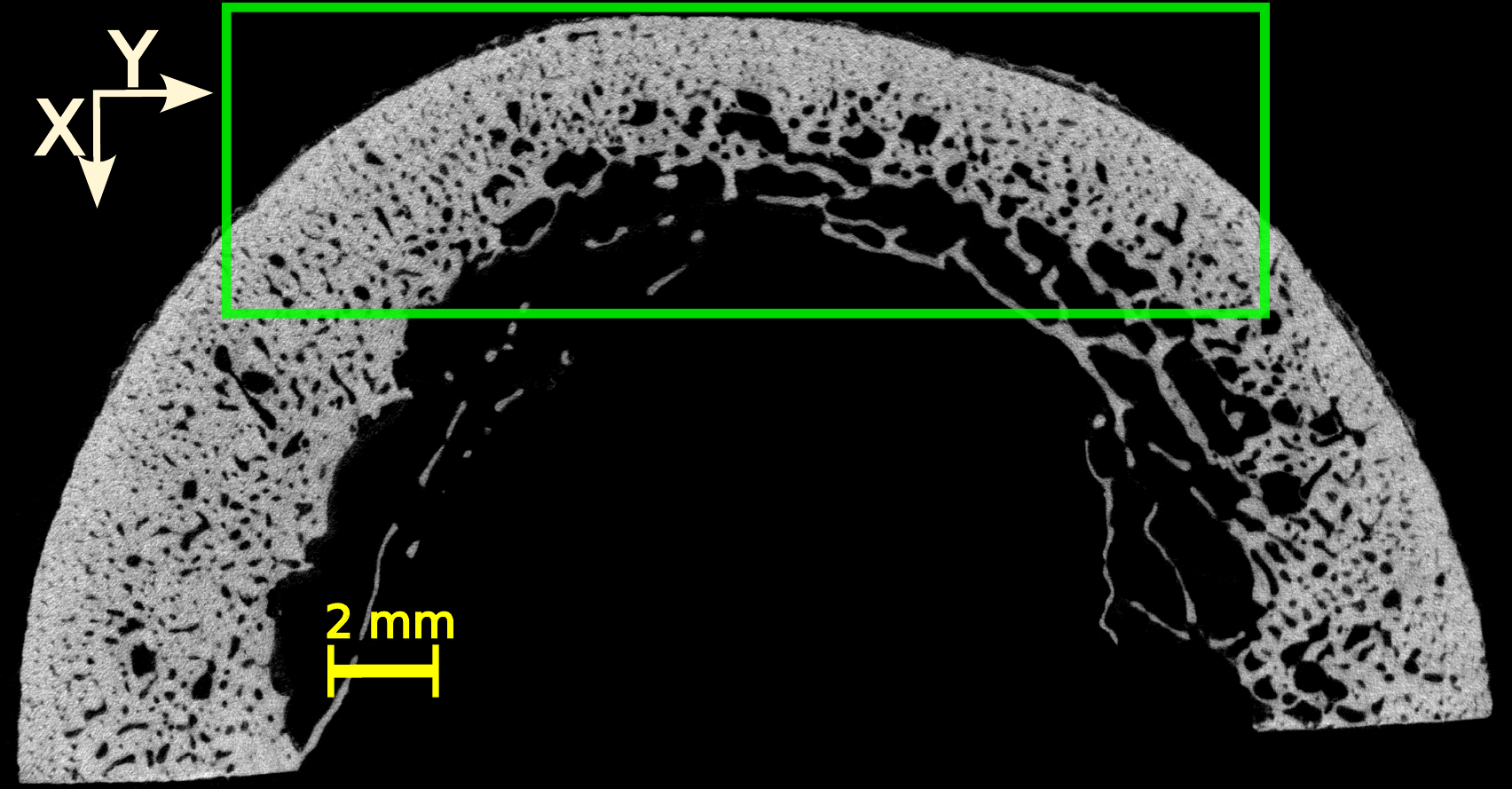}
            \caption{}
         \end{subfigure}
        \begin{subfigure}{.24\linewidth}
            \includegraphics[width=\linewidth]{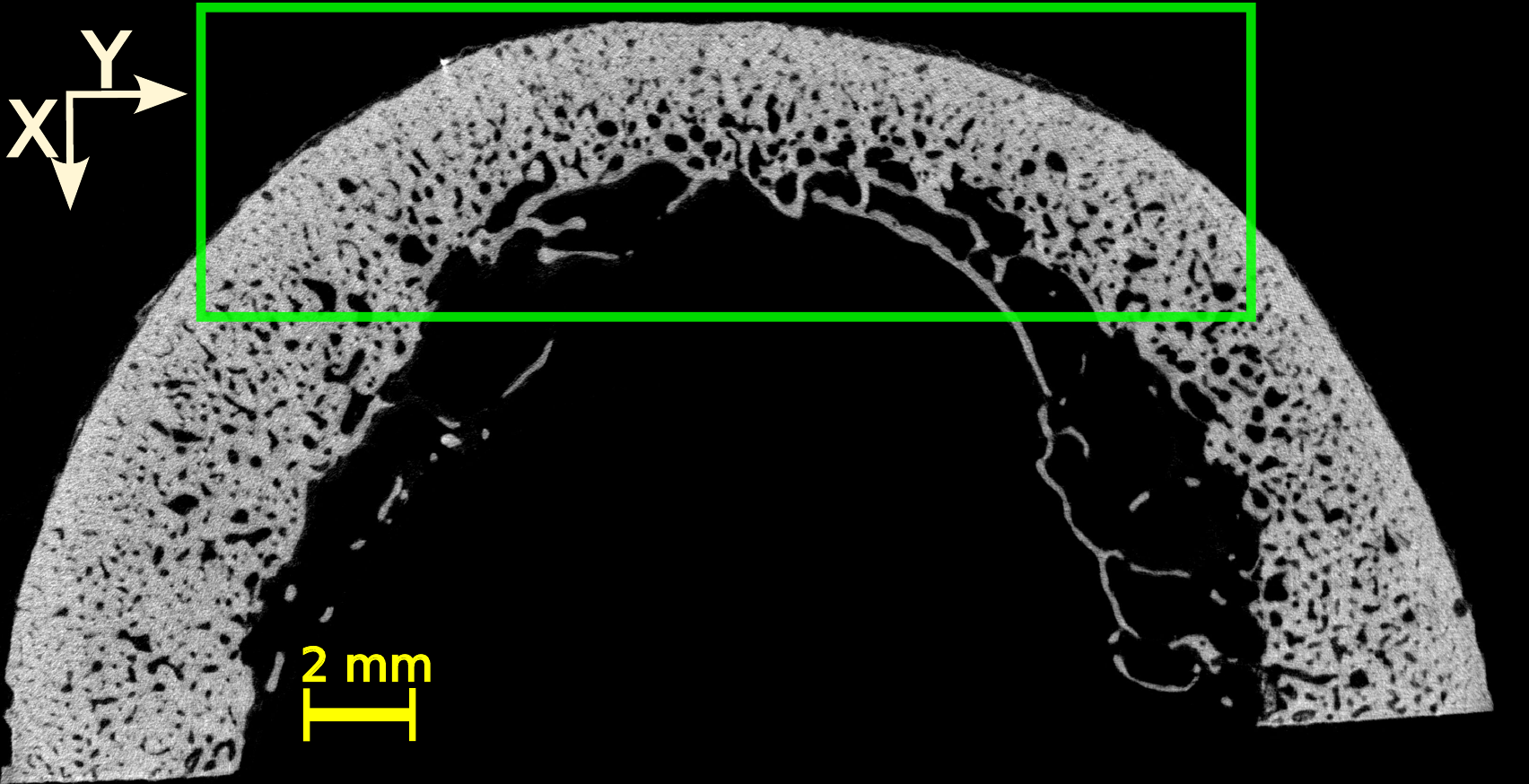}
            \caption{}
         \end{subfigure}
        \begin{subfigure}{.24\linewidth}
            \includegraphics[width=\linewidth]{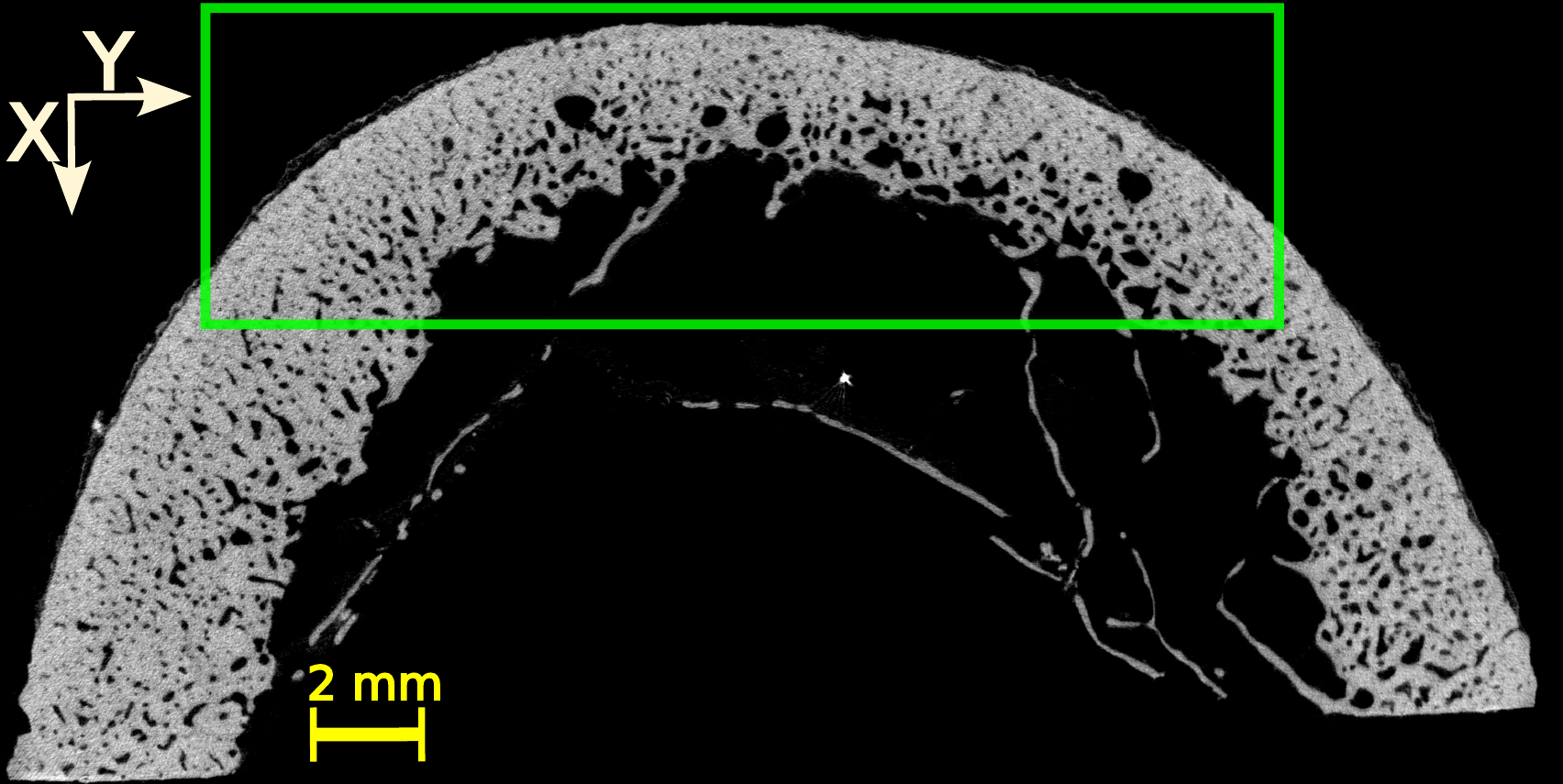}
            \caption{}
         \end{subfigure}
        \begin{subfigure}{.24\linewidth}
            \includegraphics[width=\linewidth]{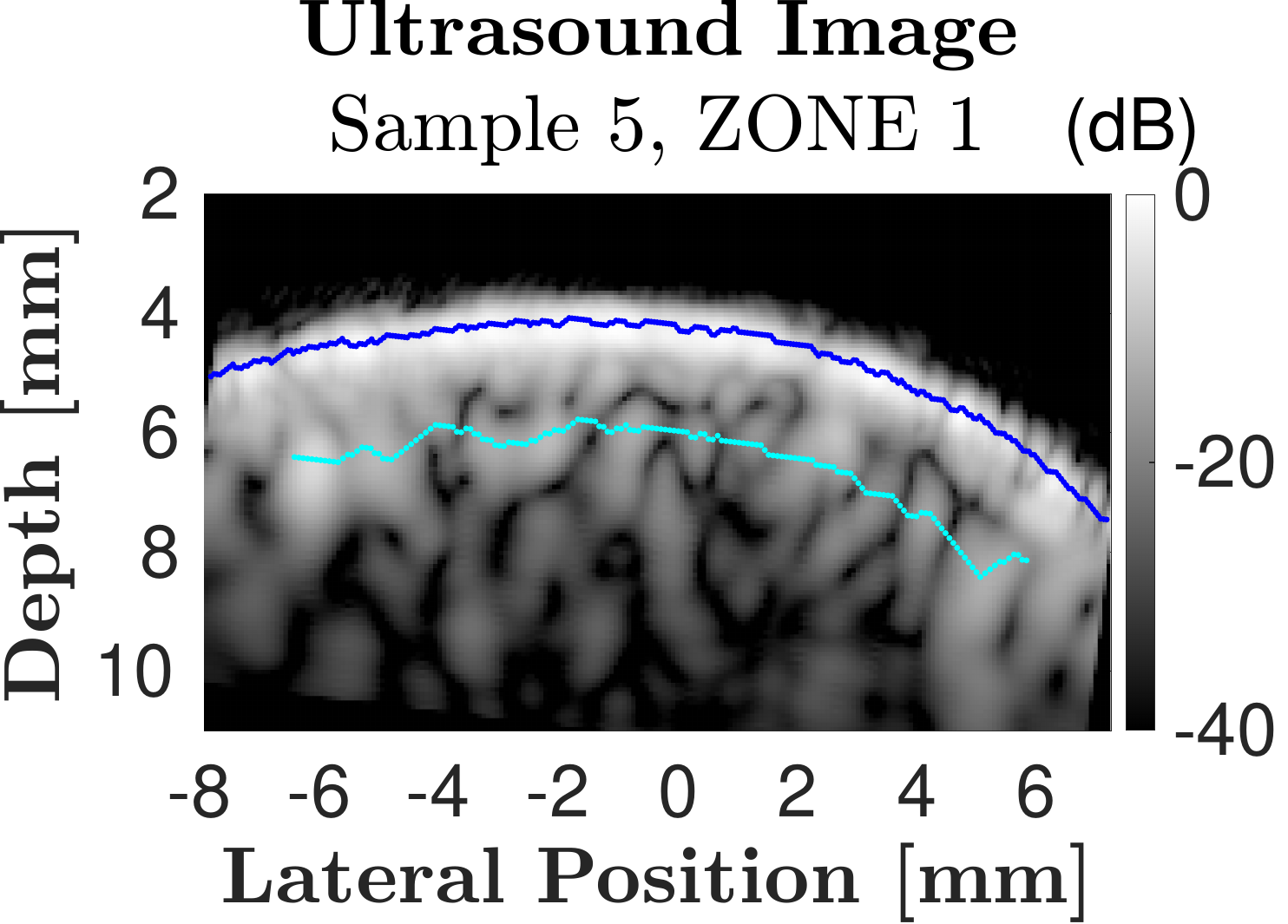}
            \caption{}
        \end{subfigure}   
        \begin{subfigure}{.24\linewidth}
            \includegraphics[width=\linewidth]{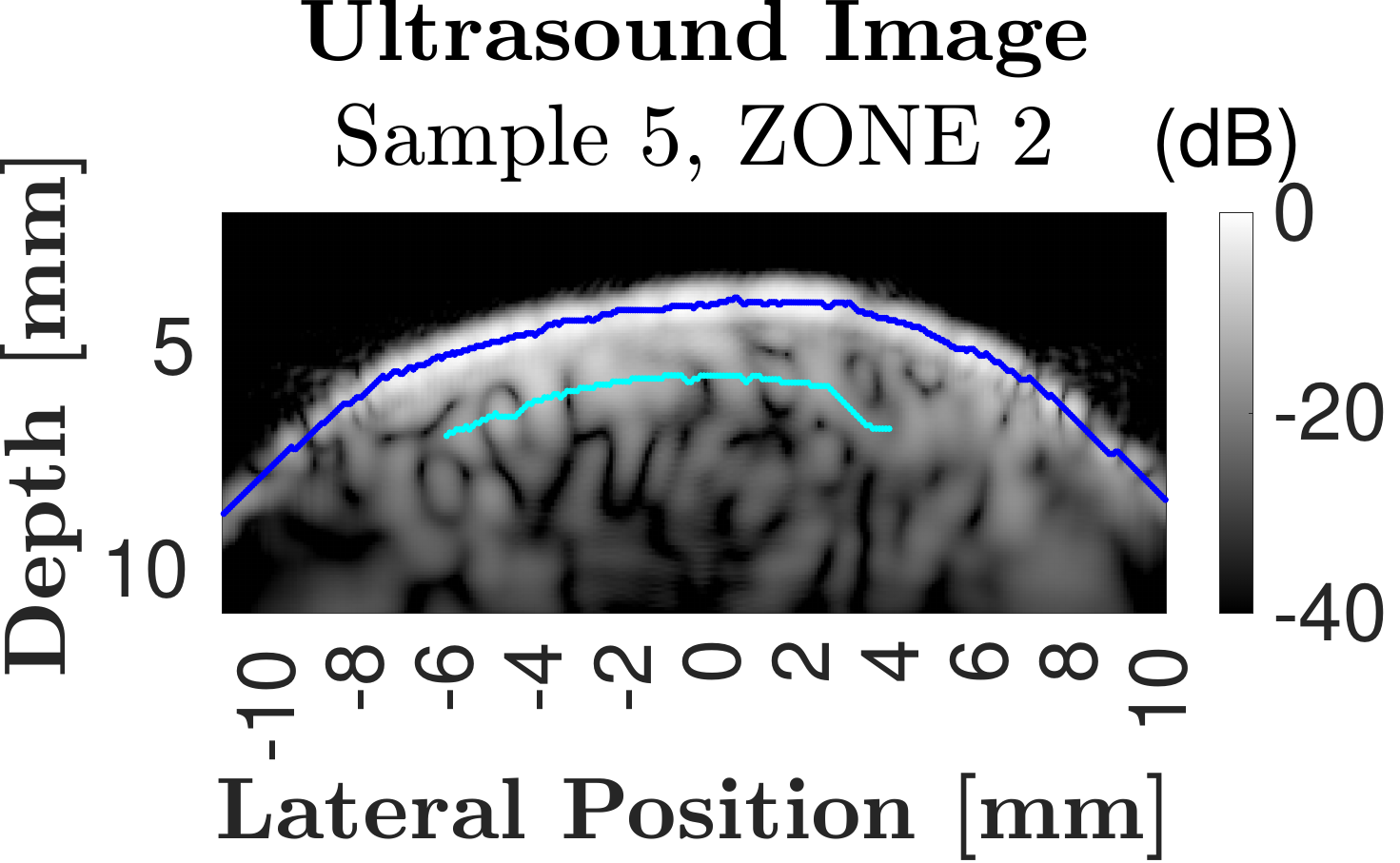}
            \caption{}
        \end{subfigure}
        \begin{subfigure}{.24\linewidth}
            \includegraphics[width=\linewidth]{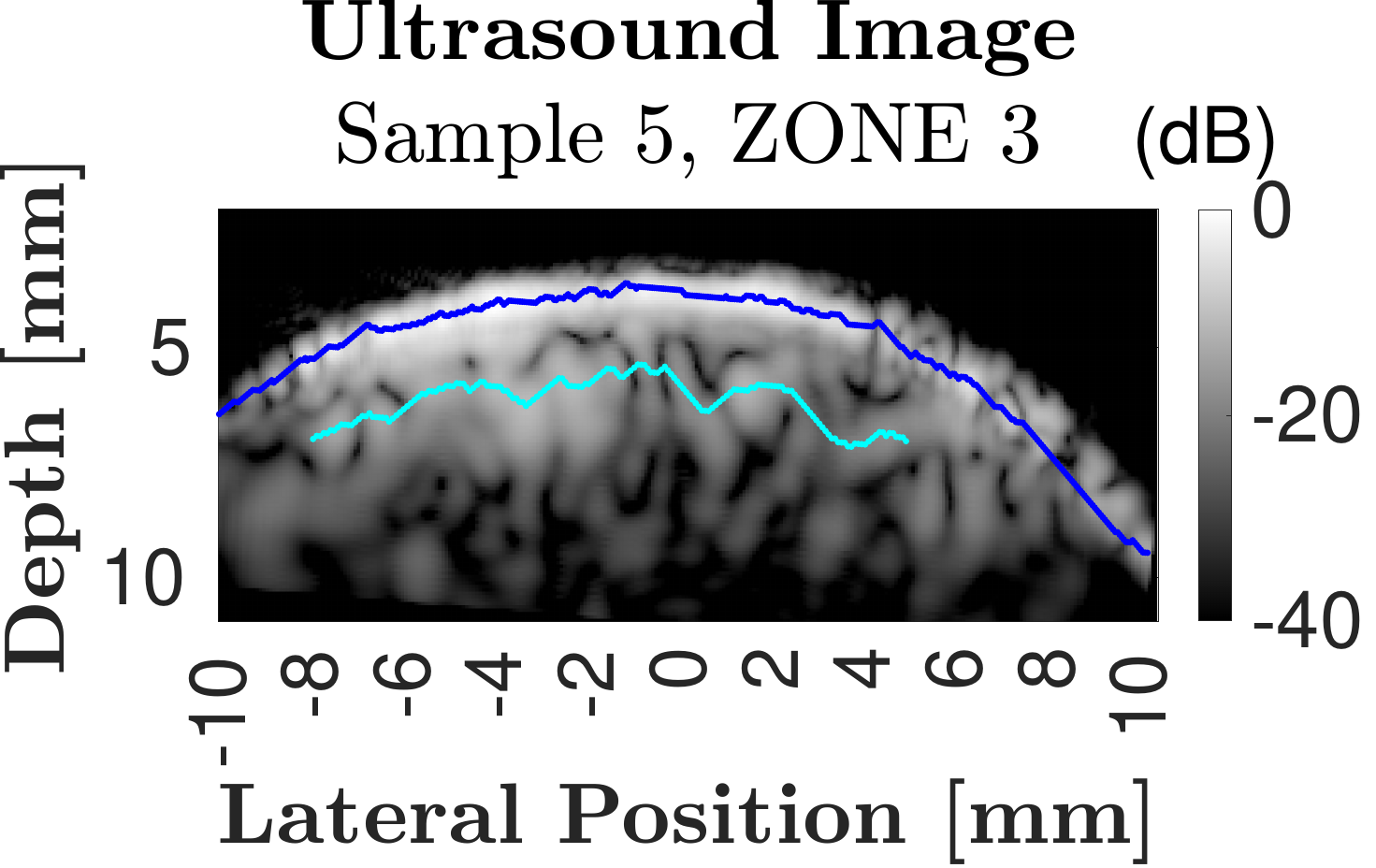}
            \caption{}
        \end{subfigure}   
        \begin{subfigure}{.24\linewidth}
            \includegraphics[width=\linewidth]{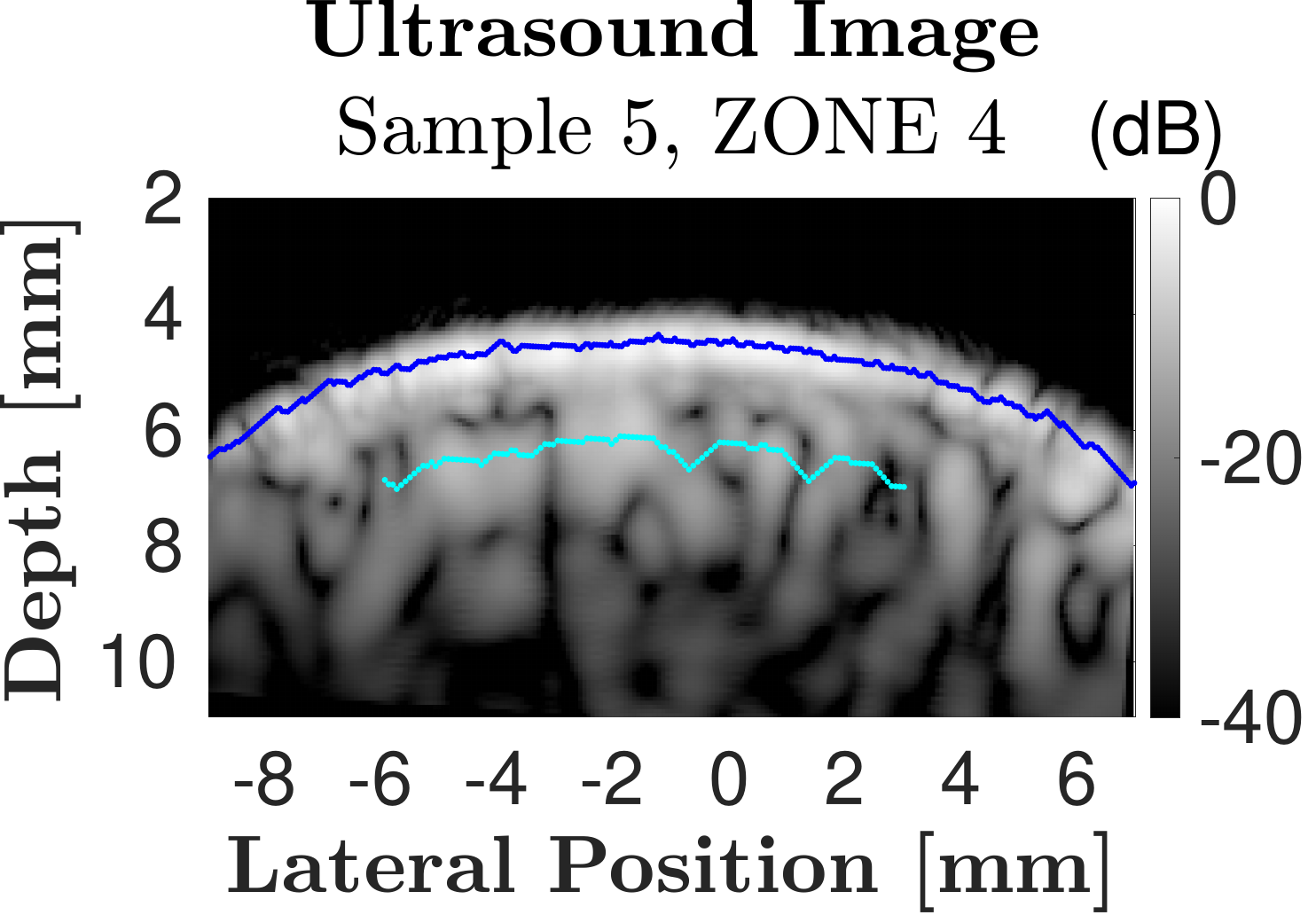}
            \caption{}
        \end{subfigure}   
        \begin{subfigure}{.24\linewidth}
            \includegraphics[width=\linewidth]{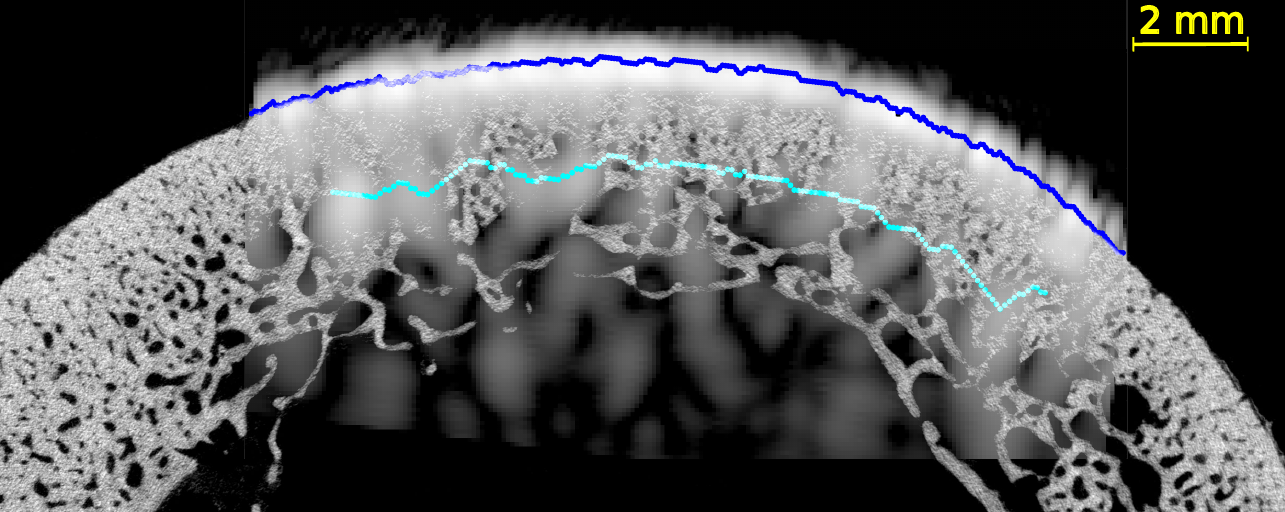}
            \caption{}
        \end{subfigure} 
        \begin{subfigure}{.24\linewidth}
            \includegraphics[width=\linewidth]{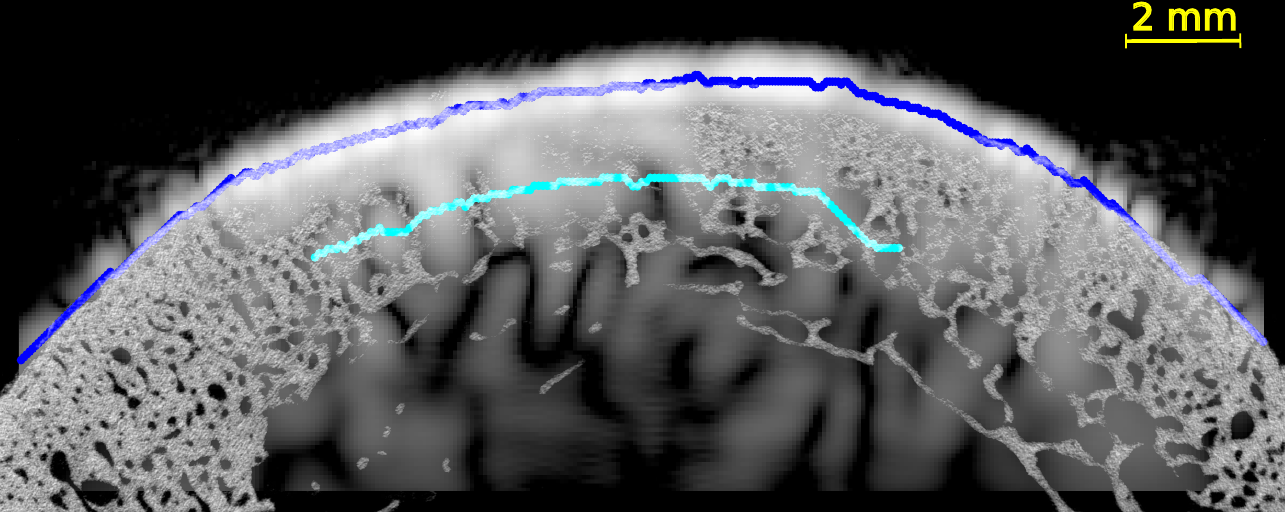}
            \caption{}
        \end{subfigure} 
        \begin{subfigure}{.24\linewidth}
            \includegraphics[width=\linewidth]{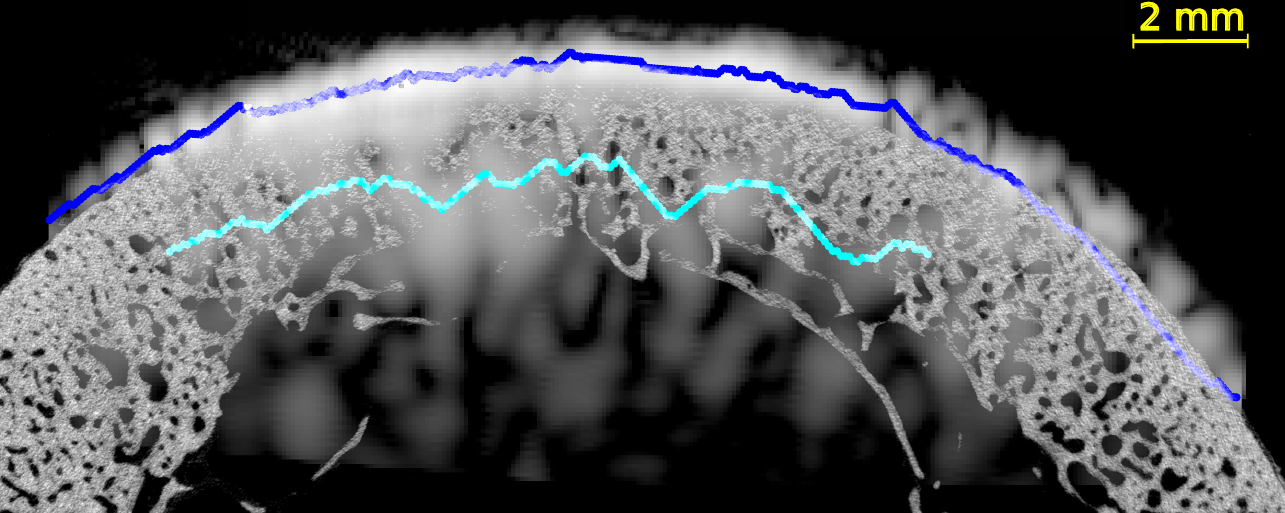}
            \caption{}
        \end{subfigure} 
        \begin{subfigure}{.24\linewidth}
            \includegraphics[width=\linewidth]{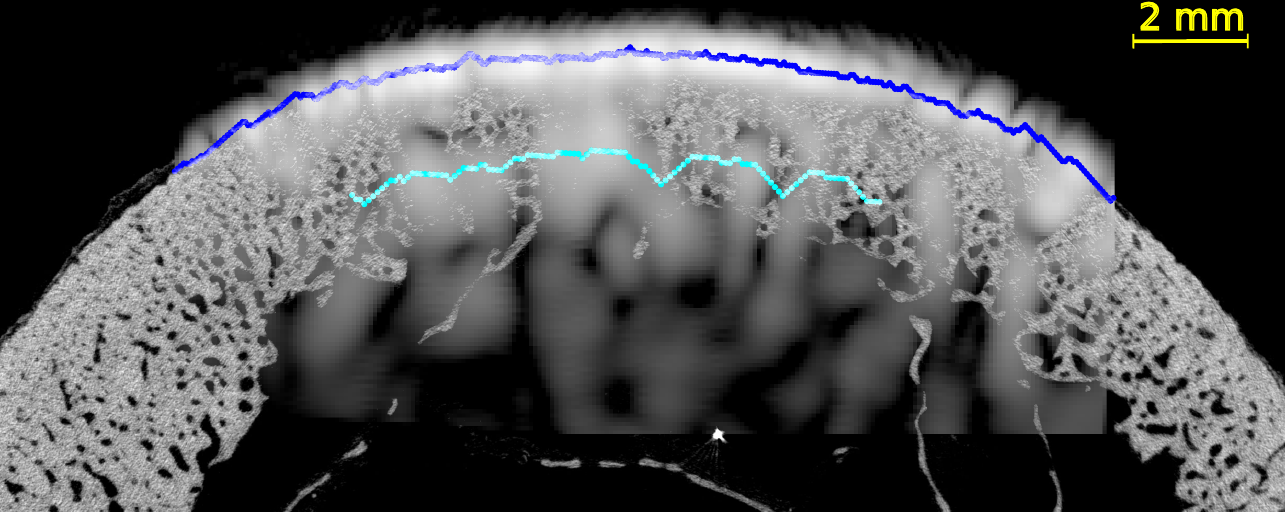}
            \caption{}
        \end{subfigure} 
        
        \caption{Same caption as Figure~\ref{supplementary_materials:sample_1}, applied to sample 5.}
        \label{supplementary_materials:sample_5}
    \end{figure}
    
    \end{landscape}
    }

\end{document}